\documentclass[final,3p,times]{elsarticle}

\usepackage{amssymb}

\usepackage{amsmath,bm}
\usepackage{caption}
\usepackage{subcaption}
\usepackage{hyperref}
\usepackage{multirow}
\usepackage{booktabs}
\usepackage{xcolor}
\usepackage[symbol]{footmisc}
\usepackage{array}
\usepackage{scalerel}
\newcommand{\EF}[1]{{\color{black}{#1}}}
\definecolor{blue2}{rgb}{0, 0.58, 1}
\usepackage{tikz}
\newcommand*\circled[1]{\tikz[baseline=(char.base)]{
            \node[shape=circle,draw,inner sep=0.5pt,draw=blue2,line width=0.5mm] (char) {#1};}}
\newcolumntype{C}[1]{>{\centering\let\newline\\\arraybackslash\hspace{0pt}}m{#1}}

\begin{document}

\begin{frontmatter}

\title{A Particle Finite Element Method based on Level--Set functions}

\author[a1]{Eduardo FERNÁNDEZ}
\author[a1]{Simon FÉVRIER}
\author[a1]{Martin LACROIX}
\author[a1]{Romain BOMAN}
\author[a1]{Luc PAPELEUX}
\author[a1]{Jean-Philippe PONTHOT}

\address[a1]{Aerospace and Mechanical Engineering Department, University of Liège, Belgium}

\begin{abstract}
Since the seminal work of Idelsohn, Oñate and Del-Pin (2004), the Particle Finite Element Method (PFEM) has relied on a Delaunay triangulation and the Alpha--Shape (AS) algorithm in the remeshing process. This approach guarantees a good quality of the Lagrangian mesh, but introduces a list of shortcomings that demand geometrical treatments tailored to each problem. In order to improve the remeshing process in PFEM, this work proposes the use of a Level--Set (LS) function instead of the Alpha--Shape algorithm. Since the Level--Set considers the boundary of the fluid and its interior, and not only a geometric criterion as does the Alpha--Shape, the proposed strategy (PFEM--LS) shows more robustness than the classical approach (PFEM--AS) owing to three main improvements. First, the LS function allows for a better control over the elements that are created during the fluid/fluid contact, which helps to reduce mass creation. Second, it helps to preserve the smoothness of the free surface and to reduce mass loss. Third, it allows the meshing of solitary particles that are detached from the free surface, which improves the representation of drops in PFEM. The methodology is presented and validated using free surface flow problems in 2D.
\end{abstract}


\end{frontmatter}


\section{Introduction}
\label{sec:Intro}

\begin{figure}[t] \captionsetup[sub]{font=normalsize}
	\centering 
	\includegraphics[width=1.00\linewidth]{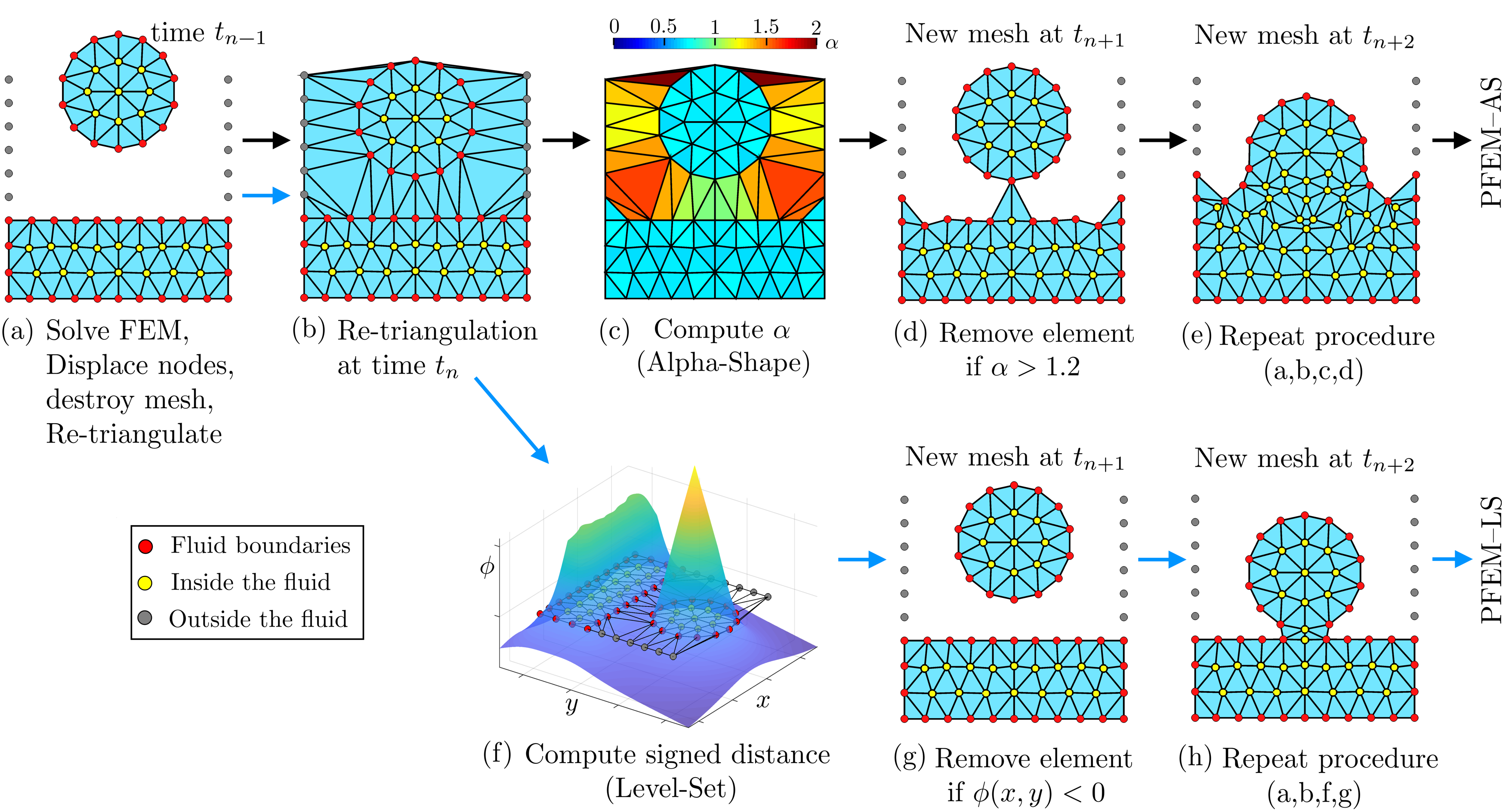}
	\caption{Illustration of the classic Particle Finite Element Method based on the Alpha--Shape (PFEM--AS), and the proposed scheme based on Level--Set functions (PFEM--LS). (a) A known state of the fluid at time $t_n$. (b) Delaunay triangulation of the updated configuration at $t_{n+1}$. (c) The $\alpha$ of each element is computed. (d) Triangulation after filtering elements using AS. (e) Solution obtained at the next time $t_{n+2}$. (f) Level--Set function built according to the boundaries and interior of the fluid. (g) Triangulation after filtering elements using LS. (h) Solution obtained at the next time step $t_{n+2}$.}
	\label{Fig:PFEM_Iluustration}
\end{figure}

The Particle Finite Element Method (PFEM) is a numerical technique for solving continuum mechanics problems dealing with large topology changes \citep{idelsohn2004particle}. PFEM uses a Lagrangian finite element mesh, which simplifies identification and tracking of body boundaries. For instance, in fluid flow simulation, PFEM allows for a simpler imposition of boundary conditions on the free surface compared to Eulerian approaches. However, PFEM requires a remeshing process to avoid degradation of the Lagrangian mesh, which is applied either after each time step or when some mesh quality criterion demands it. This remeshing operation is summarized as follows. First, the triangulation of the preceding time step is destroyed but nodal state variables are retained. Then, a Delaunay triangulation is applied on the set of nodes (or particles), as shown in Fig.~\ref{Fig:PFEM_Iluustration}b. Subsequently, elements outside the body are removed using the Alpha--Shape (AS) algorithm \citep{edelsbrunner1994three}. This algorithm computes for each element a geometric feature ($\alpha$), indicator of the size and distortion of the element, and compares it to a user-defined threshold ($\alpha_\mathrm{max}$). If larger than the threshold ($\alpha > \alpha_\mathrm{max}$), the element is removed from the triangulation, as shown in Figs.~\ref{Fig:PFEM_Iluustration}c and \ref{Fig:PFEM_Iluustration}d. Although PFEM was originally designed for the simulation of free surface flows, the seminal idea based on the Alpha--Shape (AS) has been extended to many other problems such as plasticity \citep{carbonell2020modelling}, fluid-structure interaction \citep{cerquaglia2019fully,meduri2022lagrangian} and phase change \citep{bobach2021phase} among others \citep{sengani2020review,cremonesi2020state}.

Despite the advances in PFEM, there are still challenges that have not been overcome. Among them is the mass variation of the system, whose main contributor is the remeshing due to the discrete addition and removal of finite elements from one time step to the other \citep{franci2017effect}. In this regard, two situations can be highlighted that greatly affect mass conservation. The first is detachment and contact, either between bodies or of the body with itself. Since the detachment/contact is modelled by elements vanishing/appearing in the remeshing process (as the fluid/fluid contact illustrated in Fig.~\ref{Fig:PFEM_Iluustration}d), problems with significant splashing and breaking waves exhibit large mass variations. The second situation affecting mass conservation is the flow over a dry solid surface with non-slip boundary condition, which is modelled by discrete addition/removal of elements. Given that mass variation is proportional to the element size, mesh refinement is effective in addressing such problems \citep{franci2017effect,cerquaglia2019development}. Also, free-slip boundary conditions have been shown to improve mass conservation in PFEM \citep{cerquaglia2017free}, because in such a case, the flow over a surface is defined by the displacement of nodes and not by the addition of elements.   

However, the aforementioned problems are worsened by the use of the Alpha-Shape (AS) algorithm. Given that the AS criterion focuses on the shape and size of each element individually and does not consider the topological information of the fluid (boundaries, inner and outer zones), geometrical details of the size of a finite element cannot be captured. To circumvent this issue, authors have resorted to constrained Delaunay triangulations \citep{carbonell2020modelling,rodriguez2017continuous}. However, its applicability has been limited to plasticity, since the constrained Delaunay poses difficulties for merging and separating bodies. Other authors have implemented local mesh refinement strategies to reduce the geometrical errors associated to the Alpha--Shape algorithm \citep{bobach2021phase,falla2022Mesh}. This strategy is effective in reducing mass variations, but increases computational time significantly, either because of an increase in the number of degrees of freedom or because the time step becomes smaller with mesh refinement. As an alternative idea, this work proposes the use of Level-Set functions.

The Level--Set (LS) method proposed by Osher and Sethian \citep{osher1988fronts} provides an implicit description of moving fronts. The idea has been progressively adopted in a wide variety of fields such as image processing \citep{cremers2007review}, computer vision and graphics \citep{osher2003geometric}, and structural optimisation \citep{van2013level} among many others. In the numerical simulation domain, the LS method is convenient for tracking dynamic interfaces such as cracks \citep{stolarska2001modelling}, fluid-structure interfaces \citep{becker2015unified}, and discontinuities \citep{grooss2006level,gibou2018review}. In most of these applications, LS is used as a curve (or surface) that delimits the cut of elements or that highlights the zone where discontinuities demand mesh adaptations \citep{chen2018thermomechanical}. Differently, the idea of the present work is to use the LS function to filter the Delaunay triangulation during the remeshing process of PFEM, and thus avoid the dependence on the AS algorithm, as illustrated in Figs.~\ref{Fig:PFEM_Iluustration}f-\ref{Fig:PFEM_Iluustration}h. Since LS considers the inner, outer and boundary zone of the fluid, the LS-based remeshing process is information-enriched, which permits to better capture topological details that are of the order of a finite element. 

An interesting aspect of PFEM is the possibility to simulate the generation of drops. Although the representativeness of drops remains at the macro scale, good agreement with experimental tests has been observed \citep{idelsohn2004particle,cerquaglia2017free}. The approach to modelling drops \EF{at the macro scale} in PFEM is rather standardised in the literature and consists of representing the drop by the particle (or node) that has been disconnected from the free surface. The trajectory of the drop is defined by the free--fall equations, considering as initial condition the velocity and acceleration of the particle at the moment of detachment. Although nodal drops are massless and eventually return to the fluid, they contribute to significant mass variations. On the one hand, the detachment of a particle removes mass by deleting the element that retains the particle on the free surface. On the other hand, mass is created as the particle reaches the free surface since elements are created during the drop/fluid contact \citep{franci2017effect}. The net mass variation will depend on the remeshing parameters ($\alpha_\mathrm{max}$, among others) and the imposed mesh size. However, to the best of our knowledge, the representativeness of a drop by means of a node has not been analysed, presumably because of the difficulty in capturing small geometrical details in PFEM--AS. Given that LS provides better resolution during the remeshing process than AS, this work proposes to mesh the drops once the particle is detached from the free surface. 

The LS-based remeshing process for PFEM is assessed using 2D benchmark problems that feature substantial sloshing and wave breaking, including both free-slip and no-slip boundary conditions. In addition, the treatment of drops in PFEM--LS is compared with the classical scheme that discretises the drop with a massless node. Results indicate that the proposed methodology significantly reduces the mass variation in PFEM because LS offers better control of the elements created during fluid/fluid contact, and because it helps to avoid the degradation of the free surface. In addition, meshed drops provide better mass and energy conservation than nodal drops in PFEM--LS.

The remainder of this manuscript is as follows. Section 2 introduces the classical Particle Finite Element Method based on the Alpha--Shape algorithm (PFEM--AS). Section 3 presents the Level--Set function adopted in this work and the proposed strategy for utilization in the PFEM framework. Section 4 gathers the numerical examples and discussion, while Section 5 summarizes the final conclusions of this work.

\section{PFEM framework}\label{sec:PFEM-AS}

\begin{figure}[t] \captionsetup[sub]{font=normalsize}
	\centering 
	\includegraphics[trim=2 420 10 10,clip,width=1.00\linewidth]{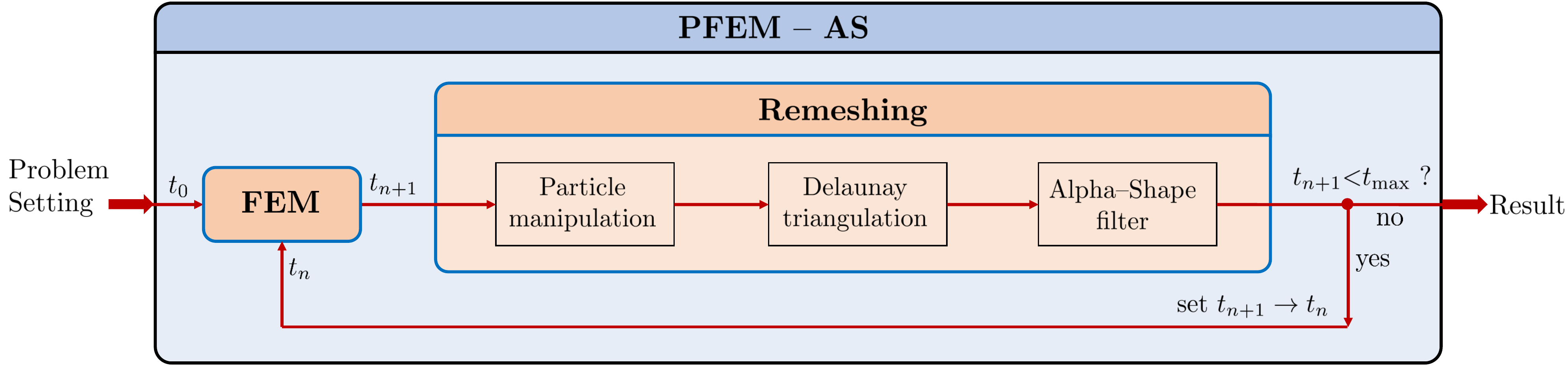}
	\caption{Workflow of the Particle Finite Element Method (PFEM) based on the Alpha--Shape (AS) algorithm during the remeshing process. The workflow is denoted as PFEM--AS in this work.}
	\label{Fig:PFEM_schemes_pdf}
\end{figure}

In a simplified fashion, PFEM is composed of two steps during the simulation of dynamic problems. Considering a temporal discretisation, where nodal state variables and positions are known at time $t_n$, the first step is to solve the Lagrangian system of equations using the Finite Element Method (FEM), and obtain nodal states and position at time $t_{n+1}$. The second step consists of remeshing the fluid domain in its updated configuration to improve mesh quality. The PFEM workflow is depicted in Fig.~\ref{Fig:PFEM_schemes_pdf}, under the name PFEM--AS to highlight that the remeshing process depends on the Alpha--Shape algorithm. The proposed strategy for PFEM using Level--Set (LS) affects the remeshing step only and does not depend on the adopted FEM. Therefore, the spatio-temporal discretisation of the governing equations is described in brief, as it has no implication in the methodology presented in this work.

\subsection{Finite element model}

The proposed remeshing scheme for PFEM is explained and assessed using free surface flow problems. Specifically, incompressible Newtonian fluids are considered in this work. The governing equations for a continuum read as follows:
\begin{subequations} \label{EQ:Continuum_Equations_VP}
\begin{align} 
\rho \: \frac{d \mathbf{v}}{dt} - \mu \Delta \mathbf{v}	+ \nabla \mathrm{p} & = \mathbf{f}  \label{EQ:Continuum_Equations_VP_a}
\\[1ex] 
\nabla \cdot \mathbf{v} & = 0  \label{EQ:Continuum_Equations_VP_b}
\end{align}
\end{subequations}
\noindent where $\rho$ is density, $d$ is the Lagrangian derivative, $t$ is time and $\mathbf{v}$ is the velocity vector. The constitutive equation for the Newtonian fluid is described by the dynamic viscosity $\mu$ and the pressure $\mathrm{p}$. The body force vector is denoted as $\mathbf{f}$. The momentum \eqref{EQ:Continuum_Equations_VP_a} and continuity \eqref{EQ:Continuum_Equations_VP_b} equations are discretised using spatial FEM-Galerkin and temporal discretisations. The system is solved in a monolithic velocity-pressure formulation that uses implicit Backward Euler as time integration scheme. Given the LBB (Ladyzhenskaya-Babu$\check{\text{s}}$ka-Brezzi) condition to be satisfied and the use of linear elements with the same order of interpolation for velocity and pressure, PSPG stabilization (Pressure-Stabilizing Petrov-Galerkin) is adopted. The discretised governing equations are omitted in this manuscript and the reader is referred to \citep{cerquaglia2019development,fernandez2022} for further details.

\subsection{Remeshing using Alpha-Shape (AS)}

\begin{figure}[t!] \captionsetup[sub]{font=normalsize}
	\centering 
	\includegraphics[width=0.90\linewidth]{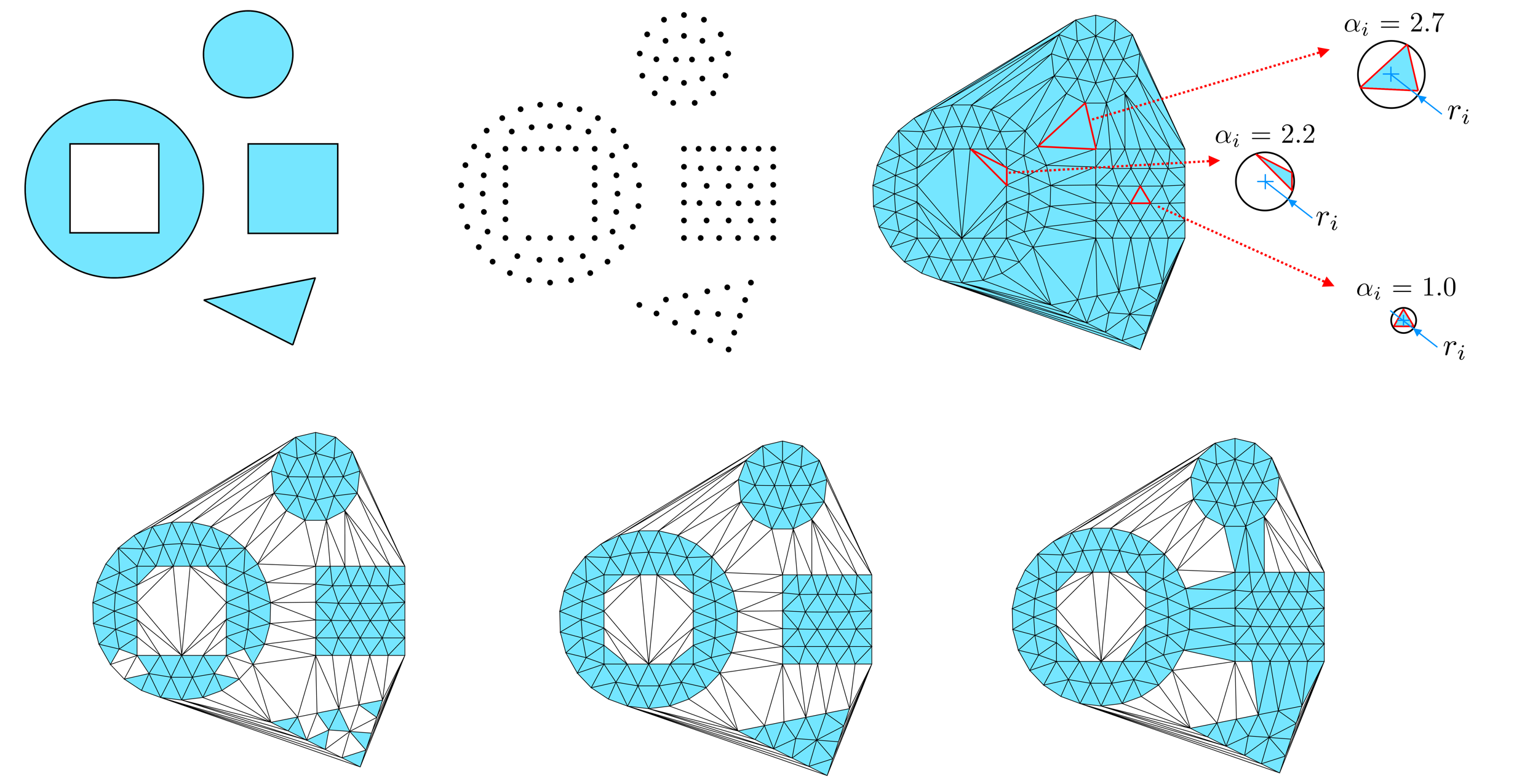}
	\\
	\vspace{-41mm}
	\hspace{-25mm} 
	\footnotesize{(\textbf{a}) Bodies to be discretised} 
	\hspace{15mm} 
	(\textbf{b}) Nodal discretisation
	\hspace{15mm} 
	(\textbf{c}) Delaunay triangulation
	\\
	\vspace{38mm}
	\hspace{-10mm} 
	(\textbf{d}) $\alpha_\mathrm{max} = 0.8$
	\hspace{22mm} 
	(\textbf{e}) $\alpha_\mathrm{max} = $ [1.1 -- 1.4]
	\hspace{20mm} 
	(\textbf{f}) $\alpha_\mathrm{max} = $ 1.6
	\caption{Remeshing process using the Alpha--Shape algorithm and the obtained discretisations using different values of $\alpha_\mathrm{max}$.}
	\label{Fig:AS_Iluustration}
\end{figure}

After solving the system of governing equations for time $t_{n+1}$ and displacing the nodes, some elements may undergo significant deformations that may compromise the convergence of the next time step. Therefore, the fluid is remeshed in its updated configuration if some mesh quality criterion requires it. The remeshing process that has proven efficiency in PFEM resorts to a Delaunay triangulation followed by the Alpha-Shape (AS) algorithm. The first creates a mesh of triangular elements that connects all particles, generating a convex hull featuring a single body, as illustrated in Fig.~\ref{Fig:AS_Iluustration}c. The Alpha--Shape algorithm is then used as a filter to only retain the elements belonging to the fluid.

The AS algorithm computes for each element a parameter $\alpha$ that corresponds to the ratio between the radius of the circumcircle of the element and the characteristic size imposed for that element. Therefore, for an element with index $i$, its $\alpha_i$ parameter is defined as follows:
\begin{equation}
\alpha_i = \frac{r_i}{h_\mathrm{u}} \;\; , \;\; i \in \mathbb{I}_\mathrm{D}
\end{equation}

\noindent where $r_i$ is the circumradius of element $i$, $h_\mathrm{u}$ is a user-defined characteristic element size, and $\mathbb{I}_\mathrm{D}$ is the set of indices of the elements generated by the Delaunay triangulation. Large values of $\alpha_i$ indicate either that the element is big with respect to $h_\mathrm{u}$ or that it is highly distorted, as shown in Fig.~\ref{Fig:AS_Iluustration}c. This parameter $\alpha$ has been used in the PFEM literature to filter the Delaunay triangulation and retain only those elements that belong to the fluid, as follows:
\begin{equation}
\mathbb{I}_\mathrm{AS} = \{\:i \in \mathbb{I}_\mathrm{D} \; \mid \; \alpha_i \leq \alpha_\mathrm{max} \: \}
\end{equation}

\noindent where $\mathbb{I}_\mathrm{AS}$ is the set of indices of the elements that are retained for discretising the fluid, and $\alpha_\mathrm{max}$ is a user-imposed threshold. Several studies suggest that $1.1 \leq \alpha_\mathrm{max} \leq 1.4$ is a good range of values to recover the fluid discretisation, as illustrated in Figs.~\ref{Fig:AS_Iluustration}d--\ref{Fig:AS_Iluustration}f. 

As $\alpha_i$ depends only on geometrical information of the element $i$, the Alpha--Shape algorithm demands a negligible computational effort and it allows efficient parallelisation. However, its effectiveness in PFEM depends on ensuring that the fluid is discretised with elements with small skewness and aspect ratio, otherwise these will be removed by the Alpha--Shape algorithm. For this reason, before performing the re-triangulation, a homogeneous distribution of the particles discretising the fluid must be guaranteed. To this end, the following two manipulations are performed on the particles: If an element is larger than $\omega \: h_u^2$, a particle is added at the centroid of the element, and if the distance between two particles is less than $\gamma \: h_u$, one is removed from the model. These two manipulations seek homogeneity of element sizes in the Delaunay triangulation to ensure good convergence in the finite element analysis of the next time step and try to avoid undue element removal by AS \citep{cerquaglia2019development,cremonesi2020state}. In this work, $\omega = 0.7$, $\gamma = 0.4$ and $\alpha_\mathrm{max} = 1.2$ in all the numerical examples.

\begin{figure}[t!]
\captionsetup[subfigure]{labelformat=empty}
\centering 
	\includegraphics[trim=0 0 0 0,clip,width=1.00\linewidth]{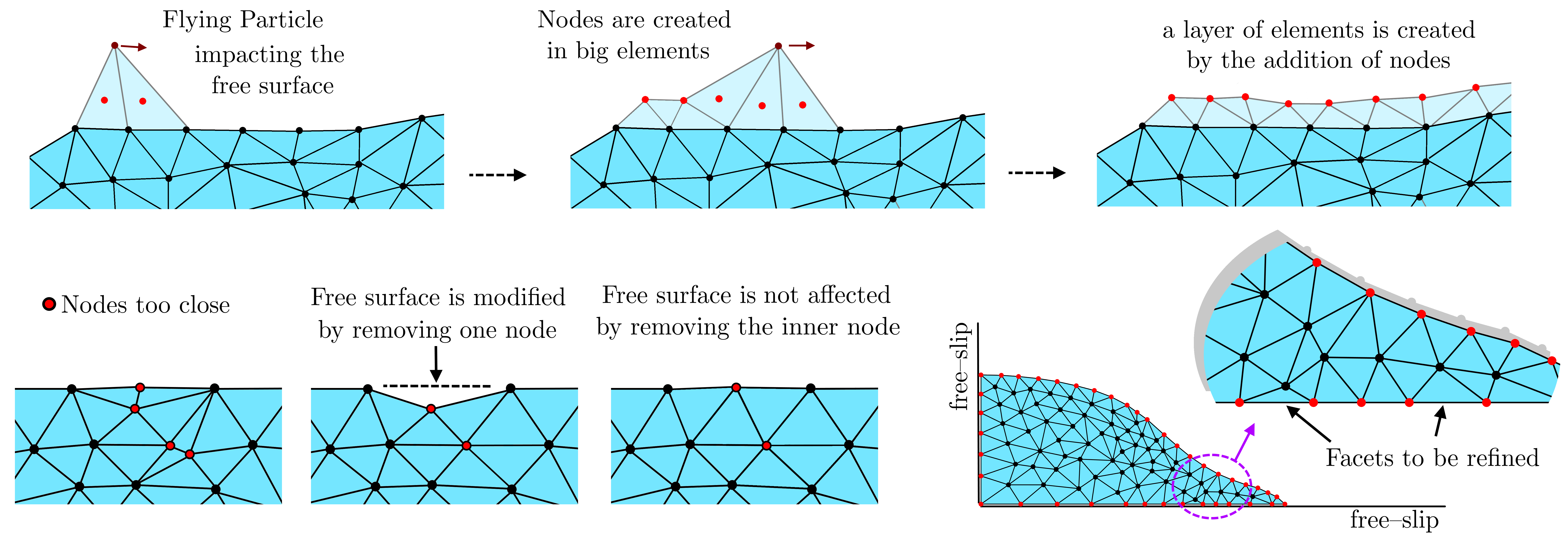}
	\begin{subfigure}[b]{0.0\textwidth}	
		\caption{}\label{Fig:mani_cares_a}
	\end{subfigure}
	~
	\begin{subfigure}[b]{0.0\textwidth}	
		\caption{}
		\label{Fig:mani_cares_b}
	\end{subfigure}
	~
	\begin{subfigure}[b]{0.0\textwidth}	
		\caption{}\label{Fig:mani_cares_c}
	\end{subfigure}
	\vspace{-65mm}\\
	\hspace{-160mm}
	\footnotesize{(\textbf{a})}
	\vspace{25mm}\\ 
	\hspace{-60mm} (\textbf{b}) \hspace{95mm} (\textbf{c})
	\vspace{25mm}
	\\
	\caption{(a) Illustration of a mass creation mechanism when particles are added at the centroid of elements lying on the free--surface. (b) Particle removal when they are too close, and mass loss mechanism when a particle is removed from the free surface. (c) A snapshot of the dam break with free-slip condition and two facets that must be refined for correct use of Alpha--Shape.}
\label{Fig:mani_cares}
\end{figure}

The two manipulations on the particles arrangement are subject to some restrictions, otherwise mass creation can be induced. For example, the addition of particles on the free surface elements must be avoided, otherwise mass addition is promoted in the system, as illustrated in Fig.~\ref{Fig:mani_cares_a}. In addition, removal of particles belonging to the free surface must be avoided, otherwise mass is removed from the system and the position of the free surface is perturbed, as illustrated in Fig.~\ref{Fig:mani_cares_b}. The free--slip condition also demands for specific arrangements. For example, when the distance between two slip particles is greater than $\beta \: h_\mathrm{u}$, a particle must be added in between to avoid excessive element distortion, as illustrated in Fig.~\ref{Fig:mani_cares_c}. In this work, all numerical examples that include a free--slip boundary condition use $\beta = 1.4$.

\begin{figure}[t!] \captionsetup[sub]{font=normalsize}
	\centering 
	\includegraphics[width=0.90\linewidth]{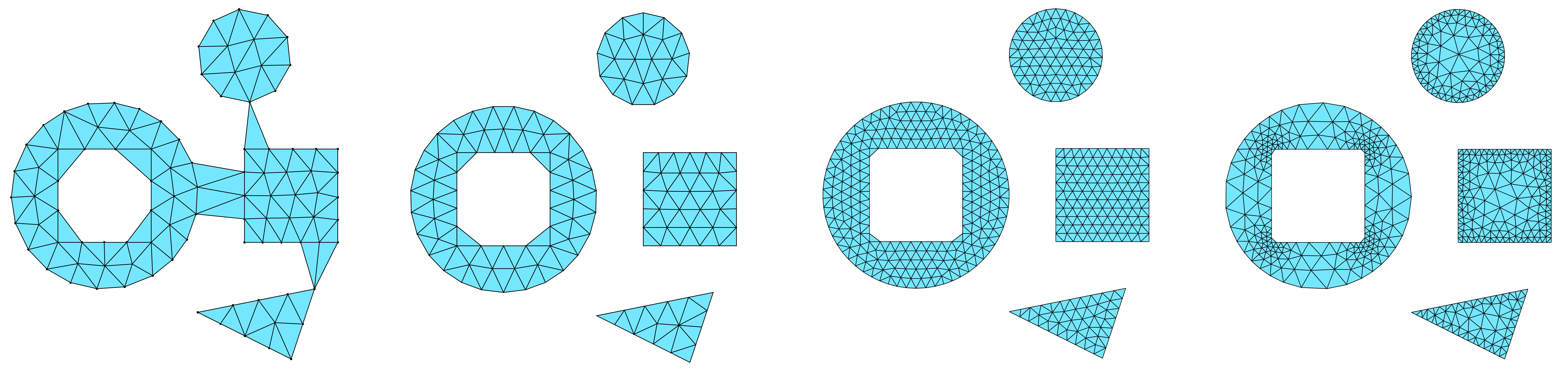}
	\\
	\vspace{-38mm}
	\hspace{-26mm} 
	\footnotesize{(\textbf{a})} 
	\hspace{34mm} 
	(\textbf{b}) 
	\hspace{34mm} 
	(\textbf{c}) 
	\hspace{34mm} 
	(\textbf{d}) 
	\\
	\vspace{33mm}
	\caption{Four discretisations obtained with Alpha--Shape. (a-c) Discretisations using a uniform element size. (d) A discretisation with mesh refinement around the reentrant corners. The fluid domain that is discretised is shown in Fig.~\ref{Fig:AS_Iluustration}a.} 
	\label{Fig:AS_Iluustration_refinement}
\end{figure}

Manipulations in particles arrangement may be numerous and some may be tailored to a specific problem in order to avoid drawbacks related to the use of AS. For this reason, it is often difficult to replicate PFEM results from the literature, as each remeshing implementation may be equipped with different conditions for an effective use of the Alpha--Shape algorithm. However, even equipped with sophisticated algorithms, current remeshing processes based on Alpha--Shape have difficulties in capturing topological details that are of the order of a finite element, such as holes, gaps, or sharp reentrant corners. The challenge is that any element that meets the distortion and size criteria is accepted by AS, even if the element does not actually belong to the body of interest. This can be seen in the re-entrant corners of the disk with a square--shaped hole depicted in Fig.~\ref{Fig:AS_Iluustration}e, which appear meshed even with low alpha values, as shown in Fig.~\ref{Fig:AS_Iluustration}d. This drawback can be addressed simply by mesh refinement, as shown in Fig.~\ref{Fig:AS_Iluustration_refinement}a--\ref{Fig:AS_Iluustration_refinement}c. Also, local mesh refinement can be applied to avoid an increase in computational resources, as shown in Fig.~\ref{Fig:AS_Iluustration_refinement}d. However, mesh refinement, even locally, raises computational resources because the time step must inevitably be reduced to ensure good convergence of the finite element analysis.

\begin{figure}[t!]
\centering 
\hspace{30mm}\includegraphics[trim=0 0 0 0,clip,width=0.70\linewidth]{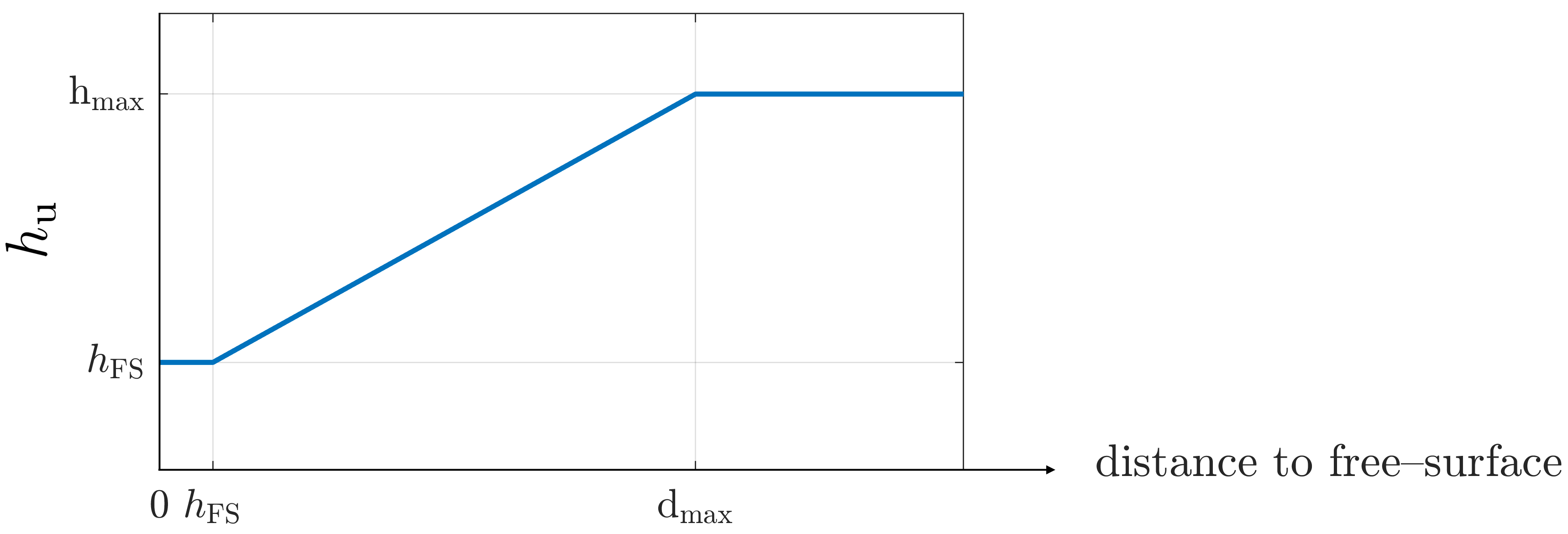}
\caption{Element size ($h$) in terms of the distance to the free--surface.}
\label{Fig:meshref}
\end{figure}

Non--uniform element size in PFEM is achieved by assigning the desired characteristic element size ($h_\mathrm{u}$) according to the zone to be refined or coarsened. Thus $h_\mathrm{u}$ is a function that depends on some refinement criterion, while $\alpha_\mathrm{max}$ is kept as a global fixed value. In this work, all numerical examples use non-uniform element sizes. To this end, minimum and maximum element sizes are prescribed, which are denoted as $h_\mathrm{FS}$ and $h_\mathrm{max}$, respectively. The minimum element size ($h_\mathrm{FS}$) is imposed on the free--surface, while the maximum element size ($h_\mathrm{max}$) is imposed from a distance $\mathrm{d}_\mathrm{max}$ to the free--surface. As illustrated in Fig.~\ref{Fig:meshref}, the element size between the free surface and  $\mathrm{d}_\mathrm{max}$ is defined by a linear interpolation, therefore $h_\mathrm{u}(h_\mathrm{FS},h_\mathrm{max},\mathrm{d}_\mathrm{max})$. For space discretisation, numerical examples in Section \ref{sec:Examples} report $h_\mathrm{FS}$, $h_\mathrm{max}$ and $\mathrm{d}_\mathrm{max}$.

Although mesh refinement reduces the geometric errors of the AS criterion, it is not possible to completely avoid them since a null element size cannot be imposed. However, it is straightforward to conceive that the drawbacks associated to the dependency of the Alpha--Shape algorithm can be addressed by accounting for the topological information of the body, i.e.~the interior, exterior and its boundaries. For example, the elements present in the 4 reetrant corners in Fig.~\ref{Fig:AS_Iluustration_refinement} should be excluded from the model because they are on the outside of the body. However this information is not known by the Alpha--Shape algorithm and thus, the use of Level--Set functions is proposed instead.

\section{Level-Set in the remeshing process of PFEM}\label{sec:PFEM-LS}

The proposed methodology for the remeshing process of PFEM is to use an implicit function that maps the topology of the computational domain, describing the interior of the fluid, its boundary and exterior, and to use this function to filter the Delaunay triangulation. The adopted function describes the domain as follows:
\begin{subequations}
\begin{align}
\phi(\mathbf{x}) < 0 & \text{\hspace{5mm},\hspace{5mm} if $\mathbf{x} \notin \Omega $}
\\
\phi(\mathbf{x}) = 0 & \text{\hspace{5mm},\hspace{5mm} if $\mathbf{x} \in \Gamma $}
\\
\phi(\mathbf{x}) > 0 & \text{\hspace{5mm},\hspace{5mm} if $\mathbf{x} \in \Omega $}
\end{align}
\end{subequations}

\noindent where $\phi$ is the implicit function and $\mathbf{x}=[x\;,\;y]$ is a vector with Cartesian coordinates. As illustrated in Fig.~\ref{Fig:PFEM_LS_build}a, $\Gamma$ and $\Omega$ represent the boundary and interior of the fluid, respectively. The function $\phi(\mathbf{x})$ is built by fitting a function to $N_\mathrm{LS}$ points describing the topology of the fluid. Following the methodology of \citep{carr2001reconstruction,belytschko2003structured}, a radial basis function is used for the fitting. This is defined as the sum of a linear polynomial and $N_\mathrm{LS}$ parameters that scale a radial function, as follows:
\begin{equation}
\phi(\mathbf{x}) = c_0 + c_x\:x + c_y\:y + \displaystyle\sum_{i=1}^{N_\mathrm{LS}} \lambda_i \: \| \mathbf{x} - \bar{\mathbf{x}}_i\|
\end{equation}
\noindent where $\lambda_i$ and $\bar{\mathbf{x}}_i$ are, respectively, arbitrary parameters and coordinates chosen to best fit the function. The coefficients of the low-order polynomial are $c_0$, $c_x$ and $c_y$, and $\| \cdot \|$ denotes the Euclidean norm. Thus, the chosen radial function is a biharmonic spline \citep{carr2001reconstruction}. The advantage of this function is that $\phi(\mathbf{x})$ provides two useful pieces of information for the remeshing process, the topology (given by the sign), and the distance to the boundaries (given by the magnitude). 

The accuracy of the Level--Set function $\phi(\mathbf{x})$ to represent the topological details of the domain will depend on the position and \EF{amount} ($N_\mathrm{LS}$) of points chosen for the fitting. The prescribed values at the coordinates of the fitting points ($\bar{\mathbf{x}}$) are denoted as $\bar{\phi}$. Thus, the Level--Set function must satisfy:
\begin{equation}\label{EQ:phi_cond_1}
\phi(\bar{\mathbf{x}}_i) = \bar{\phi}_i \;\;\;,\;\;\; i=1,\;...\;,N_\mathrm{LS}
\end{equation}

Furthermore, orthogonality conditions must be satisfied \citep{carr2001reconstruction}:
\begin{equation}\label{EQ:phi_cond_2}
\displaystyle\sum_{i}^{\EF{N_\mathrm{LS}}} \lambda_i = 
\sum_{i}^{\EF{N_{LS}}} \lambda_i x_i
= 
\sum_{i}^{\EF{N_\mathrm{LS}}} \lambda_i y_i = 0
\end{equation}

Eqs.~\eqref{EQ:phi_cond_1} and \eqref{EQ:phi_cond_2} can be assembled in matrix form as follows:

\begin{equation}\label{EQ:LS_system}
\left[
\:
\begin{matrix}
\mathbf{A} \;&\; \mathbf{B}
\\[2ex]
\mathbf{B}^\intercal \;&\; \bm{0}
\end{matrix}
\:
\right]
\;
\left[
\:
\begin{matrix}
\bm{\lambda}
\\[2ex]
\bm{c}
\end{matrix}
\:
\right]
=
\left[
\:
\begin{matrix}
\bar{\bm{\phi}}
\\[2ex]
\bm{0}
\end{matrix}
\:
\right]
\end{equation}

\noindent where the first row represents the condition of Eq.~\eqref{EQ:phi_cond_1} and the second row contains the orthogonality conditions. For that, the vectors must be defined as $\bm{\lambda}=$ [$\lambda_1$, ... , $\lambda_{N_\mathrm{LS}}$]$^\intercal$, $\bm{c}$ = [$c_0$ , $c_x$ , $c_y$]$^\intercal$, $\bar{\bm{\phi}}=$ [$\bar{{\phi}}_1$, ... , $\bar{{\phi}}_{N_\mathrm{LS}}$]$^\intercal$, and the components of $\mathbf{A}$ and rows of $\mathbf{B}$ must be defined as: 
\begin{subequations}
\begin{align}
\mathrm{A}_{i,j} &=  \| \bar{\mathbf{x}}_i - \bar{\mathbf{x}}_j\| \;\;\;\;\; , \;\; i,j=1, \: ... \: , N_\EF{\mathrm{LS}}
\\[2ex]
\mathrm{B}_{i} &= [1 \;,\; \bar{x}_i \;,\; \bar{y}_i] \;\; , \;\;\;\; i = 1,\: ... \: , N_\EF{\mathrm{LS}}
\end{align}
\end{subequations}
\vspace{1mm}

\begin{figure}[t] \captionsetup[sub]{font=normalsize}
	\centering 
	\includegraphics[trim=2 90 10 350,clip,width=1.00\linewidth]{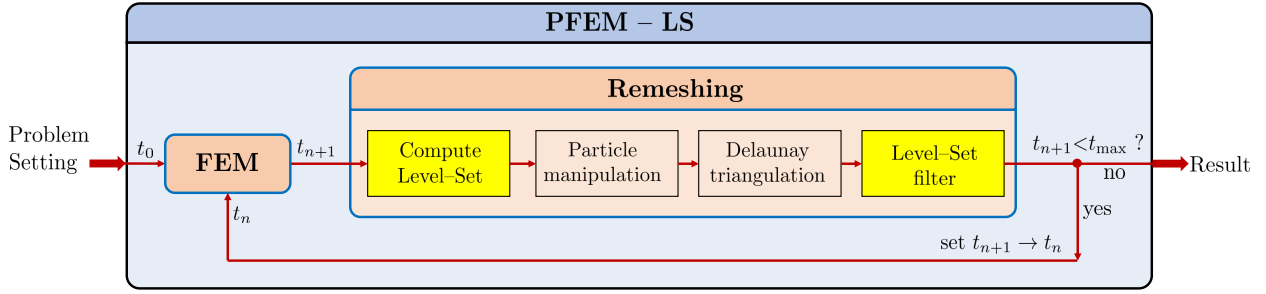}
	\caption{Workflow of the Particle Finite Element Method (PFEM) based on a Level--Set (LS) function during the remeshing process. The workflow is denoted as PFEM--LS in this work. The stages in yellow highlight the difference with respect to PFEM--AS.}
	\label{Fig:PFEM_schemes_pdf_LS}
\end{figure}

Solving the linear system \eqref{EQ:LS_system} yields $\bm{\lambda}$ and $\bm{c}$, and hence $\phi(\mathbf{x})$. Then, $\phi(\mathbf{x})$ can be used to filter the Delaunay triangulation and obtain a body featuring smaller topological changes than those obtained with AS. For this, an accurate representation of the fluid topology in its updated configuration is required. So the function $\phi(\mathbf{x})$ is built in the updated configuration (at time $t_{n+1}$) using the deformed mesh, prior to the manipulations on the particles. Thus, the first stage in the remeshing process is to obtain $\phi(\mathbf{x})$, as illustrated in Fig.~\ref{Fig:PFEM_schemes_pdf_LS}. The different stages during the LS--based remeshing process are detailed below.

\subsection{Building the Level--Set function in PFEM--LS}\label{sec:PFEM-LS_build}

\begin{figure}[t] \captionsetup[sub]{font=normalsize}
	\centering 
	\includegraphics[trim=0 0 0 0,clip,width=1.00\linewidth]{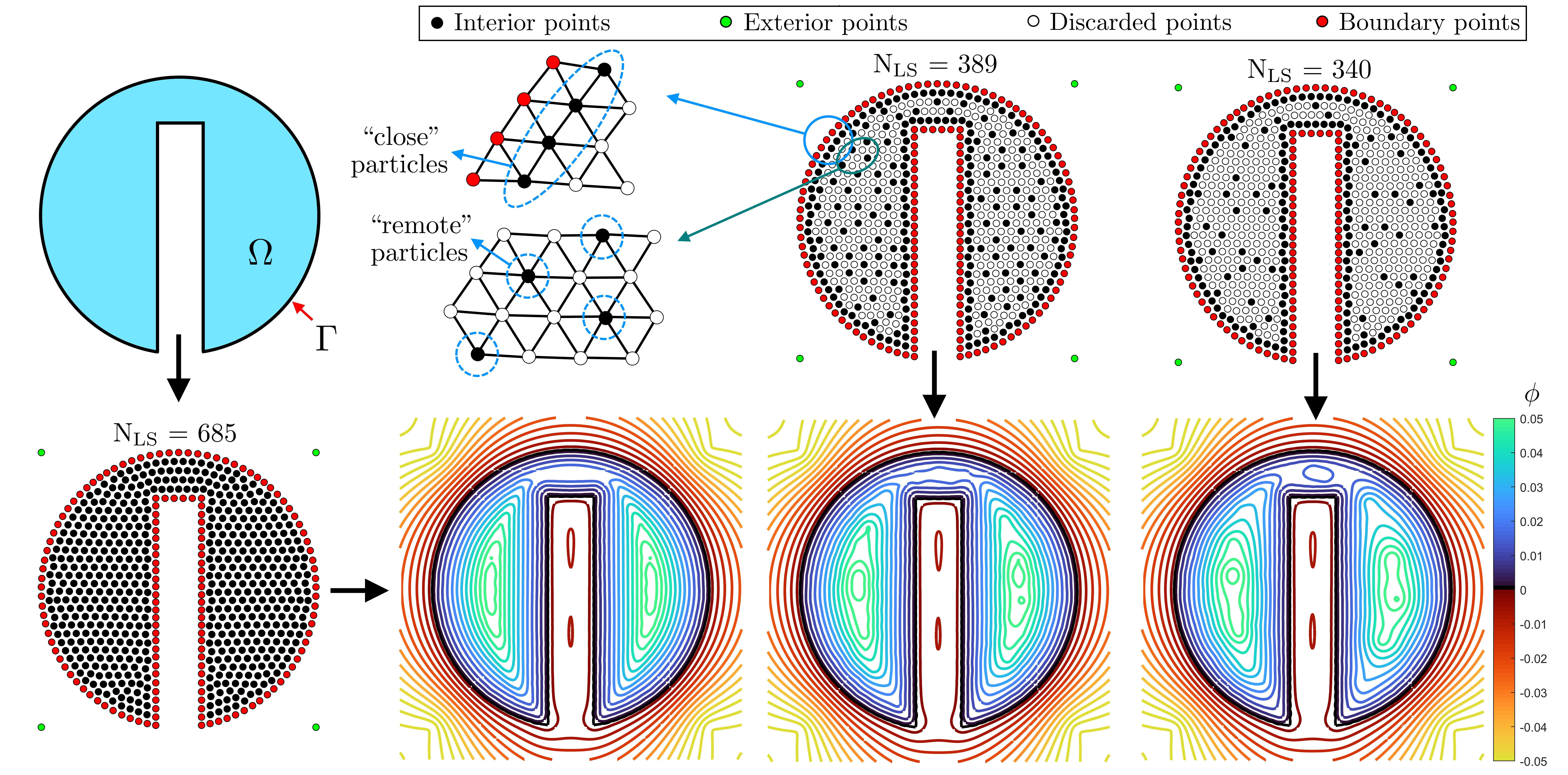}
	\vspace{-81mm}\\
	\hspace{-40mm} \footnotesize{(\textbf{a})} \hspace{35mm} (\textbf{b}) \hspace{32mm} (\textbf{c}) \hspace{35mm} (\textbf{d})	
	\vspace{32mm}\\
	\hspace{-40mm} \footnotesize{(\textbf{e})} \hspace{35mm} (\textbf{f}) \hspace{32mm} (\textbf{g}) \hspace{35mm} (\textbf{h})
	\vspace{36mm}\\
	\caption{\EF{(a) Zalesak's disk \citep{zalesak1979fully} to be discretized for building the Level--Set function $\phi(\mathbf{x})$. (b) Triangulation is used to define the inner points for the $\phi(\mathbf{x})$ function. (c) Only a set of internal points are chosen based on a criteria dictated by the triangulation. (d) Only half of the internal points of subfigure c are chosen. (e) All points discretising the domain are used to build $\phi(\mathbf{x})$. (f-h) Contour plots of $\phi(\mathbf{x})$ obtained from the set of points indicated by the black arrows.}}
	\label{Fig:PFEM_LS_build}
\end{figure}

To build $\phi(\mathbf{x})$, it is assumed that the topology of the body is known at time $t_n$ and it does not change after solving the governing equations, i.e., even though bodies undergo deformations, the amount of bodies and boundaries does not change within the time step of the FEM. Therefore, a simple way to build the LS function is to use nodal coordinates in the updated configuration as fitting points. As external points, the 4 corners of the bounding box that limits the computational domain are used. \EF{However, given the dense linear system to be solved in Eq.~\eqref{EQ:LS_system}, building the function $\phi(\mathbf{x})$ at each time step can rise computation time significantly if all the particles in the model are used as fitting points. Instead, only a few particles are chosen. The procedure is explained using the Zalesak's disk \citep{zalesak1979fully} illustrated in Fig.~\ref{Fig:PFEM_LS_build}a. Having a finite element discretization of the body, all the particles (nodes) discretising the boundary of the body are used as fitting points, as illustrated in Fig.~\ref{Fig:PFEM_LS_build}c (red points). To select the interior points, the triangulation is analysed. The nodes belonging to the elements located at the boundaries of the domain are chosen and labelled as ``close". Also, only one node per element located in the interior of the domain is chosen and labelled as ``remote", as illustrated in Fig.~\ref{Fig:PFEM_LS_build}b. This procedure significantly reduces the number of particles used for the fitting, and thus the size of the linear system in Eq.~\eqref{EQ:LS_system}. In case of highly refined discretisations, the number of inner points can be further reduced, as is done in Fig.~\ref{Fig:PFEM_LS_build}d. In such a case, half of the array containing the ``remote" points is used.
}

\begin{figure}[h!] \captionsetup[sub]{font=normalsize}
	\centering 
	\includegraphics[trim=0 0 0 0,clip,width=0.98\linewidth]{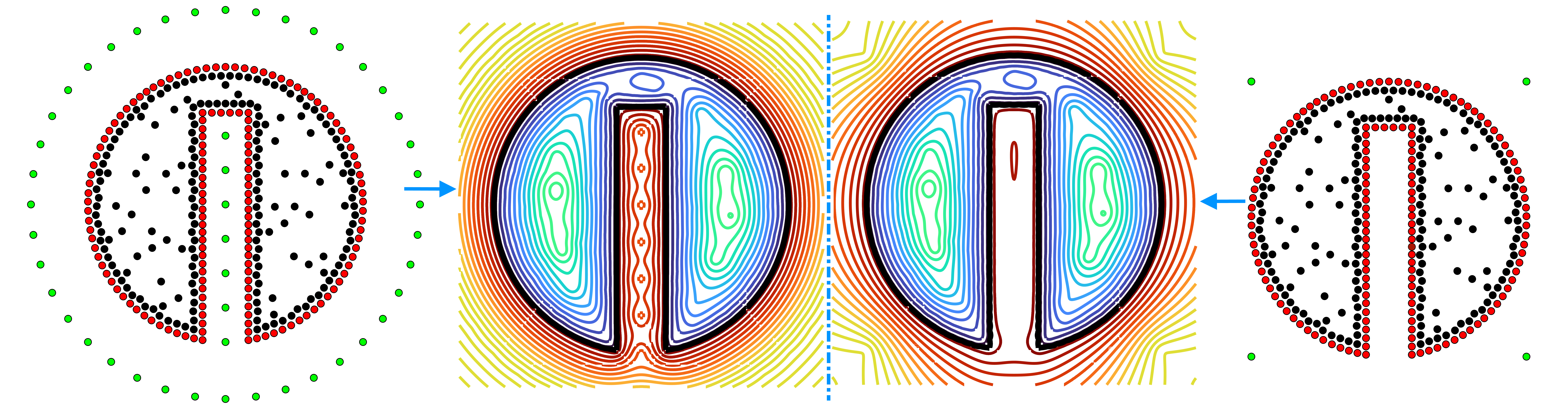}
	\vspace{-45mm}\\
	\hspace{5mm} \footnotesize{(\textbf{a})} \hspace{80mm} (\textbf{b})	 
	\vspace{39mm}\\
	\caption{\EF{Comparison of two Level--Set functions that are built using (a) several and (b) few external points. The legend of the black, red and green dots is given in Fig.~\ref{Fig:PFEM_LS_build}.}}
	\label{Fig:PFEM_Exterior}
\end{figure}

\begin{figure}[h]
\centering 
	\includegraphics[trim=0 0 0 0,clip,width=1.00\linewidth]{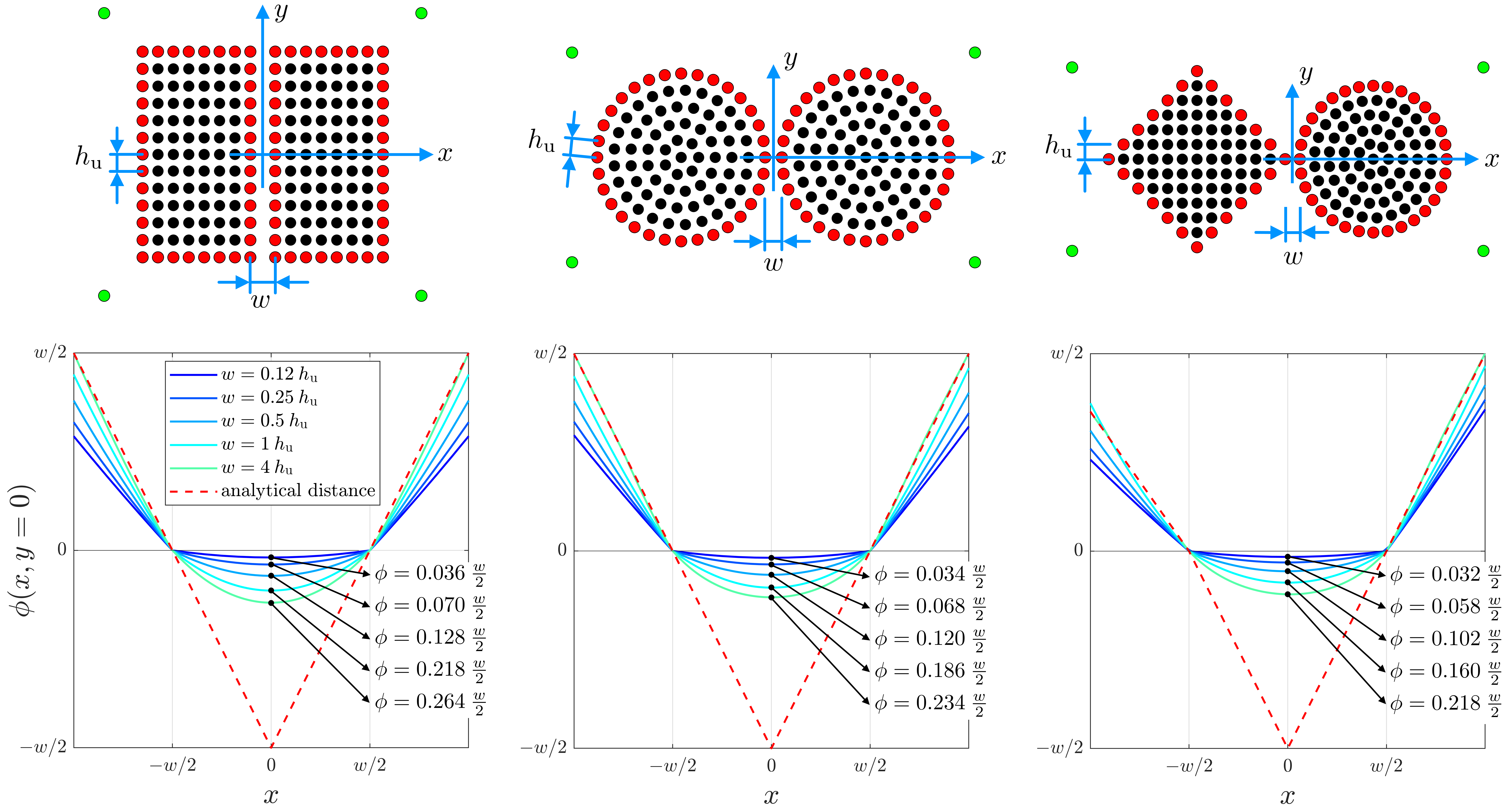}
	\vspace{-95mm}\\
	\hspace{-51mm} \footnotesize{(\textbf{a})} \hspace{50mm} (\textbf{b}) \hspace{53mm} (\textbf{c})
	\vspace{ 35mm}\\
	\hspace{-52mm} \footnotesize{(\textbf{d})} \hspace{50mm} (\textbf{e}) \hspace{53mm} (\textbf{f})  
	\\
	\vspace{46mm}
\caption{\EF{ Level--Set function in the gap between two bodies. (a--c) The points used to build $\phi(\mathbf{x})$, where $w$ and $h_\mathrm{u}$ represent the gap and element size, respectively. (d--f) The LS functions around the gap. These functions are built for different gap distances, which are given in the legend of subfigure (d).}}
\label{Fig:LS_gap}
\end{figure}
\begin{figure}[h!]
\captionsetup[subfigure]{labelformat=empty}
\centering 
	\begin{subfigure}[b]{0.3\textwidth}
		\includegraphics[trim=70 0 90 20,clip,width=1.00\linewidth]{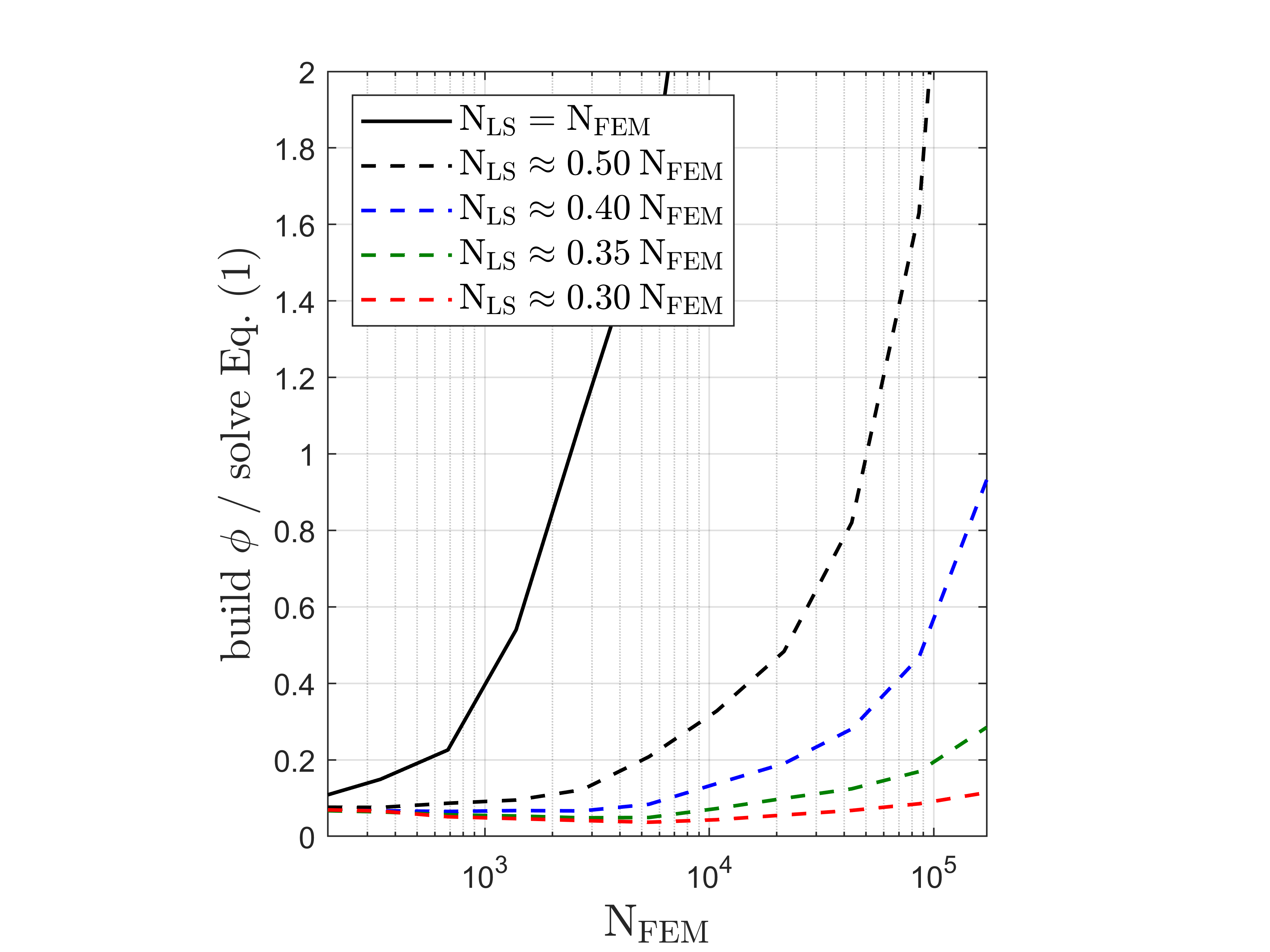}
		\caption{}
		\label{Fig:ZD_time_a}
	\end{subfigure}
	~\hspace{0mm}
	\begin{subfigure}[b]{0.3\textwidth}	
		\includegraphics[trim=70 0 90 20,clip,width=1.00\linewidth]{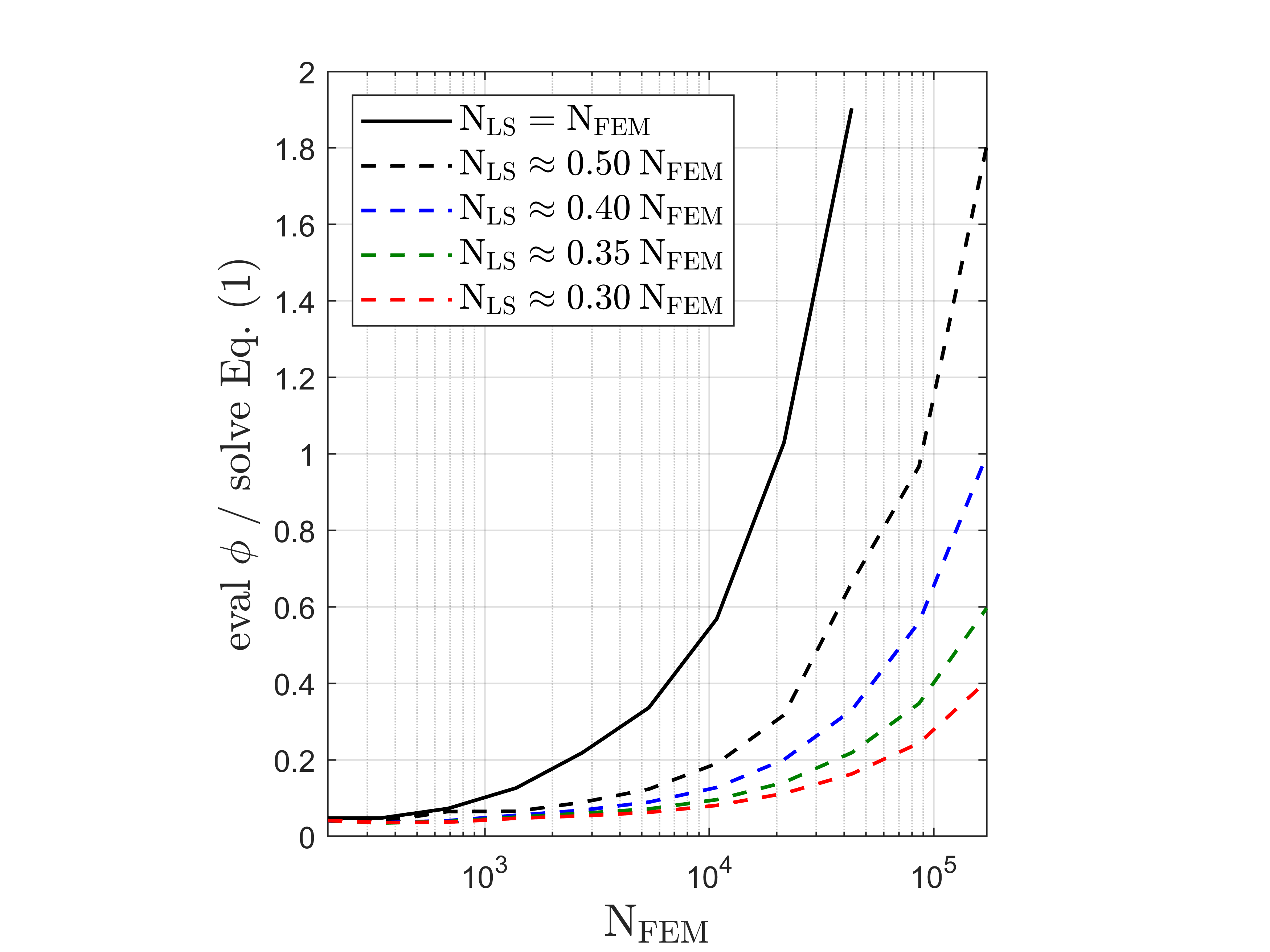}
		\caption{}
		\label{Fig:ZD_time_b}
	\end{subfigure}
	~\hspace{0mm}
	\begin{subfigure}[b]{0.3\textwidth}	
		\includegraphics[trim=70 0 90 20,clip,width=1.00\linewidth]{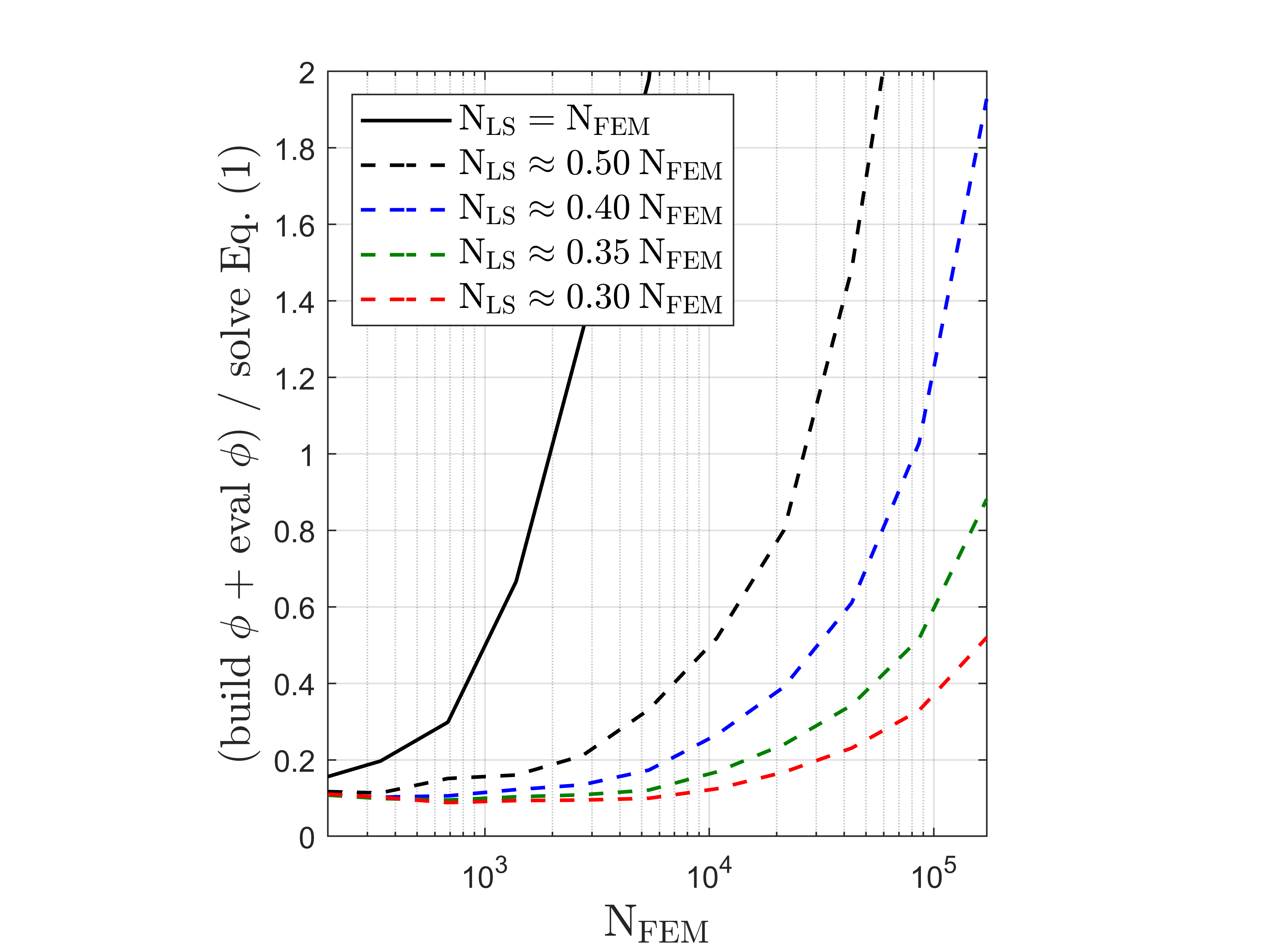}
		\caption{}
		\label{Fig:ZD_time_c}
	\end{subfigure}
	\\
	\vspace{-62mm}
	\hspace{-49mm} \footnotesize{(\textbf{a})} \hspace{47mm} (\textbf{b}) \hspace{47mm} (\textbf{c}) 
	\\
	\vspace{52mm}
\caption{\EF{ Time required to (a) build, (b) evaluate, and (c) build and evaluate the Level--Set function for different discretisations of the Zalesak's disc. $\mathrm{N_{FEM}}$ denotes the number of particles discretising the model and $\mathrm{N_{LS}}$ the number of particles used to build $\phi$. The time is normalised with respect to the time required to solve the governing equations (1) for one time step.}}
\label{Fig:ZD_time}
\end{figure}

\EF{
The reduction of internal points degrades the accuracy of the Level--Set (LS) function as a distance gauge, notably in the inner zone far from the boundaries (``remote" zone). However, this work uses $\phi(\mathbf{x})$ mainly as a topology indicator (given by the sign) and as an estimator of the distance to the body boundaries. Both criteria are well preserved when reducing the number of internal points, as can be appreciated in Figs.~\ref{Fig:PFEM_LS_build}f--\ref{Fig:PFEM_LS_build}h, when comparing the contour lines around the boundaries (``close" zone). Similarly, using a few exterior points limits the accuracy of the LS function as a distance gauge in the exterior zone. Yet in the vicinity of the boundaries, $\phi(\mathbf{x})$ still provides correct information regarding the topology (sign). This can be noticed in Figs.~\ref{Fig:PFEM_Exterior}a and \ref{Fig:PFEM_Exterior}b, where contour plots obtained with many and few exterior points are shown, respectively. There, the main difference between LS curves is found in the notch of the disk. This geometrical feature is further analyzed in the following because the interest of this work is to capture small geometrical details, in particular those that are about the size of a finite element.

The notch is now represented by two rectangular bodies spaced by $w$. Likewise, other shapes are examined, as illustrated in Figs.~\ref{Fig:LS_gap}a--\ref{Fig:LS_gap}c. Different distances $w$ are set in terms of the finite element size ($h_\mathrm{u}$). For each case, the $\phi(\mathbf{x})$ function is built and plotted around the gap zone. The  graphs are shown in Figs.~\ref{Fig:LS_gap}d--\ref{Fig:LS_gap}f. It is observed that the accuracy of the LS function as a distance gauge decreases drastically as the gap ($w$) narrows. In the extreme case of $w=0.12\:h_\mathrm{u}$, $\phi(\mathbf{x})$ underestimates the distance by $\approx$30 times. Despite the large error, LS is nevertheless consistent in the three geometric scenarios (Figs.~\ref{Fig:LS_gap}a-\ref{Fig:LS_gap}c), as the distance is underestimated in the same order in all 3 cases. As the LS function is reliable in the sign and consistent in the magnitude, we did not seek to increase the accuracy of the LS function and therefore did not require more external points when building the LS function. 
}

\EF{
To give an idea of the computational cost of building the LS function and the impact of reducing the amount of interior points, $\phi(\mathbf{x})$ is computed for different discretisations of the Zalesak's disk shown in Fig.~\ref{Fig:PFEM_LS_build}a. To report the computing time in a comparative scale, the required time to build $\phi(\mathbf{x})$ is normalised with respect to the time needed to solve the system of governing equations \eqref{EQ:Continuum_Equations_VP} for one time step, considering the disc as a fluid in free fall. The code routines are compared in MATLAB and the results are shown in Fig.~\ref{Fig:ZD_time_a}. There, 5 curves are drawn, which are obtained with different amount of internal points. In the graph, $N_\mathrm{FEM}$ represents the number of particles discretising the body. Fig.~\ref{Fig:ZD_time_a} reveals that, when no reduction is applied ($N_\mathrm{LS} = N_\mathrm{FEM}$) and $N_\mathrm{FEM}>2000$, the time for obtaining $\phi(\mathbf{x})$ is much higher than that needed for solving the system of equations. With a 50$\%$ reduction of points ($N_\mathrm{LS} \approx 0.50\:N_\mathrm{FEM}$), the times match at $N_\mathrm{FEM} = 50000$ particles. Beyond that amount, building the LS function is more expensive. Larger reductions of points ($N_\mathrm{LS} \approx 0.3\:N_\mathrm{FEM}$) result in a negligible building time. Notably, under the proposed sampling scheme for the fitting points, the reduction of computation time depends on the Perimeter-To-Area (PTA) ratio of the model. Indeed, domains with low PTA ratio will allow a greater reduction of internal points than those with a high PTA ratio. Thus, Fig.~\ref{Fig:ZD_time_a} would not be representative of such cases. In all academic examples of this manuscript, the one-point-per-element reduction is used (procedure illustrated in Figs.~\ref{Fig:PFEM_LS_build}b and \ref{Fig:PFEM_LS_build}c, which leads to $N_\mathrm{LS} \approx 0.50\:N_\mathrm{FEM}$).} Although there are reinitialization algorithms for LS functions that avoid a complete reconstruction of $\phi(\mathbf{x})$ \citep{hartmann2010constrained}, they have not been considered in this work to avoid undesired effects that may be related to a degradation of the LS function that successive reinitializations entail.

\subsection{Particle manipulation in PFEM--LS}

\begin{figure}[t!]
\captionsetup[subfigure]{labelformat=empty}
\centering 
	\includegraphics[trim=0 0 0 0,clip,width=0.85\linewidth]{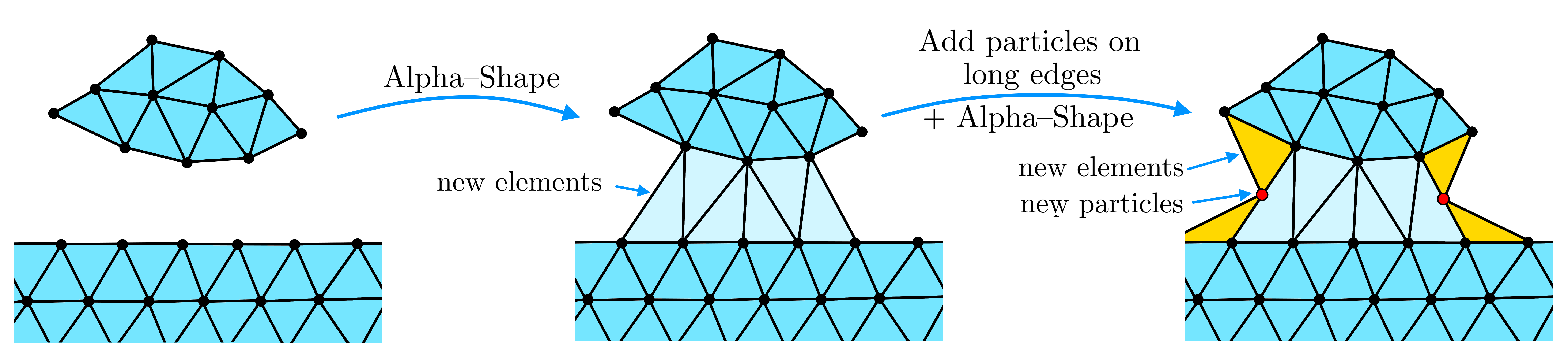}
	\\
	\vspace{-33mm}
	\hspace{-32mm} \footnotesize{(\textbf{a})} \hspace{44mm} (\textbf{b}) \hspace{50mm} (\textbf{c}) 
	\\
	\vspace{28mm}
	\caption{\EF{ Mechanism of mass creation due to the refinement of facets present on the free surface. (a) The domain to be remeshed. (b) The remeshing process creates new elements with large facets on the free surface. (c) The facets are refined, but new elements are created, adding mass to the system.}}
\label{Fig:refine_FS_AS}
\end{figure}

As it will be explained in the following subsection, the filter of the Delaunay triangulation based on the Level--Set function considers the topology of the body and not the shape of the elements. Therefore, from the remeshing perspective, it is not essential to discretise the fluid with elements of low aspect ratio and low skewness. However, the addition and removal of particles is still convenient for ensuring robustness of the finite element analysis. For this reason, manipulations presented for PFEM--AS are equally implemented in PFEM--LS, with exactly the same set of parameters ($\omega$, $\gamma$ and $\beta$). However, an additional procedure is applied in PFEM--LS. When a facet belonging to the free surface is stretched beyond $\beta \: h_\mathrm{u}(\Gamma)$, a particle is placed in the middle of the facet. Here, $h_\mathrm{u}(\Gamma)$ is the user-defined element size at the fluid boundary and $\beta$ is the same stretching parameter for facets on free--slip surfaces. Note that this manipulation that seeks for smoothness of the free surface can be equally implemented in PFEM--AS, but additional conditions must be added to the AS criterion to avoid mechanisms that lead to mass creation, such as mesh refinement \citep{falla2022Mesh}. \EF{The problem is that by adding particles on the free surface, it is more likely to obtain elements satisfying the geometrical condition of the Alpha--Shape, as illustrated in Fig.~\ref{Fig:refine_FS_AS}. For this reason, the smoothing of the free surface is not adopted in our PFEM--AS implementation, otherwise, huge mass variations are obtained.}

\begin{figure}[t!]
\captionsetup[subfigure]{labelformat=empty}
\centering 
	\includegraphics[trim=0 0 0 0,clip,width=0.75\linewidth]{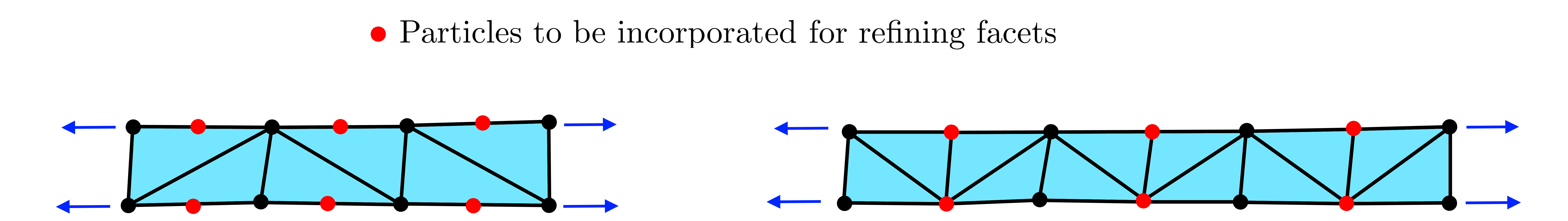}
	\caption{Illustration of a fluid layer that cannot be split due to the addition of particles on the free surface.}
\label{Fig:thinLayer}
\end{figure}

Refining the facets helps to avoid degradation of the free surface, to keep its smoothness and a homogeneous discretisation, and to guarantee a good accuracy of the LS function in the boundary description. On the other hand, this strategy can promote mass creation by preventing the detachment of small fluid layers, as illustrated in Fig.~\ref{Fig:thinLayer}. To avoid this, a facet is refined only if the thickness of the fluid layer is greater than $3\: h_\mathrm{u}$. Note that this thickness--based condition can be easily checked using the LS function.

\subsection{Filtering the Delaunay triangulation in PFEM--LS}

Once the Level--Set function has been built, and the particles distribution has been improved and triangulated, the next important stage in PFEM--LS is to filter the Delaunay triangulation and obtain the discretisation that will be used in the next time step. This stage is the one that changes the topology of the simulated body, since it defines the bodies that come into contact, those that detach, and the new boundaries that are generated. 

\begin{figure}[t!]
\captionsetup[subfigure]{labelformat=empty}
\centering 
	\includegraphics[trim=0 0 0 0,clip,width=1.00\linewidth]{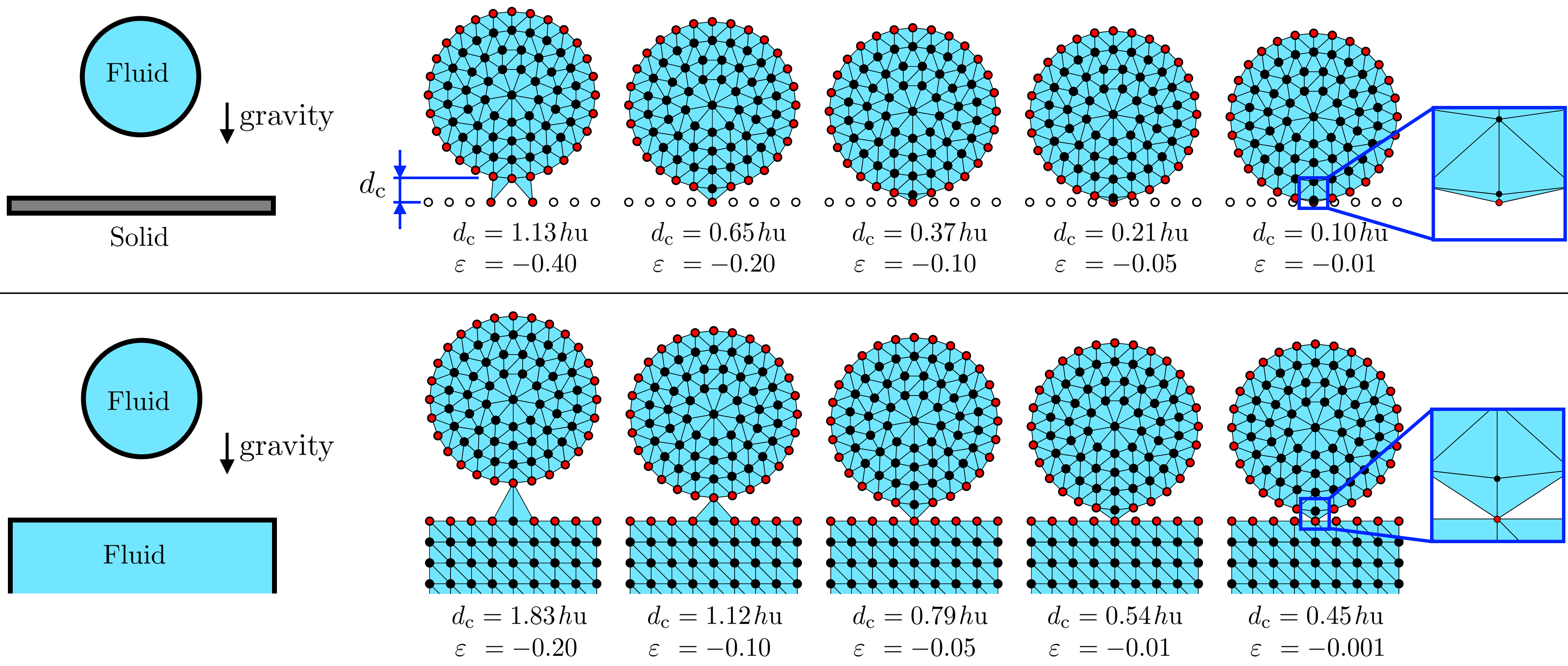}
	\\
	\vspace{-70mm}
	\hspace{-120mm} \footnotesize{(\textbf{a})} \hspace{36mm} (\textbf{b}) 
	\vspace{28mm}\\
	\hspace{-120mm} \footnotesize{(\textbf{c})} \hspace{36mm} (\textbf{d}) 
	\\
	\vspace{34mm}
	\caption{\EF{ A disc of fluid in free fall that impacts (a) a rigid surface, (c) another fluid at rest. (b) and (d) show the instant of the contact between the bodies for different values of $\varepsilon$. The distance at which contact is established is defined by $d_\mathrm{c}$.}}
\label{Fig:study_epsi}
\end{figure}

\begin{figure}[t] \captionsetup[sub]{font=normalsize}
	\centering 
	\includegraphics[trim=0 0 0 0,clip,width=1.00\linewidth]{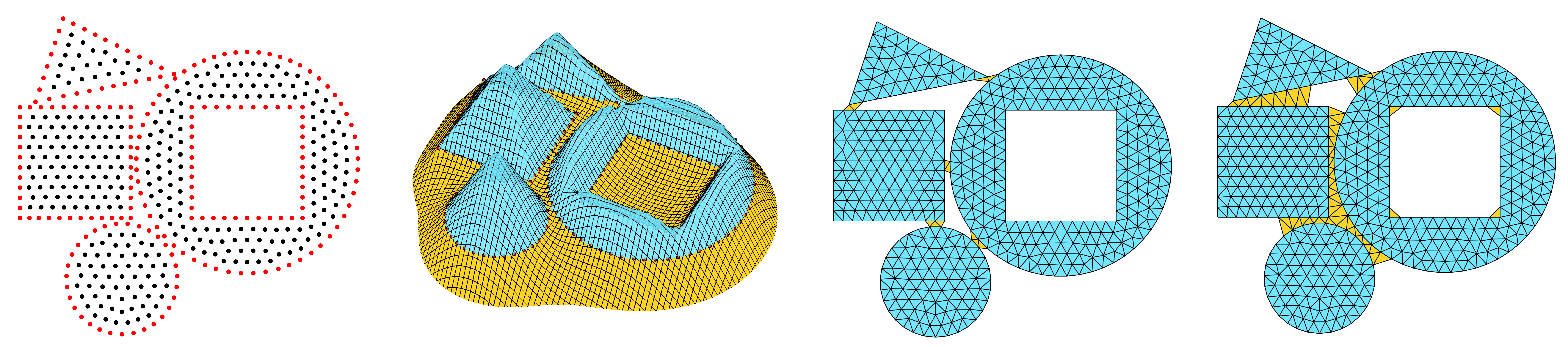}
	\vspace{-40mm}\\
	\hspace{-33mm}
	\footnotesize{(\textbf{a})} 
	\hspace{37mm} (\textbf{b}) \hspace{39mm} (\textbf{c}) \hspace{36mm} (\textbf{d})	
	\vspace{30mm}\\
	\caption{Comparison of filtered triangulations using Alpha--Shape (AS) and Level--Set (LS). (a) The nodal discretisation of the fluid domain, where boundary nodes are shown in red and interior nodes in black. (b) The LS function, where the interior is shown in blue and the exterior in yellow. (c) The filtered triangulation using LS and $\phi(\mathbf{x}) \geq -0.001\:h_\mathrm{u}$. (d) The filtered triangulation using AS and $\alpha_\mathrm{max} = 1.2$. In (c) and (d), the yellow elements are those that do not belong to the initial domain, but are incorporated during the meshing process and are required to model the contact in PFEM.}
	\label{Fig:PFEM_LS_filter}
\end{figure}

Given that the contact between bodies is modelled by the introduction of elements in PFEM, it is convenient to condition the addition of the elements between bodies in terms of the distance between them. In this work, elements of the Delaunay triangulation that are outside the fluid domain are accepted by the LS filter if the distance from the centroid of the element to the fluid boundary is less than some imposed distance. Although the function $\phi(\mathbf{x})$ does not guarantee precision as a distance gauge in the outer zone of the fluid since only 4 outer points are used for the fitting (the bounding box), numerical examples indicate that $\phi(\mathbf{x})$ provides \EF{reliable topological information and a consistent distance to the boundaries}. Thus, the condition that models the contact between bodies, or self-contact as in the breaking of a wave, is given by:
\begin{equation} \label{EQ:IcLS}
\mathbb{I}_\mathrm{LS} = \{\:i \in \mathbb{I}_\mathrm{D} \; \mid \; \phi(\mathbf{x}_i) \geq \varepsilon \: h_\mathrm{u}(\Gamma) \}
\end{equation}

\noindent where $\mathbb{I}_\mathrm{LS}$ is \EF{the set of element indices that are retained for discretising the fluid and $\varepsilon$ is a parameter chosen by the user. To analyze the influence of this parameter, its effect on the detection of fluid--fluid and fluid--structure contact is discussed. For this, a disc of fluid falling on a rigid floor and on a reservoir containing the same fluid are used as test cases. The problems are illustrated in Fig.~\ref{Fig:study_epsi}. The geometry and physical parameters of these problems are omitted since they are not relevant in this example, as the interest here is merely in detecting the contact. In both problems, only 4 external points are used to build the LS function. That is, the particles on the solid surface are not included in $\phi(\mathbf{x})$. The Fig.~\ref{Fig:study_epsi} shows snapshots of the disk impacting the solid surface and the fluid at rest for different values of $\varepsilon$. The snapshots show the instant at which elements are created between the disc and the impacted surface, according to the criterion of Eq.~\eqref{EQ:IcLS}. It is noted that for the fluid-structure contact case, the size of the created elements is consistent with the desired size ($\varepsilon \: h_\mathrm{u}$), which is measured from the centroid of the element. As expected, the smaller the value of $\varepsilon$, the smaller the size of the contact element. In the extreme case of $\varepsilon \approx 0$, a very small contact distance is obtained, which could be missed if the displacement of the body from the previous time step ($\Delta t \mathbf{v}_n$) is larger. For this reason, we recommend a value of $\varepsilon$ large enough to guarantee the contact, but small enough to avoid significant mass creation. In this work, we define $\varepsilon = -0.1$ when dealing with fluid-structure contact. In the fluid-fluid contact case, the same pattern is observed but on a different scale. This is because the LS function underestimates the distance from the element to the free surface in narrow areas, so lower values of epsilon are required. In this work, $\varepsilon = -0.001$ is used when dealing with fluid-fluid contact. As can be seen in Fig.~\ref{Fig:PFEM_LS_filter}c, this criterion results in a lower addition of elements in the fluid-fluid contact than the AS-based criterion, i.e.~$\mathbb{I}_\mathrm{LS} \subseteq \mathbb{I}_\mathrm{AS}$.}

\begin{figure}[t] \captionsetup[sub]{font=normalsize}
	\centering 
	\includegraphics[trim=0 0 0 0,clip,width=1.00\linewidth]{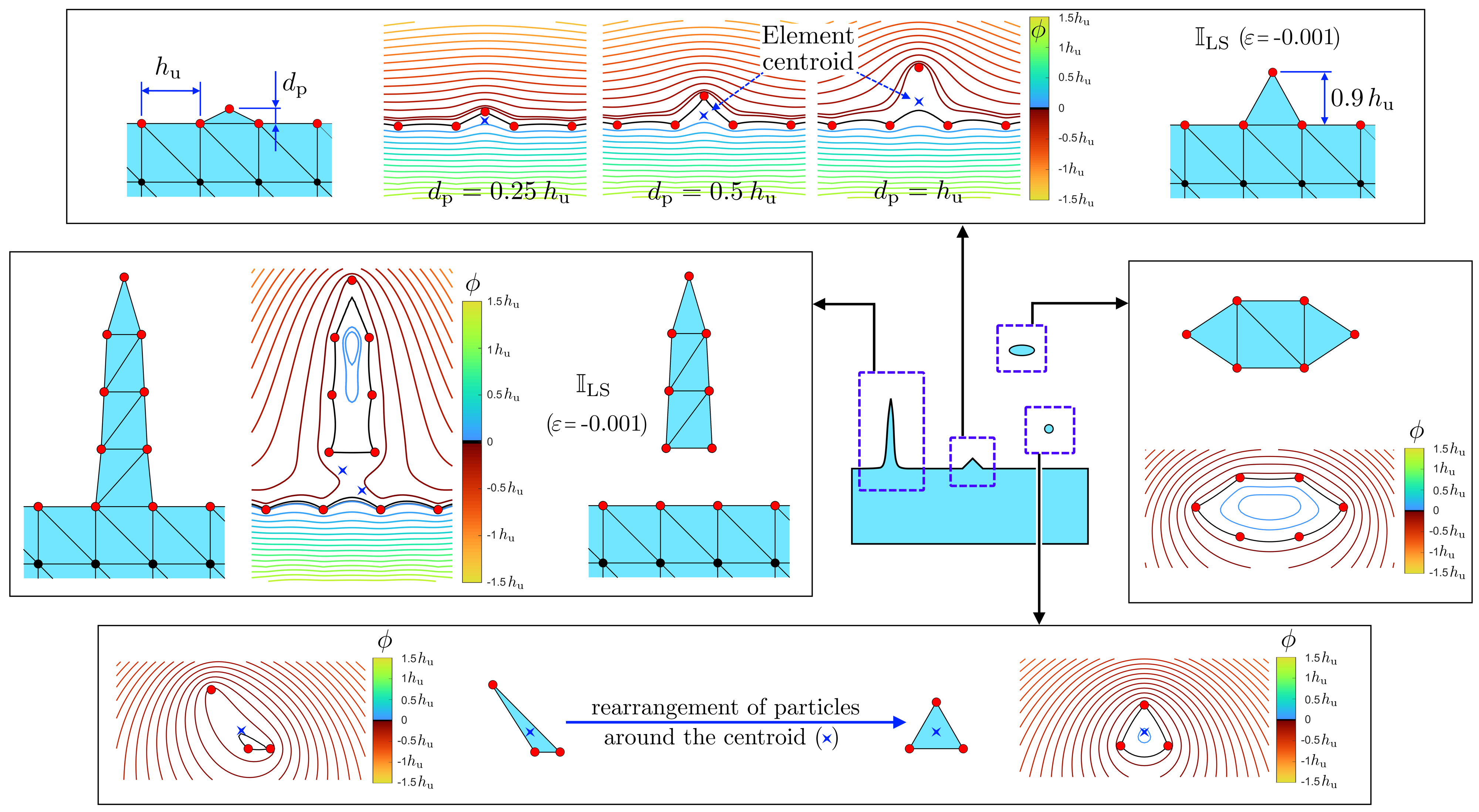}
	\vspace{-94mm}\\
	\hspace{-27mm}
	\footnotesize{(\textbf{a})} \hspace{24mm} (\textbf{b}) \hspace{83mm} (\textbf{c})	
	\vspace{25mm}\\
	\hspace{-32mm} \footnotesize{(\textbf{d})} \hspace{16mm} (\textbf{e}) \hspace{36mm} (\textbf{f}) \hspace{58mm} (\textbf{g})	
	\vspace{38mm}\\
	\hspace{-5mm} \footnotesize{(\textbf{h})} \hspace{53mm} (\textbf{i})	
	\vspace{16mm}\\
	\caption{\EF{Illustration of element removal in PFEM--LS. (a) A small perturbation of the free surface modelled by one element, (b) the LS function around the element, and (c) the largest element allowed by LS criterion using $\varepsilon = -0.001$. (d) A fluid spike, (e) the LS function, and (f) the mesh obtained by the LS criterion with $\varepsilon = -0.001$. (g) A small body detached from the bulk body and its LS function. (h) A tiny body modelled by a highly distorted element. (i) The element is rearranged as an equilateral triangle.}}
	\label{Fig:PFEM_disconection}
\end{figure}

\EF{
It was shown that the LS--based contact criterion allows for a smaller addition of elements than the AS--based criterion. However, the separation of bodies must also be possible. Below we show that Eq.~\ref{EQ:IcLS} also allows for the separation of bodies. The reason is that the LS function is not accurate in the presence of elements composed only of boundary nodes. In particular, if an element on the free surface of a large body has its 3 nodes at the boundary, then the centroid of the element will be outside the body according to the LS function. This is illustrated in Fig.~\ref{Fig:PFEM_disconection}a, where a particle on the free surface is displaced vertically. In Fig.~\ref{Fig:PFEM_disconection}b is noted that the centroid of the element falls outside the body (negative sign) if the displaced particle is $h_\mathrm{u}$ away from the free surface. In a remeshing process using $\varepsilon = -0.001$, the maximum height of the element holding the particle is $\mathrm{d}_e = 0.9 h_\mathrm{u}$, as shown in Fig.~\ref{Fig:PFEM_disconection}c.
}

\EF{
In the previous case, the LS function indicates that the element is outside a body if it has 3 nodes at the boundary, if one of them is far enough away from the free surface, and if the main body is composed of internal nodes. Although the above analysis has been carried out for a particular geometry, we have noticed a similar behaviour for other shapes. This can be seen in Fig.~\ref{Fig:PFEM_disconection}d, which illustrates a fluid spike attached to the body by two elements consisting of boundary nodes. This spike is detached from the bulk body if $\varepsilon = -0.001$, as shown in Fig.~\ref{Fig:PFEM_disconection}f. Finally, it is worth noting that in a small body composed of only boundary nodes, the inner zone is identified as positive by the LS function, as the one shown in Fig.~\ref{Fig:PFEM_disconection}g. In that case, the elements belonging to the small body would not be removed if $\varepsilon < 0 $. A special case are tiny bodies discretised by a single element. which may arise from splashes. Depending on their shape, they could be removed by the LS-based criterion. This is because the centroid of highly distorted triangles lies in the outer zone of the LS function (negative sign), as shown in Fig.~\ref{Fig:PFEM_disconection}h. To avoid mass loss by removal of isolated elements, these are arranged as an equilateral triangle. For this, the nodes of the element are displaced to achieve the desired triangle, but keeping the element centroid. In addition, the nodal velocities of the isolated element are equalised so that the element is not deformed during its flight. This action can be interpreted as a coarse discretisation of a fluid drop that seeks to preserve the mass of the system, but it does not constitute a faithful representation of a drop, as has been done in other PFEM works \citep{ryzhakov2017application}.
}

\begin{figure}[t!]
\captionsetup[subfigure]{labelformat=empty}
\centering 
	\begin{subfigure}[b]{0.35\textwidth}	
		\includegraphics[trim=0 0 0 0,clip,width=1.00\linewidth]{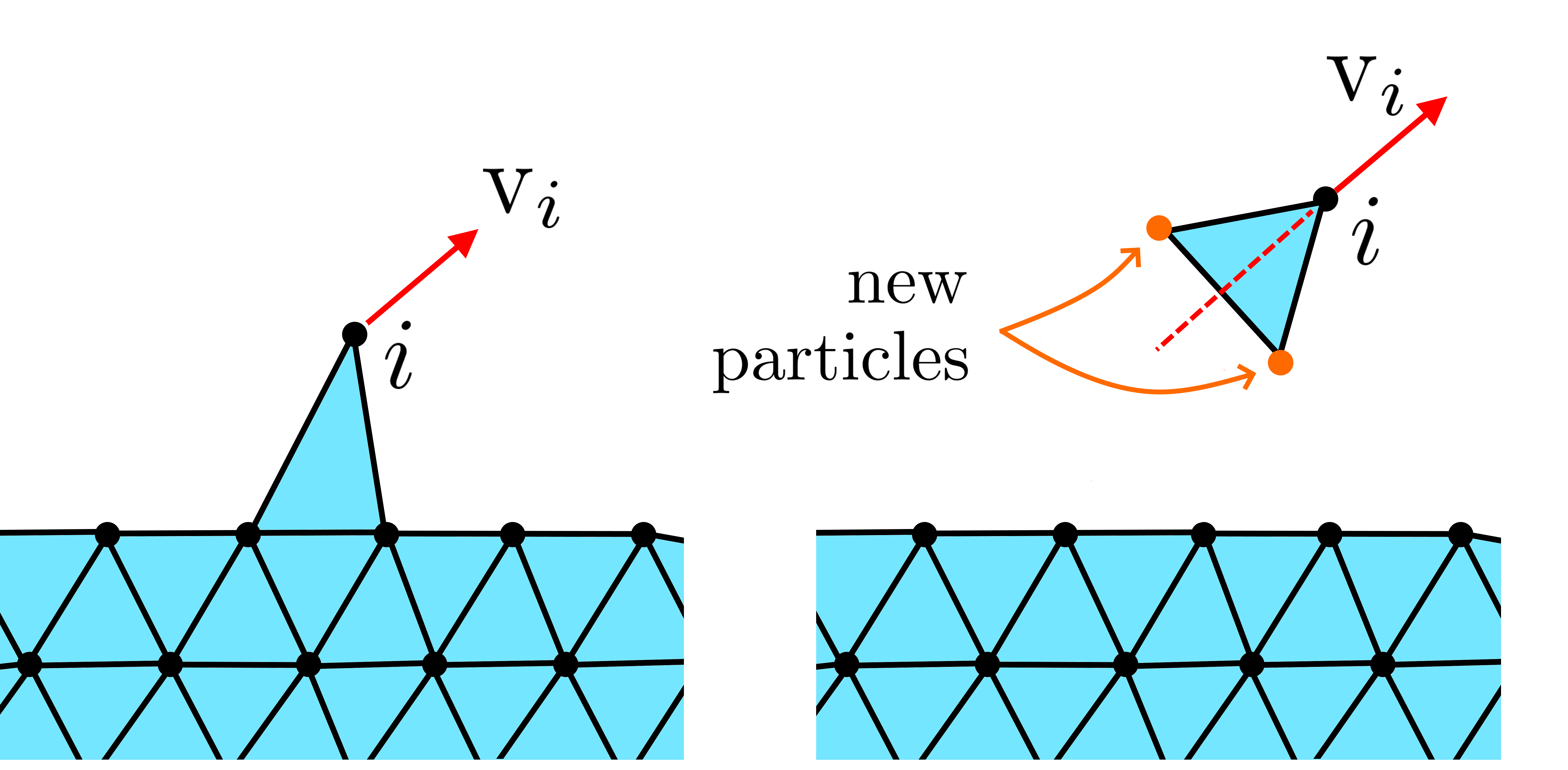}
		\caption{}
		\label{Fig:mani_cares_LS_c}
	\end{subfigure}
	~\hspace{4mm}
	\begin{subfigure}[b]{0.57\textwidth}	
		\includegraphics[trim=0 0 0 0,clip,width=1.00\linewidth]{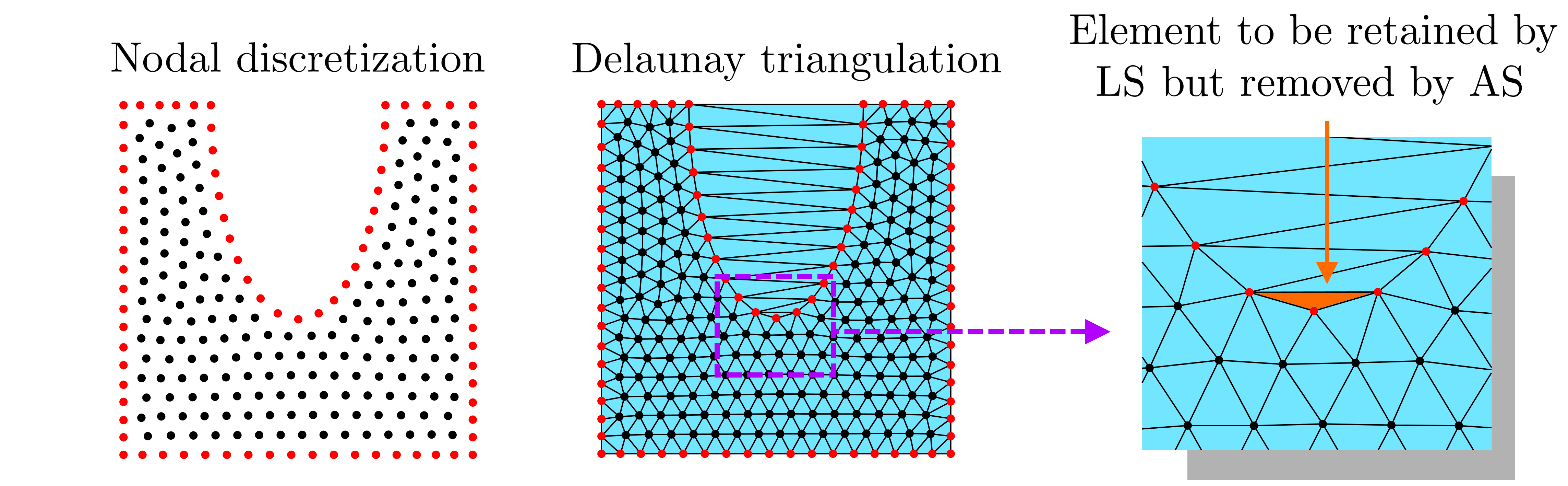}
		\caption{}
		\label{Fig:mani_cares_LS_b}
	\end{subfigure}
	\vspace{-35mm}\\ 
	\hspace{-95mm} \footnotesize{(\textbf{a})} \hspace{60mm} \footnotesize{(\textbf{b})}
	\vspace{25mm}\\
	\caption{(a) Illustration of the detachment of a particle from the free surface and the generation of a triangular drop. (b) Concave free surface that shows an element from the Delaunay triangulations that is allowed by the LS function but removed by AS.}
\label{Fig:mani_cares_LS}
\end{figure}

\EF{
It is recalled that the aim of this work is to reduce the mass variation from to the remeshing process of PFEM. Therefore, for the sake of mass conservation,} once a particle is detached from the free surface, this is meshed with an \EF{equilateral} triangular element of \EF{side} such that its volume equals that of the element that held the particle on the free surface. The orientation of the triangle is given by the velocity of the particle, as illustrated in Fig.~\ref{Fig:mani_cares_LS_c}. The \EF{states of the} two added particles are initialized as a copy of the released particle to keep the volume of the drop constant during its flight. Importantly, drops are meshed only if two conditions are met. First, drops must be at a distance greater than $2\:h_\mathrm{u}(\Gamma)$ from the free surface, otherwise the meshed drop risks being reattached to the free surface. Second, drops must be released from elements having a volume that is representative of the discretisation, otherwise tiny drops may be generated. To this end, a drop is considered to be meshed if the element that held it on the free surface had a size greater than 30$\%$ of the imposed characteristic size.

\EF{
Although the LS function helps to identify boundaries and topology, and thus reduces the mass variation with respect to the AS criterion, there are still challenging situations. For example, in the case of concave surfaces, the Delaunay triangulation may result in highly distorted elements that are close to the boundary, and hence be accepted according to the LS-based criterion, as illustrated in Fig.~\ref{Fig:mani_cares_LS}b. To avoid this, we resort to the AS algorithm, which is applied only on elements whose 3 nodes lie on the free surface. Although the AS algorithm could also remove elements belonging to the interior of the body, this is unlikely, as the manipulations on the particles result in fairly smooth free surfaces with few distorted elements. In addition, the Alpha--Shape criterion does not influence the fluid--fluid contact because, as stated previously, the LS-based criterion leads the contact condition by being more restrictive than the AS algorithm ($\mathbb{I}_\mathrm{LS} \subseteq \mathbb{I}_\mathrm{AS}$). 
}

\EF{
To finish this section, the computation time of PFEM--LS is discussed. For this purpose, the Zalesak's disk \citep{zalesak1979fully} is used, which was previously employed to measure the time to build the LS function. Now the time to evaluate $\phi(\mathbf{x})$ at the centroid of each element is measured for different discretisations of the disk. The results are shown in Fig.~\ref{Fig:ZD_time}b. These indicate that the time to evaluate the LS function on the triangulation is similar to the time to build the LS function (Figs.~\ref{Fig:ZD_time}a). However, the MATLAB code used in this example does not account for parallelisation, which would lower the evaluation time considerably. Fig.~\ref{Fig:ZD_time}c indicates that the total time (build + evaluate) allocated to the remeshing criterion based on Level--Set functions is similar to the cost of the Finite Element Analysis (FEA), when the amount of sampling points is reduced ($N_\mathrm{LS} < N_\mathrm{FEM}$) and when the number of particles discretising the disc is about 30000. Note that the computational cost of the AS algorithm is negligible, so the computational time in PFEM--AS is led by the FEA analysis. Thus, the graphs in Fig.~\ref{Fig:ZD_time} can also be seen as the additional time of PFEM--LS with respect to PFEM--AS. Although these results have been obtained for a particular geometry, we highlight that they are fairly representative of the computational time of the academic examples reported in this work.

Remarkably, the graphs in Fig.~\ref{Fig:ZD_time} display a cubic behaviour. The reason is that the system in Eq.~\eqref{EQ:LS_system} leads to a highly dense matrix, of size $N_\mathrm{LS} \times N_\mathrm{LS}$, and with a diagonal of zeros (hollow matrix). This requires solvers for dense linear systems, which have lower scalability performances than solvers for sparse systems. This implies that to increase the dimension of the problem (3D) or the scale, where $N_\mathrm{FEM} \approx 10^6$ particles, it is necessary to devise methods to mitigate the computational time. This can be done by setting criteria to trigger the remeshing only when necessary and not at every time step; by applying more efficient re-initialisation methods \citep{hieber2005lagrangian,henri2022geometrical}; and by implementing strategies that minimise the number of sampling points at the boundary of the domain, given that the zones of interest are the surfaces that come into contact. These items are beyond the scope of this work and will be addressed in future research.  
}
 
\begin{figure}[t] \captionsetup[sub]{font=normalsize}
	\centering 
	\includegraphics[trim=2 370 3 12,clip,width=1.00\linewidth]{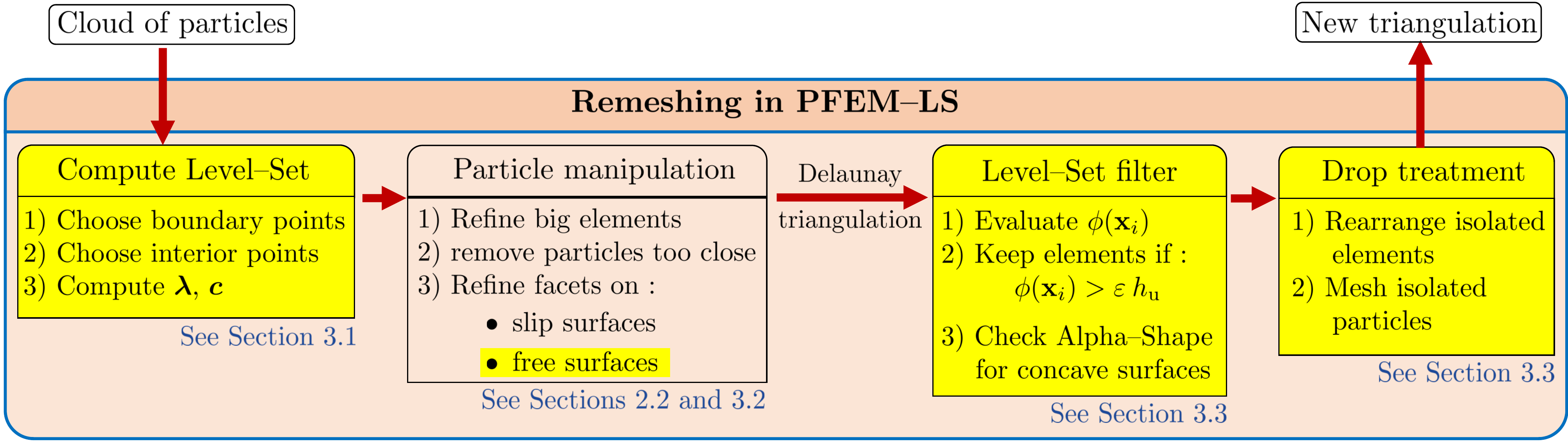}
	\caption{\EF{Summary of the remeshing process in PFEM--LS. The steps in yellow are not present in PFEM--AS.}}
	\label{Fig:PFEM_remeshing_pdf}
\end{figure}

The following numerical examples compare PFEM--AS and PFEM--LS. Both schemes use exactly the same FEM, so the only difference lies in the remeshing process. \EF{To illustrate this difference and provide a summary of the remeshing process of PFEM--LS, Fig.~\ref{Fig:PFEM_remeshing_pdf} lists the required steps and highlights in yellow the ones that are not present in PFEM--AS.}

\section{Numerical examples}\label{sec:Examples}

The proposed PFEM--LS scheme is compared against PFEM--AS using four problems. The first one is a water dam break considering both, free-slip and no-slip boundary conditions. The second problem is the fall and impact of a fluid drop that is discretised by a triangular element. The aim of this problem is to represent the drops arising in PFEM simulations featuring splashing, and to quantify the volume and energy variation resulting from the coarse discretisation of a drop. The third example is the mixing of a viscous fluid, following the implementation of Franci and Cremonesi \citep{franci2017effect}. The objective is to compare the achieved volume variation with the reference, and to evaluate PFEM--LS in a problem involving the impact of two fluid bodies. The fourth is an original example involving the pouring of a fluid. The reservoir from which the fluid is poured has two nozzles, which induces flows featuring waves and small gaps that are difficult to capture during a remeshing process. This example seeks to highlight the advantage of PFEM-LS over PFEM-AS. 

The chosen numerical examples are intended to highlight the differences between PFEM--AS and PFEM--LS and to demonstrate the merits of the latter. The focus of the analysis is on the mass (or volume) conservation, and the issues that influence it. Several snapshots have been included in the manuscript to support the observations, but to facilitate the interpretation of these and the ideas postulated in this section, animations of each problem have been posted in \citep{YoutubeAll}. 

Importantly, the remeshing process is applied at each time step in all the problems of this section. The time steps are defined by the maximum CFL number, which is set to 0.1 in all examples.  As the focus of this work is on the remeshing process, the volume (or mass) variation reported in this section is due to the remeshing process only. All examples in this section use $9.81$ m/s$^2$ (downwards) as gravitational acceleration.

\begin{figure}[t!]
\captionsetup[subfigure]{labelformat=empty}
\centering 
	\hspace{-10mm}
	\begin{subfigure}[b]{0.32\textwidth}
		\includegraphics[trim=0 0 0 0,clip,width=1.00\linewidth]{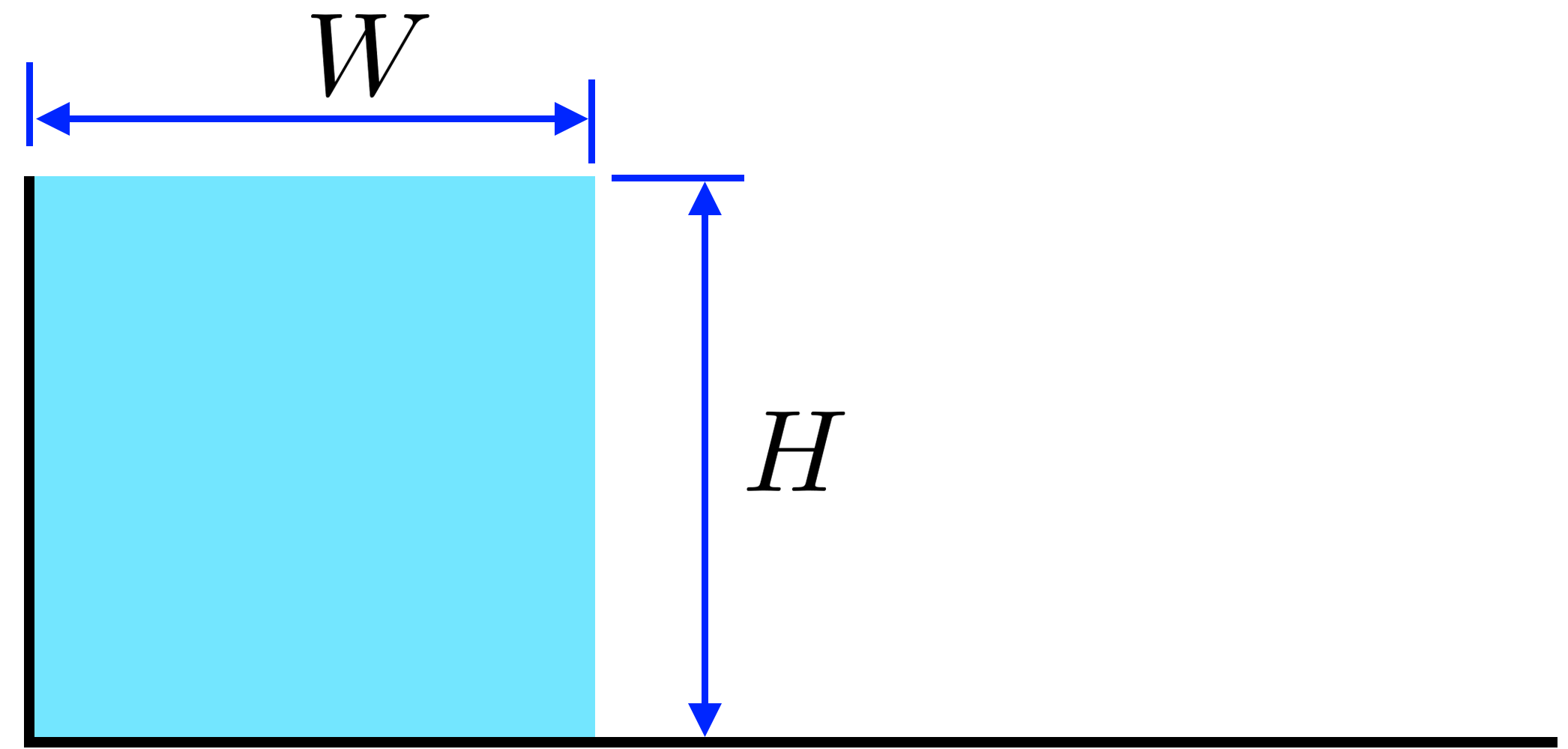}
		\caption{}
		\label{Fig:dbf_geometry_a}
	\end{subfigure}
	~ \hspace{5mm}
	\begin{subfigure}[b]{0.32\textwidth}	
		\includegraphics[trim=0 0 0 0,clip,width=1.00\linewidth]{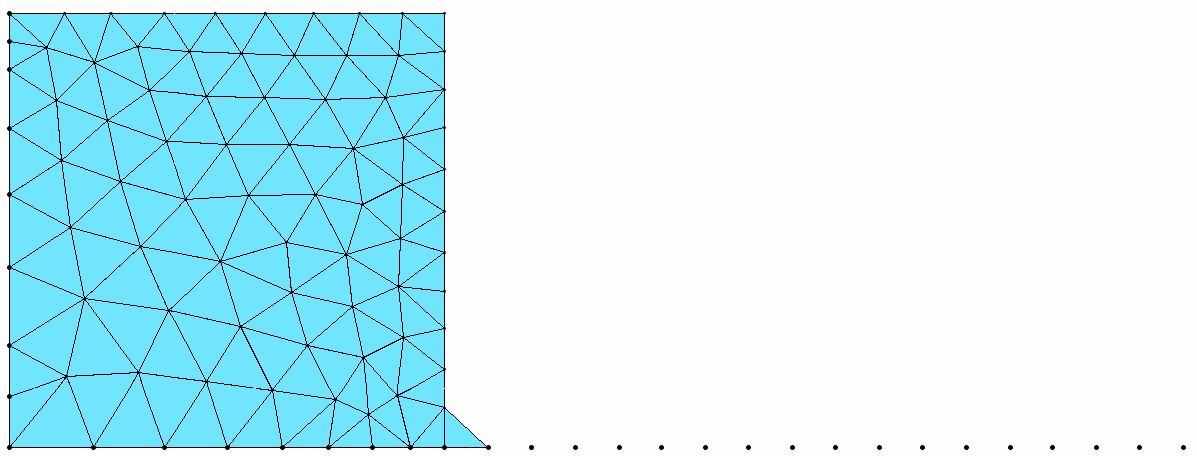}
		\caption{}
		\label{Fig:dbf_geometry_b}
	\end{subfigure}
	~ \hspace{5mm}
	\begin{subfigure}[b]{0.12\textwidth}	
		\includegraphics[trim=0 0 0 0,clip,width=1.00\linewidth]{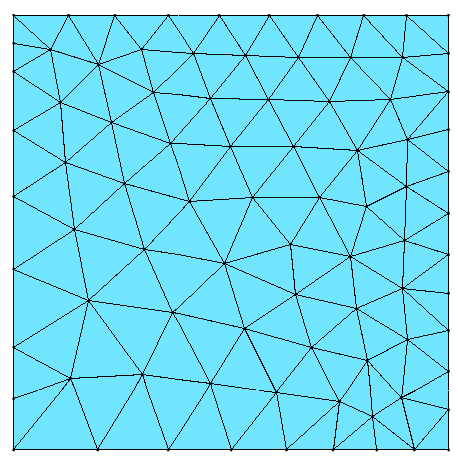}
		\caption{}
		\label{Fig:dbf_geometry_c}
	\end{subfigure}
	\\
	\vspace{-25mm}
	\hspace{-37mm} \footnotesize{(\textbf{a})} 
	\hspace{55mm} (\textbf{b})
	\hspace{56mm} (\textbf{c}) 
	\\
	\vspace{17mm}
	\caption{Water dam break problem. (a) Initial geometry. Initial discretisation using PFEM--AS with (b) no-slip condition, with (c) free-slip condition.}
\label{Fig:dbf_geometry}
\end{figure}

\subsection{Water dam break}

\begin{figure}[t!]
\captionsetup[subfigure]{labelformat=empty}
\centering 
	\begin{subfigure}[b]{0.48\textwidth}
		\includegraphics[trim=20 5 20 10,clip,width=1.00\linewidth]{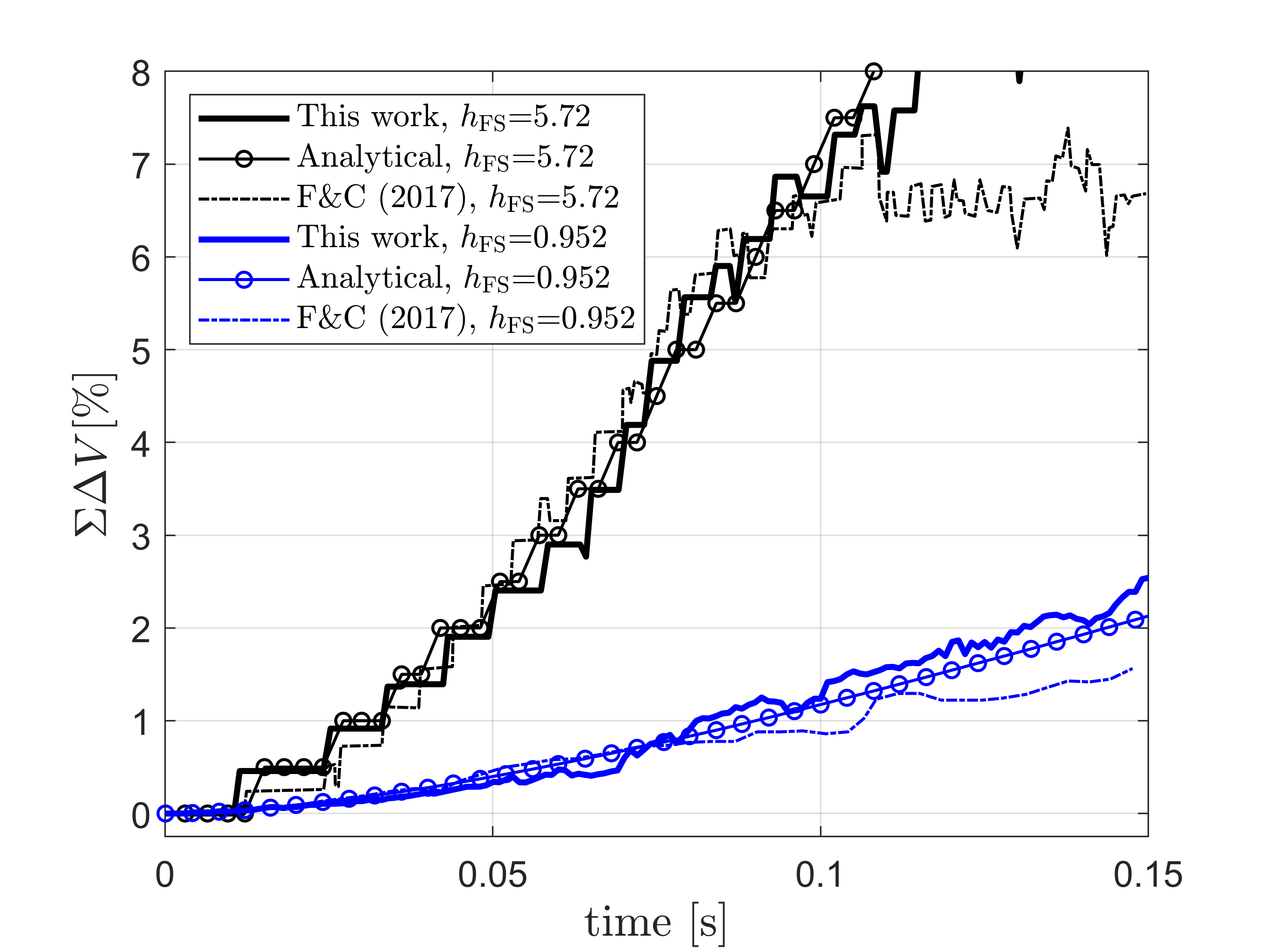}
		\caption{}
		\label{Fig:dbf_results1_a}
	\end{subfigure}
	~
	\begin{subfigure}[b]{0.48\textwidth}	
		\includegraphics[trim=20 5 20 10,clip,width=1.00\linewidth]{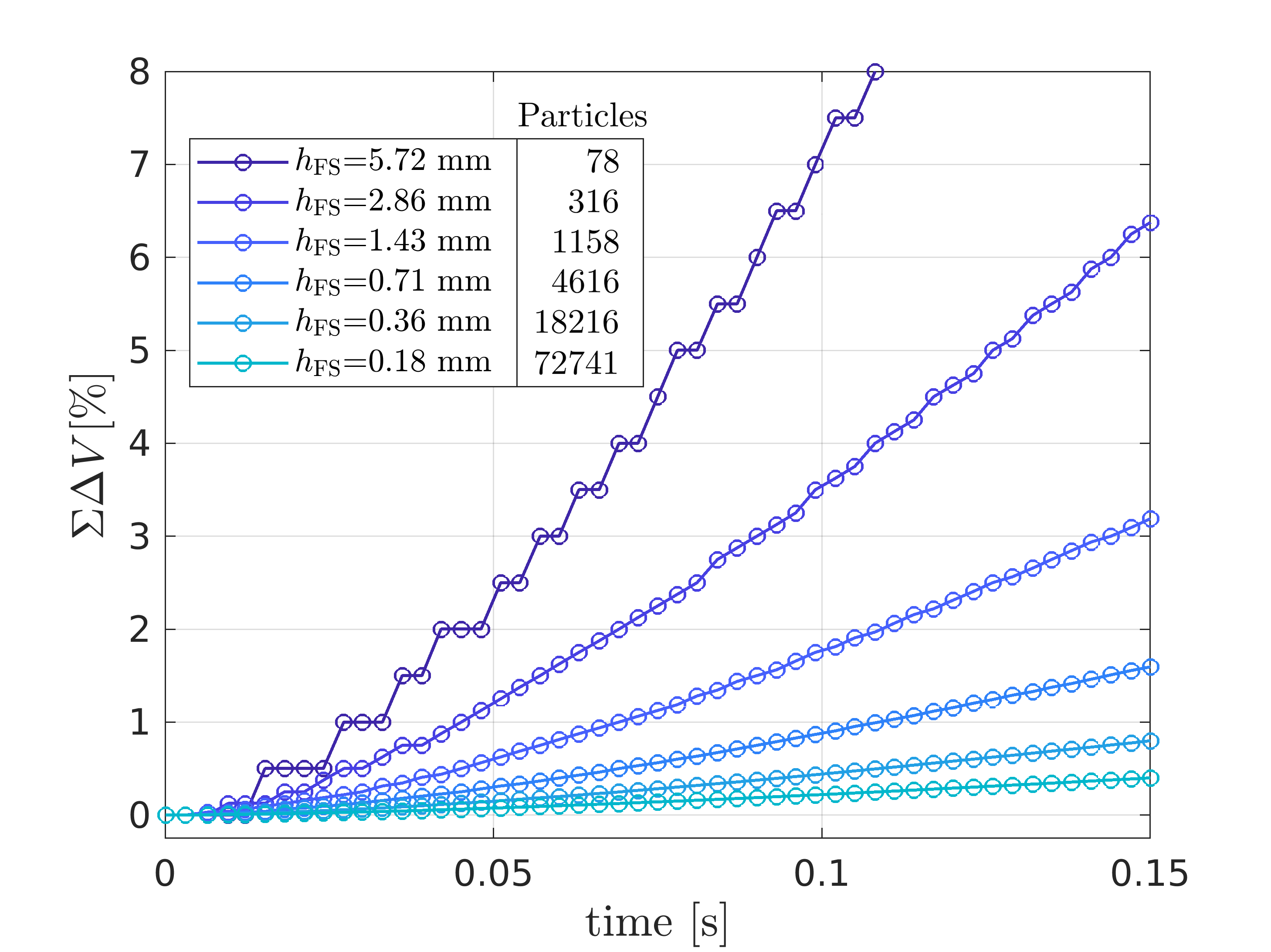}
		\caption{}
		\label{Fig:dbf_results1_b}
	\end{subfigure}
	\\
	\vspace{-67mm}
	\hspace{-80mm} \footnotesize{(\textbf{a})} \hspace{75mm} (\textbf{b}) 
	\\
	\vspace{57mm}
	\caption{Volume variation of the dam break problem. (a) Comparison of results againts those obtained by Franci and Cremonesi \citep{franci2017effect} (abbreviated as F$\&$C (2017)). Here, $h_\mathrm{FS}$ is expressed in mm and ``This work" means PFEM--AS and no--slip. (b) Analytical volume variation for different discretisations. For a better perception of the discretisation, the initial number of particles in a PFEM simulation is indicated in the right hand side of the legend.}
\label{Fig:dbf_results1}
\end{figure}

\begin{figure}[t!]
\captionsetup[subfigure]{labelformat=empty}
\centering 
	\begin{subfigure}[b]{1.0\textwidth}
		\includegraphics[trim=0 0 0 0,clip,width=0.90\linewidth]{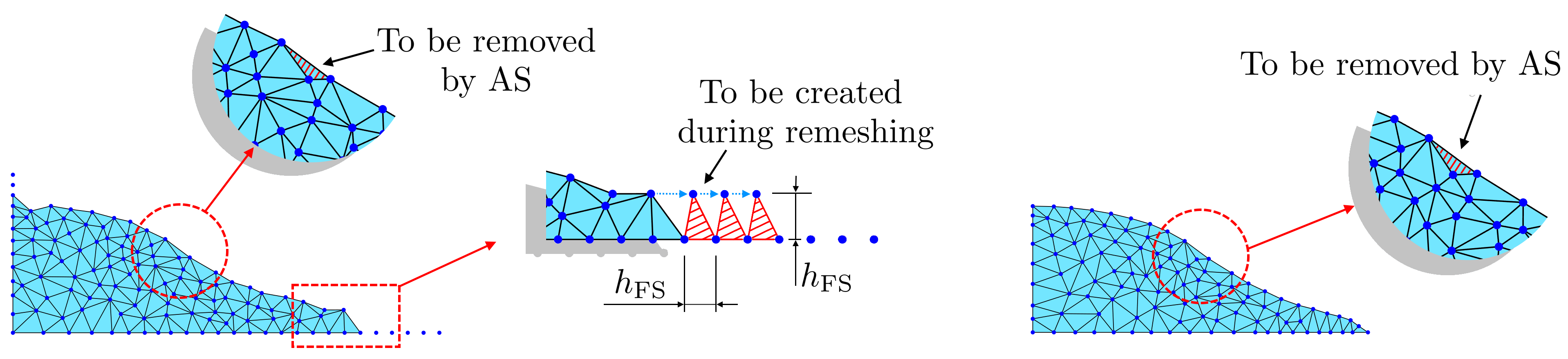}
		\caption{}
		\label{Fig:dbf_addtriangle_a}
	\end{subfigure}
	~
	\begin{subfigure}[b]{0.0\textwidth}
		\caption{}
		\label{Fig:dbf_addtriangle_b}
	\end{subfigure}
	\\
	\vspace{-43mm}
	\hspace{-65mm} \footnotesize{(\textbf{a})} 
	\hspace{90mm} (\textbf{b}) 
	\\
	\vspace{27mm}
	\caption{Two snapshots of the dam-break using PFEM--AS with (a) no--slip and (b) free--slip boundary condition. Close-up views in circular sections illustrate the loss of volume due to a stretching of the free--surface and the use of AS. The hatched elements in the rectangular close-up view illustrate the volume addition during remeshings due to the no-slip model.}
	\label{Fig:dbf_addtriangle}
\end{figure}

\begin{figure}[t!]
\captionsetup[subfigure]{labelformat=empty}
\centering 
	\begin{subfigure}[b]{0.48\textwidth}
		\includegraphics[trim=15 5 20 10,clip,width=1.00\linewidth]{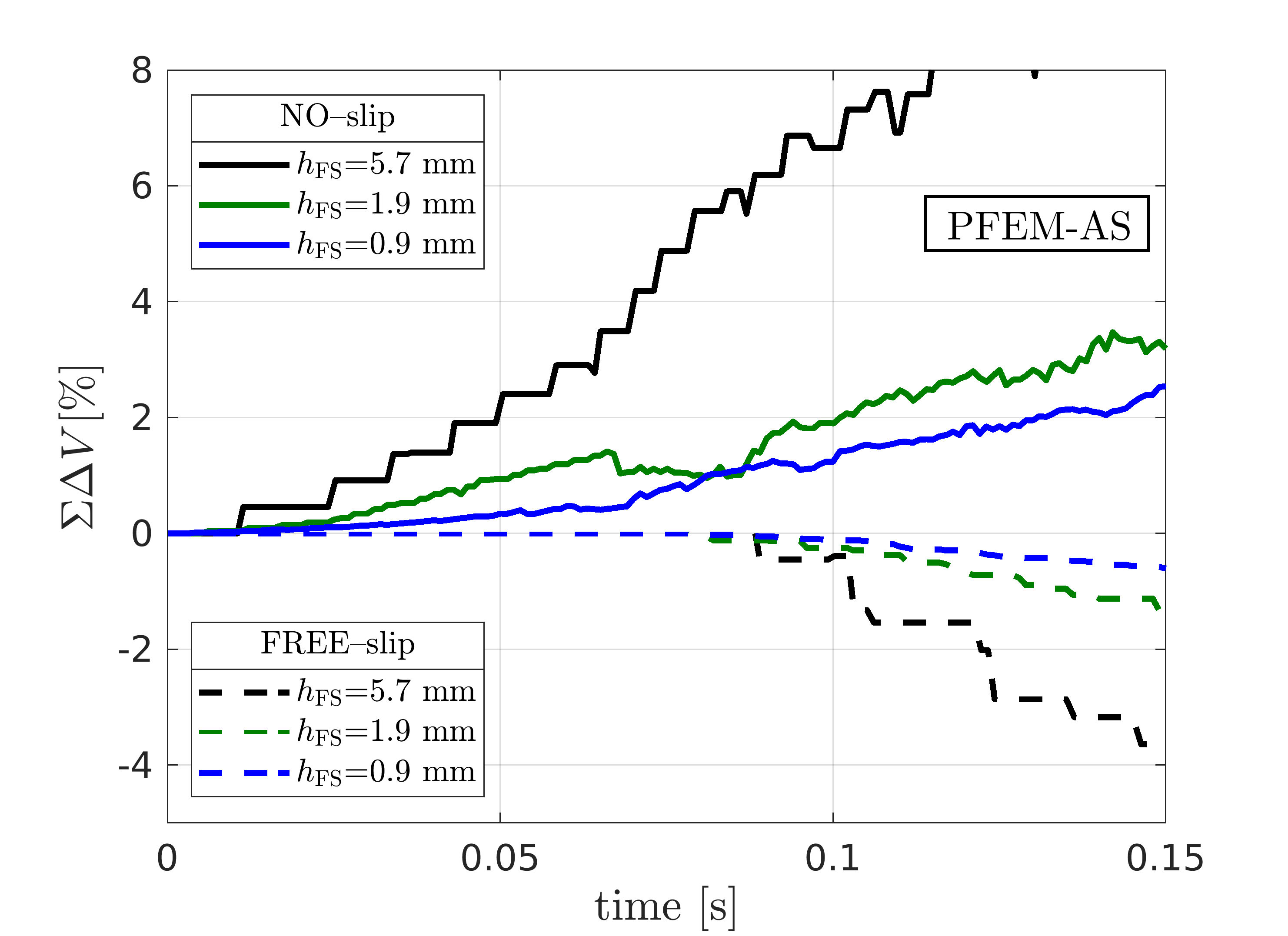}
		\caption{}
		\label{Fig:dbf_resultsASLS_a}
	\end{subfigure}
	~
	\begin{subfigure}[b]{0.48\textwidth}	
		\includegraphics[trim=15 5 20 10,clip,width=1.00\linewidth]{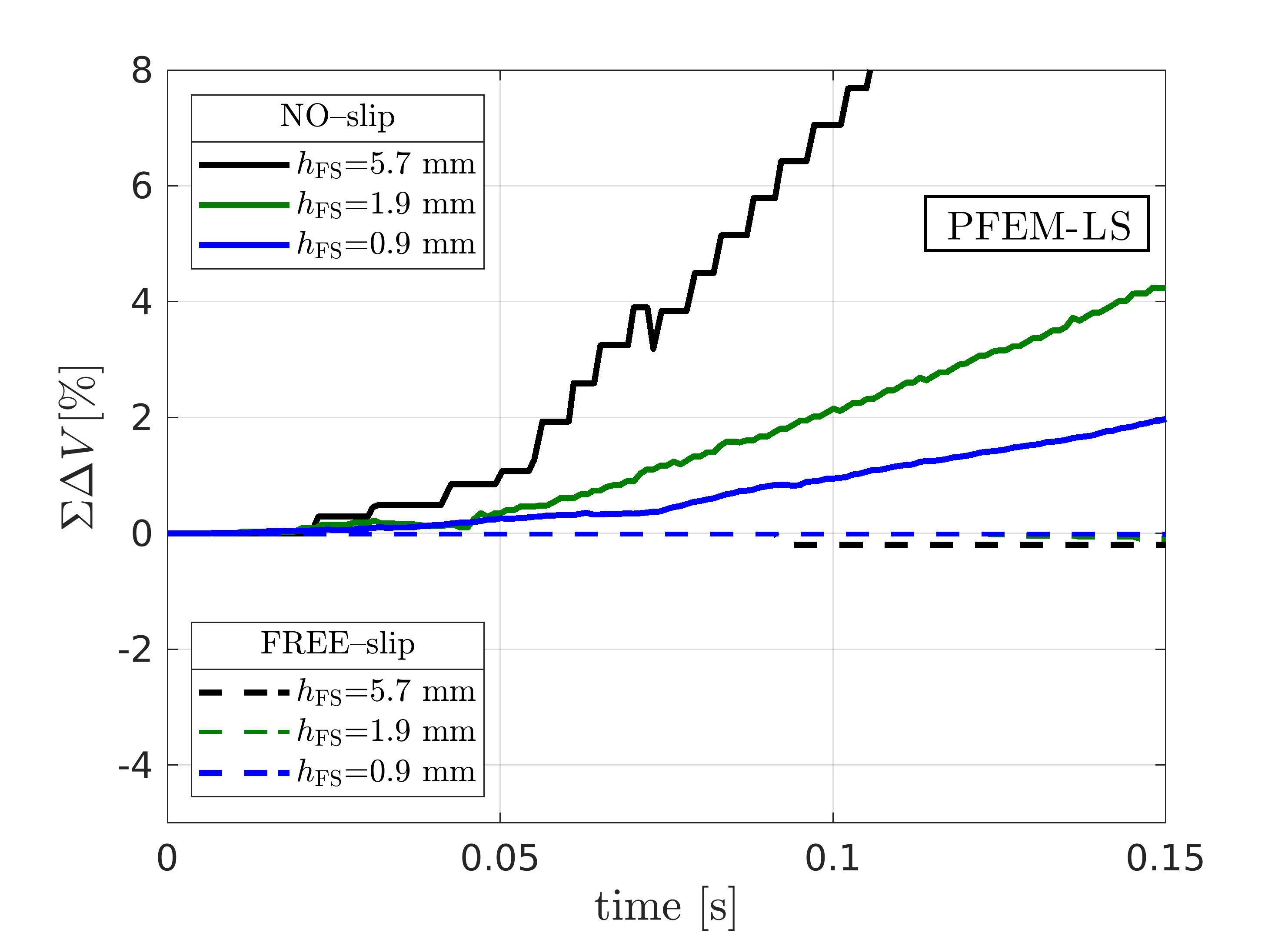}
		\caption{}
		\label{Fig:dbf_resultsASLS_b}
	\end{subfigure}
	\\
	\vspace{-65mm}
	\hspace{-80mm} \footnotesize{(\textbf{a})} \hspace{75mm} (\textbf{b}) 
	\\
	\vspace{55mm}
	\caption{Volume variation of the dam break problem using (a) PFEM--AS and (b) PFEM--LS. Each graph compares results with no--slip and free--slip boundary conditions.}
\label{Fig:dbf_resultsASLS}
\end{figure}

The problem concerns the collapse of a water column that flows on a horizontal plane. The geometry of the problem is shown in Fig.~\ref{Fig:dbf_geometry} and follows the configuration of Franci and Cremonesi \citep{franci2017effect}. Therefore, $H = W = 0.05715$ m and the fluid is defined by $\rho = 1000$ kg/m$^3$ and $\mu = 0.001$ Pa s. Following the study of \citep{franci2017effect}, three discretisations are considered, which are defined by $h_\mathrm{FS} =$ [5.72 , 1.90 , 0.95] mm, $h_\mathrm{max} = 2h_\mathrm{FS}$ and $\mathrm{d}_\mathrm{max} = H$. The problems are solved for two boundary conditions, free--slip and no--slip. For the latter, the solid surface is discretised with nodes spaced in $h_\mathrm{FS}$, as shown in Fig.~\ref{Fig:dbf_geometry_b}. Each case is solved with PFEM--AS and PFEM--LS. That is, 12 problems are solved in total (3 discretisations $\times$ 2 boundary conditions $\times$ 2 PFEM schemes). The volume variation due to remeshing is computed in each problem and the results are summarised in Figs.~\ref{Fig:dbf_results1}--\ref{Fig:dbf_resultsASLS}.

Fig.~\ref{Fig:dbf_results1_a} compares the volume variation obtained with PFEM--AS and no--slip boundary condition for two discretisations, $h_\mathrm{FS}$ = 5.72 (curves in black) and $h_\mathrm{FS}$ = 0.95 mm (curves in blue). The figure also includes the results of Franci and Cremonesi \citep{franci2017effect}, which are labelled as F$\&$C (2017) in the figure. The reference and obtained curves show good agreement, which validates the PFEM--AS implementation of our work. The graph also includes an analytical approximation of the volume variation. This approximation is built by assuming that the displacement of the wavefront creates triangular elements of size $h_\mathrm{FS}$, as illustrated in Fig.~\ref{Fig:dbf_addtriangle_a}. The wavefront position over time is obtained from a simulation using a refined mesh ($h_\mathrm{FS} = 0.5$ mm) and free--slip boundary condition. This information is used to estimate the wavefront displacement in terms of triangular finite elements and to compute the volume variation due to the adition of elements. The good agreement between the analytical curves and those obtained with PFEM--AS reinforces the observation that part of the volume variation in the remeshing process is due to a deficient modelling of the fluid--solid contact when dealing with no-slip boundary conditions \citep{cerquaglia2017free}. Although mesh refinement reduces the volume variation, this does not solve efficiently the non-conservation of volume in the dam break problem, since as shown in Fig.~\ref{Fig:dbf_results1_b}, an extremely refined discretisation would be needed to reach errors of $0.1\%$ for 0.15 seconds of simulation. 

In the following, results between PFEM--AS and PFEM--LS are compared, taking into account the 3 discretisations and both boundary conditions, free--slip and no--slip. The results are summarized in Fig.~\ref{Fig:dbf_resultsASLS}. By comparing the solid lines of both graphs (no--slip), no major differences are observed between volume variations obtained with PFEM--AS and PFEM--LS. This is because the main mechanism of volume variation is the non-slip flow over the solid surface, which is modelled by the addition of elements as the wave front moves. Since PFEM--LS does not involve a new fluid/solid contact model, it presents the same volume variation as PFEM--AS and that estimated analytically. Regarding the free-slip boundary condition in PFEM--AS (dashed lines), a smaller volume variation is clearly seen with respect to the no-slip condition, but now the remeshing process removes volume from the system. This is in line with the observations of Cerquaglia et al.~\citep{cerquaglia2017free}, who attribute the mass loss to the residual of the system of conservation equations. However, graphs in Fig.~\ref{Fig:dbf_resultsASLS} consider the volume variation due only to the remeshing process. Here, the negative volume variation is due to a degradation of the free surface as it stretches, which generates highly distorted elements that are eventually removed by AS, as illustrated in Figs.~\ref{Fig:dbf_addtriangle_a} and \ref{Fig:dbf_addtriangle_b}. Since PFEM--LS avoids the free--surface degradation, it exhibits small volume variations when used with free-slip boundary conditions.

\begin{figure}[t!]
\captionsetup[subfigure]{labelformat=empty}
\centering 
	\begin{subfigure}[b]{0.48\textwidth}
		\includegraphics[trim=35 110 40 100,clip,width=1.00\linewidth]{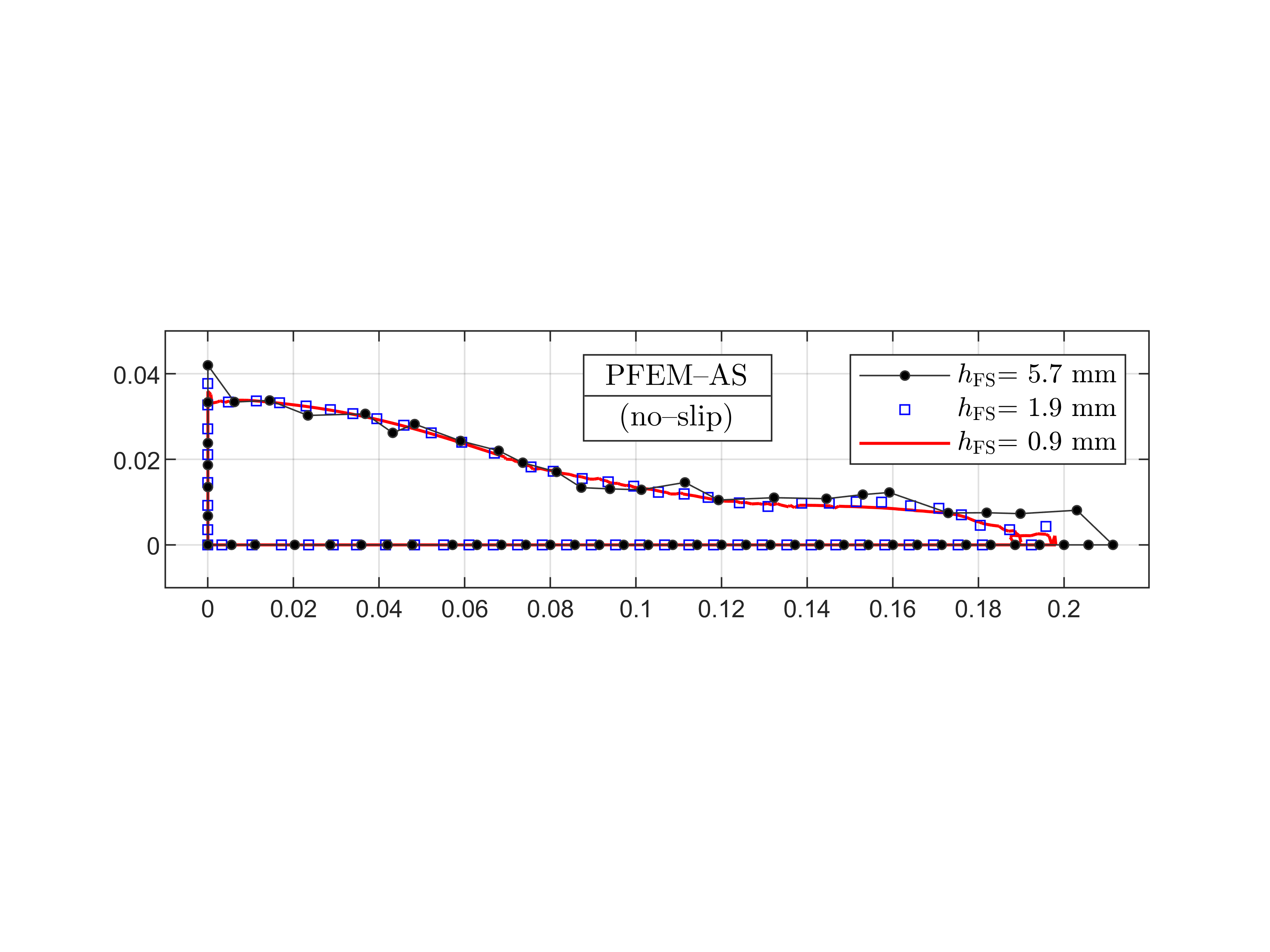}
		\caption{}
		\label{Fig:dbf_contour_a}
	\end{subfigure}
	~
	\begin{subfigure}[b]{0.48\textwidth}	
		\includegraphics[trim=35 110 40 100,clip,width=1.00\linewidth]{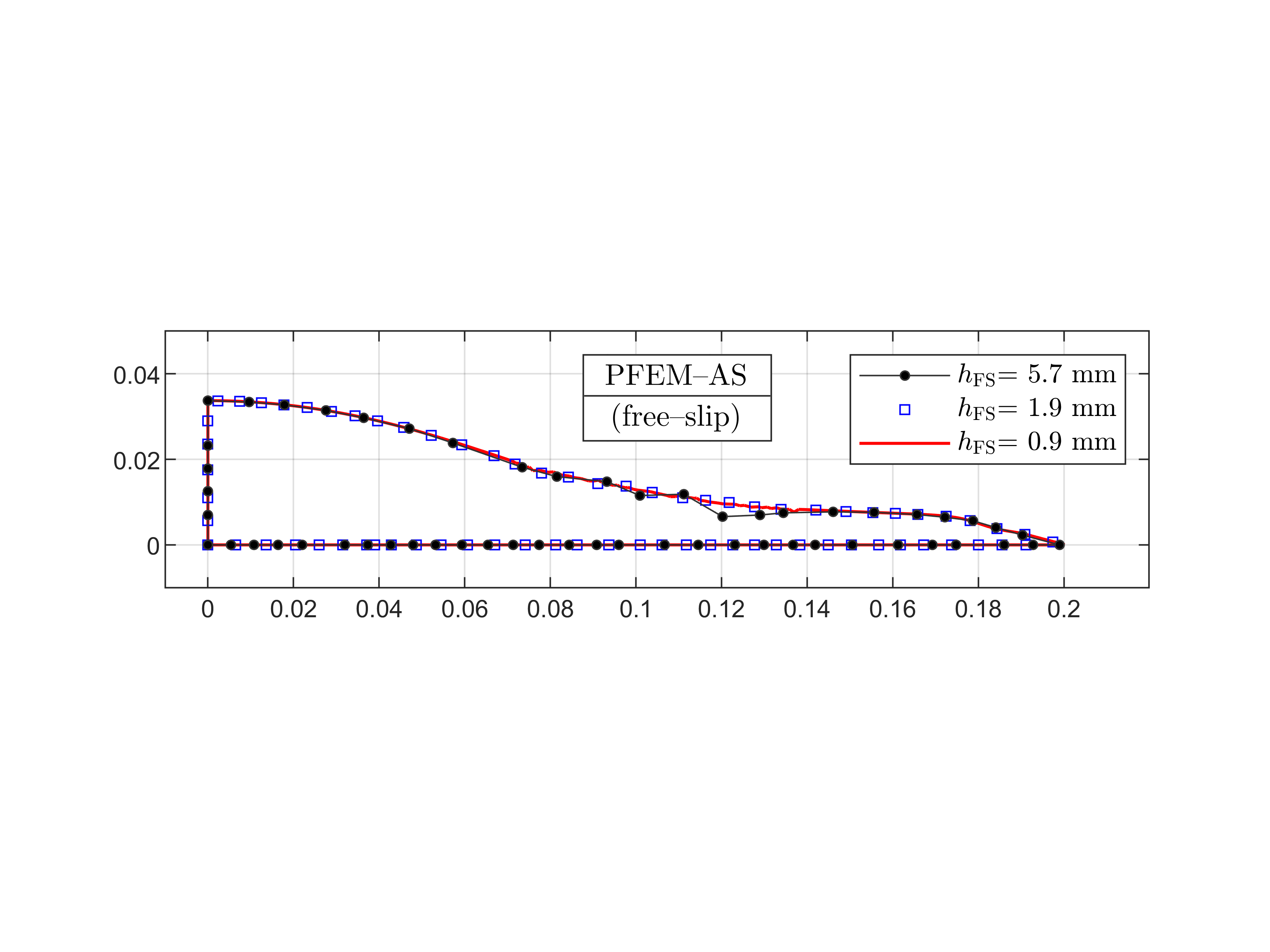}
		\caption{}
		\label{Fig:dbf_contour_b}
	\end{subfigure}
	\\[-1ex]
	\begin{subfigure}[b]{0.48\textwidth}
		\includegraphics[trim=35 110 40 100,clip,width=1.00\linewidth]{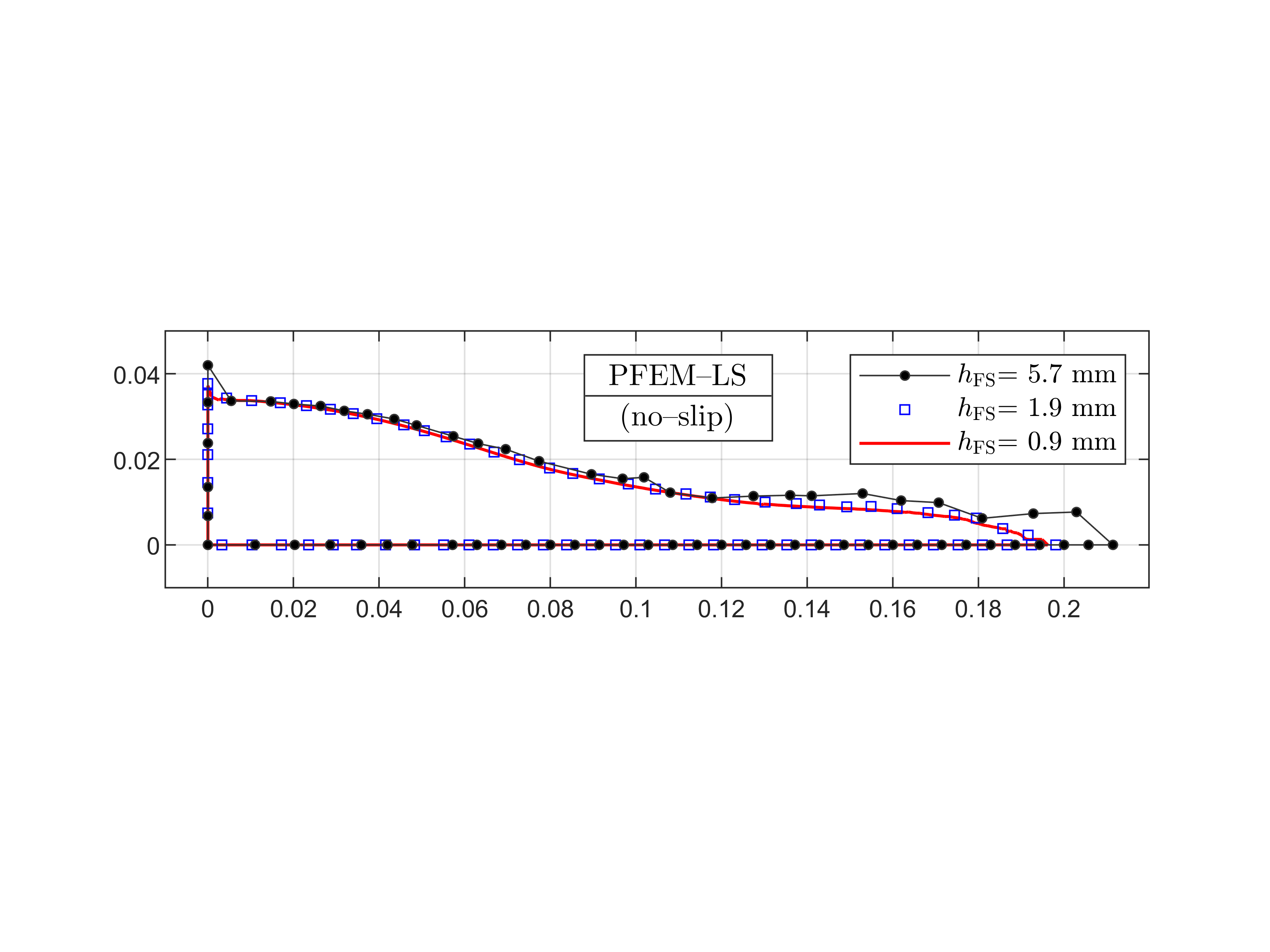}
		\caption{}
		\label{Fig:dbf_contour_c}
	\end{subfigure}
	~
	\begin{subfigure}[b]{0.48\textwidth}	
		\includegraphics[trim=35 110 40 100,clip,width=1.00\linewidth]{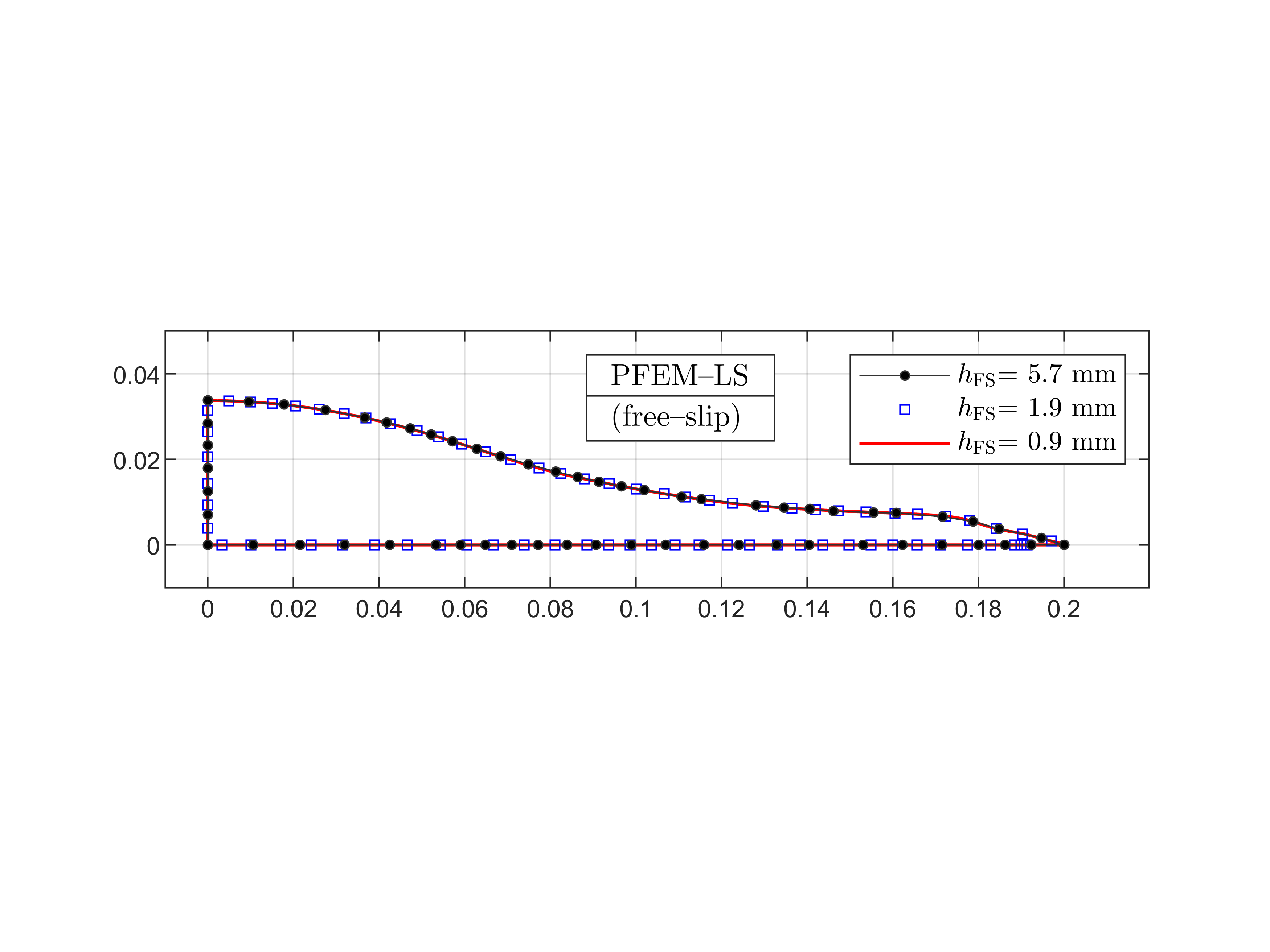}
		\caption{}
		\label{Fig:dbf_contour_d}
	\end{subfigure}
	\\
	\vspace{-57mm}
	\hspace{-79mm} \footnotesize{(\textbf{a})} \hspace{77mm} (\textbf{b}) 
	\\
	\vspace{25mm}
	\hspace{-79mm} \footnotesize{(\textbf{c})} \hspace{77mm} (\textbf{d}) 
	\\
	\vspace{19mm}
	\caption{Fluid boundaries at time 0.148 s for the 12 simulations of the dam break problem. Animations of these simulations can be found in \citep{YoutubeAll}.} 
\label{Fig:dbf_contour}
\end{figure}

In this dam break problem, the free-slip and no-slip boundary conditions give similar results, especially in the refined discretisations. As can be seen in Fig.~\ref{Fig:dbf_contour}, for both boundary conditions and for both PFEM formulations, the wavefronts have practically the same position after 0.148 s of simulation. As remarked by Cerquaglia et al.~\citep{cerquaglia2017free}, the slip boundary condition is still representative of a no--slip condition when the fluid/solid contact involves a thin boundary layer, which occurs in the presence of high Reynolds number caused either by a high characteristic velocity or low dynamic viscosity. In other words, although the free--slip boundary condition allows to reduce the volume variation in the remeshing process of PFEM, its applicability is not general and is limited to some problems. However, the free--slip condition will be used in some of the following examples regardless of its representativeness of the fluid/solid contact phenomenon, because it allows to isolate the sources of volume variation during remeshing.

\subsection{Impact of a coarse fluid drop}

\begin{figure}[t!]
\captionsetup[subfigure]{labelformat=empty}
\centering 
	\begin{subfigure}[b]{0.31\textwidth}
		\includegraphics[trim=0 0 0 0,clip,width=1.00\linewidth]{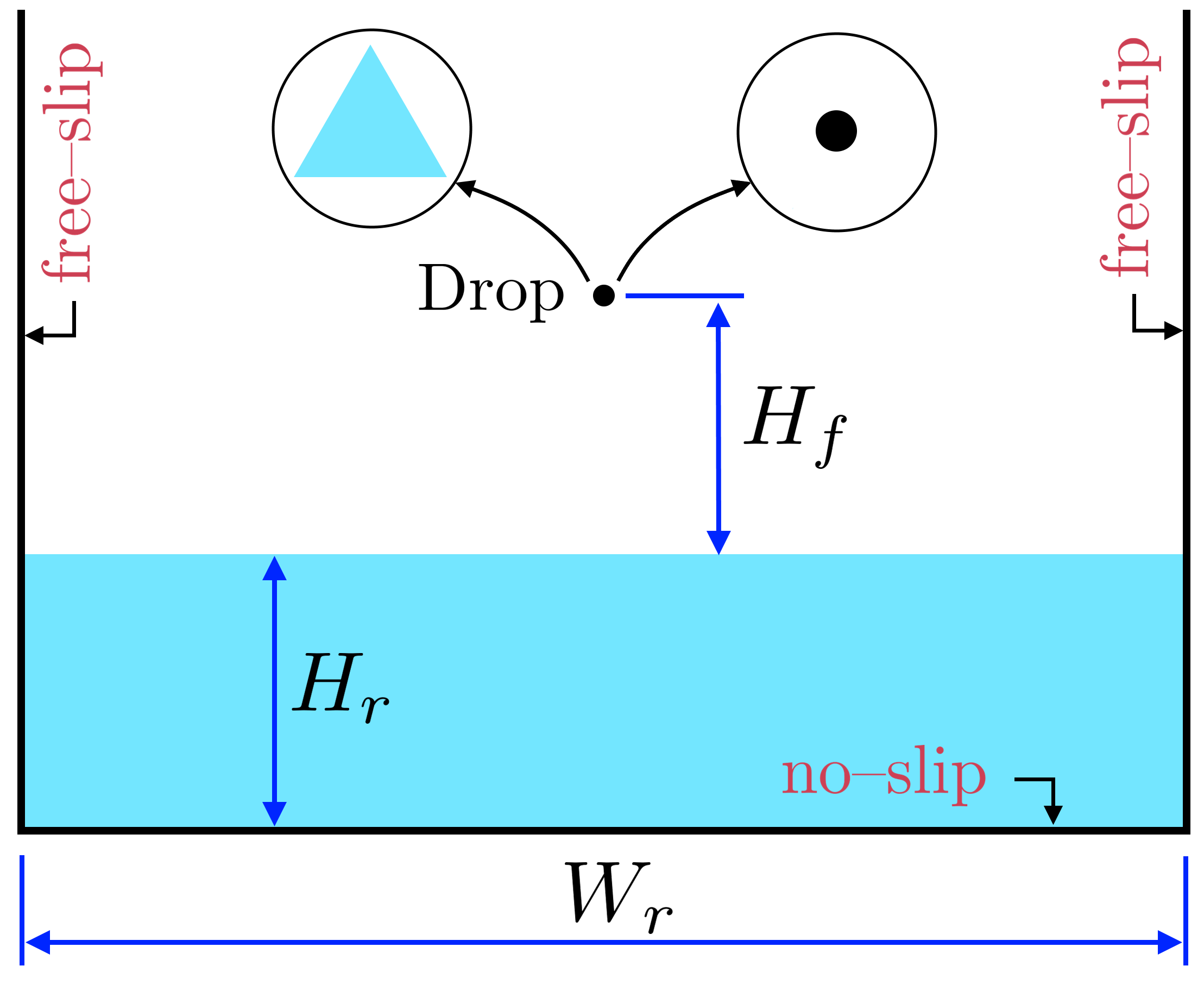}
		\caption{}
		\label{Fig:drop_geometry_a}
	\end{subfigure}
	~
	\begin{subfigure}[b]{0.63\textwidth}	
		\includegraphics[trim=0 0 0 0,clip,width=1.00\linewidth]{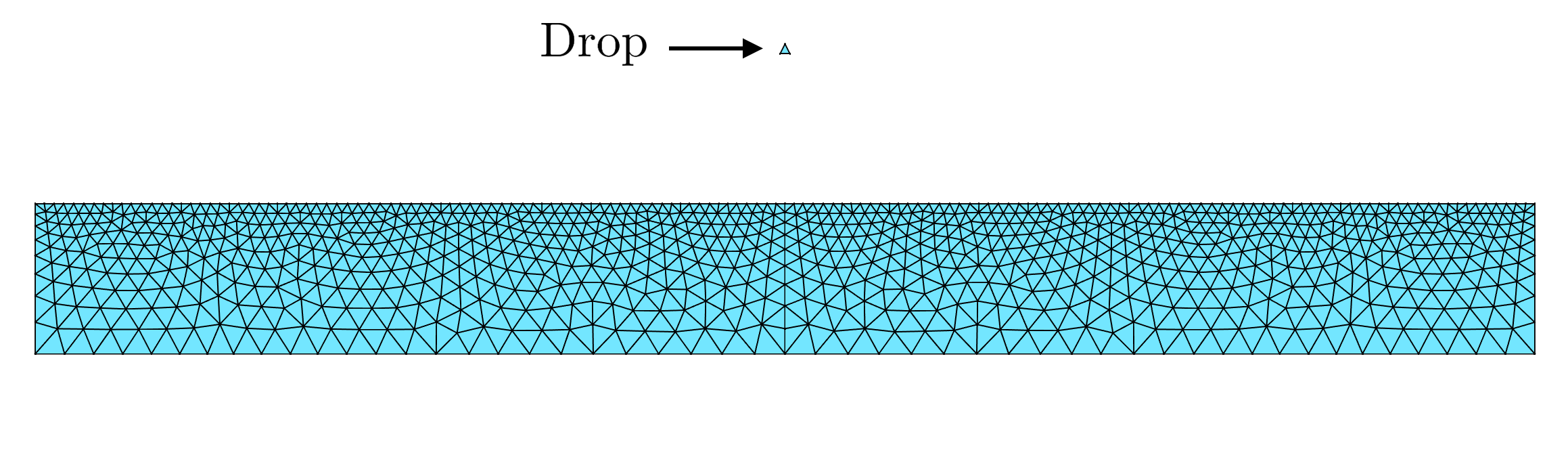}
		\caption{}
		\label{Fig:drop_geometry_b}
	\end{subfigure}
	\\
	\vspace{-48mm}
	\hspace{-165mm} \footnotesize{(\textbf{a})} 
	\vspace{10mm}
	\\ 
	\hspace{-44mm} (\textbf{b}) 
	\\
	\vspace{27mm}
\caption{Fall of a triangular water drop. (a) Geometry and illustration of the two triangular drops that are analysed, and (b) space discretisation.
}
\label{Fig:drop_geometry}
\end{figure}

\EF{
This problem consists of simulating the fall and impact of a drop of water in PFEM. The idea is to analyse the influence on the mass and energy variations when a drop is discretised by a single element in PFEM--AS and PFEM--LS. It can be assumed that the drop originates from splashing, i.e.~it may represent an element or a particle that is detached from the free surface. For simplicity, in this work} the drop is released with zero initial velocity at height $H_f$ from the free surface, as shown in Fig.~\ref{Fig:drop_geometry_a}. The free surface elevation is $H_r=0.25$ m from the bottom of the reservoir, which is $W_r=2.50$ m wide. The fluid properties are $\rho = 1000$ kg/m$^3$ and $\mu = 0.001$ Pa s. The domain is discretised with a non--uniform mesh size following the expression given in Fig.~\ref{Fig:meshref}, with $h_\mathrm{FS}$ = 0.016 m, $h_\mathrm{max}$ = 3$h_\mathrm{FS}$, and $\mathrm{d}_\mathrm{max} = H_r$. The discretised domain is shown in Fig.~\ref{Fig:drop_geometry_b}.

\begin{figure}[t!]
\captionsetup[subfigure]{labelformat=empty}
\centering 
	\begin{subfigure}[b]{0.48\textwidth}
		\includegraphics[trim=20 72 20 90,clip,width=1.00\linewidth]{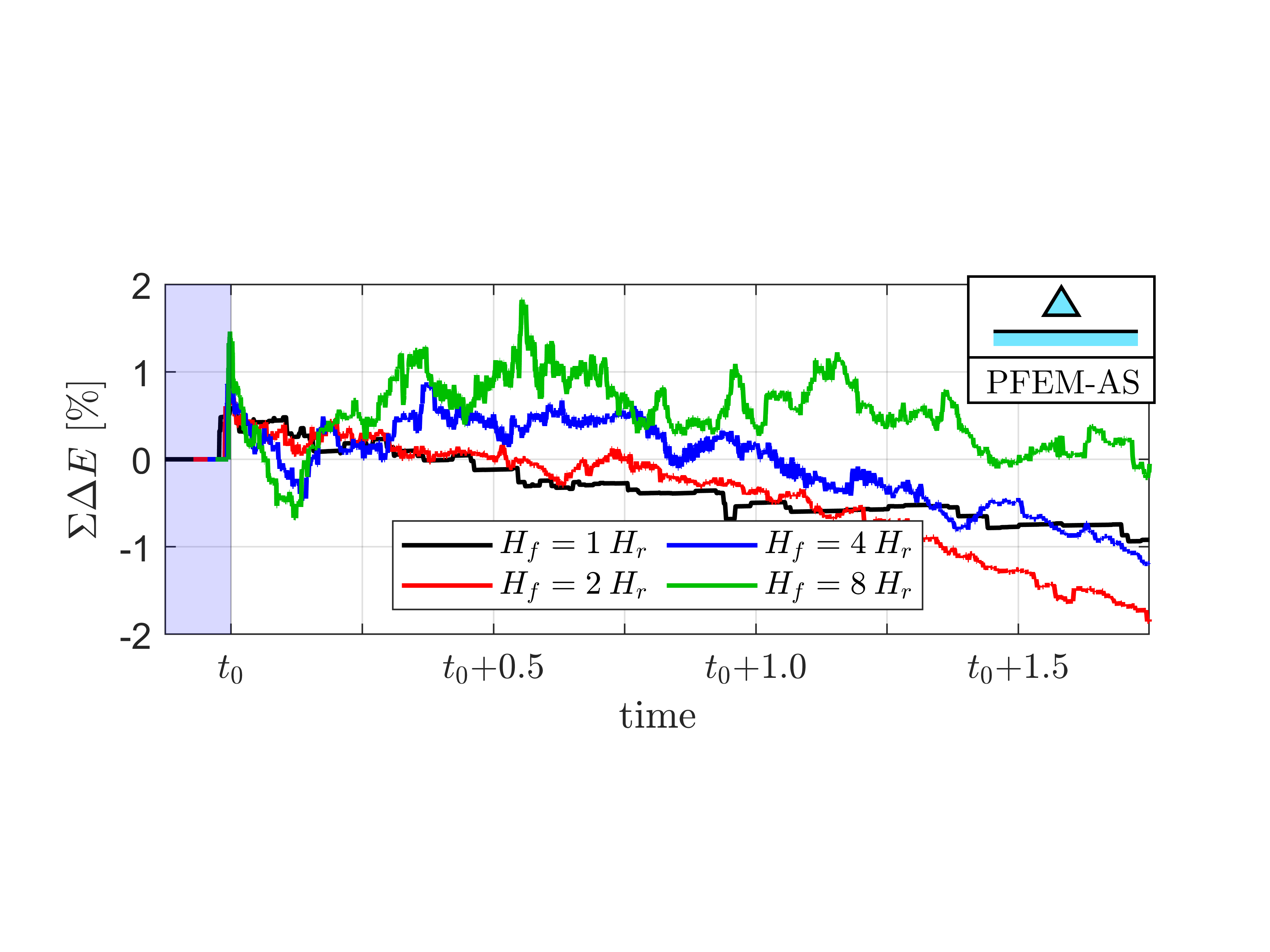}
		\caption{}
		\label{Fig:dropE_a}
	\end{subfigure}
	~
	\begin{subfigure}[b]{0.48\textwidth}	
		\includegraphics[trim=20 72 20 90,clip,width=1.00\linewidth]{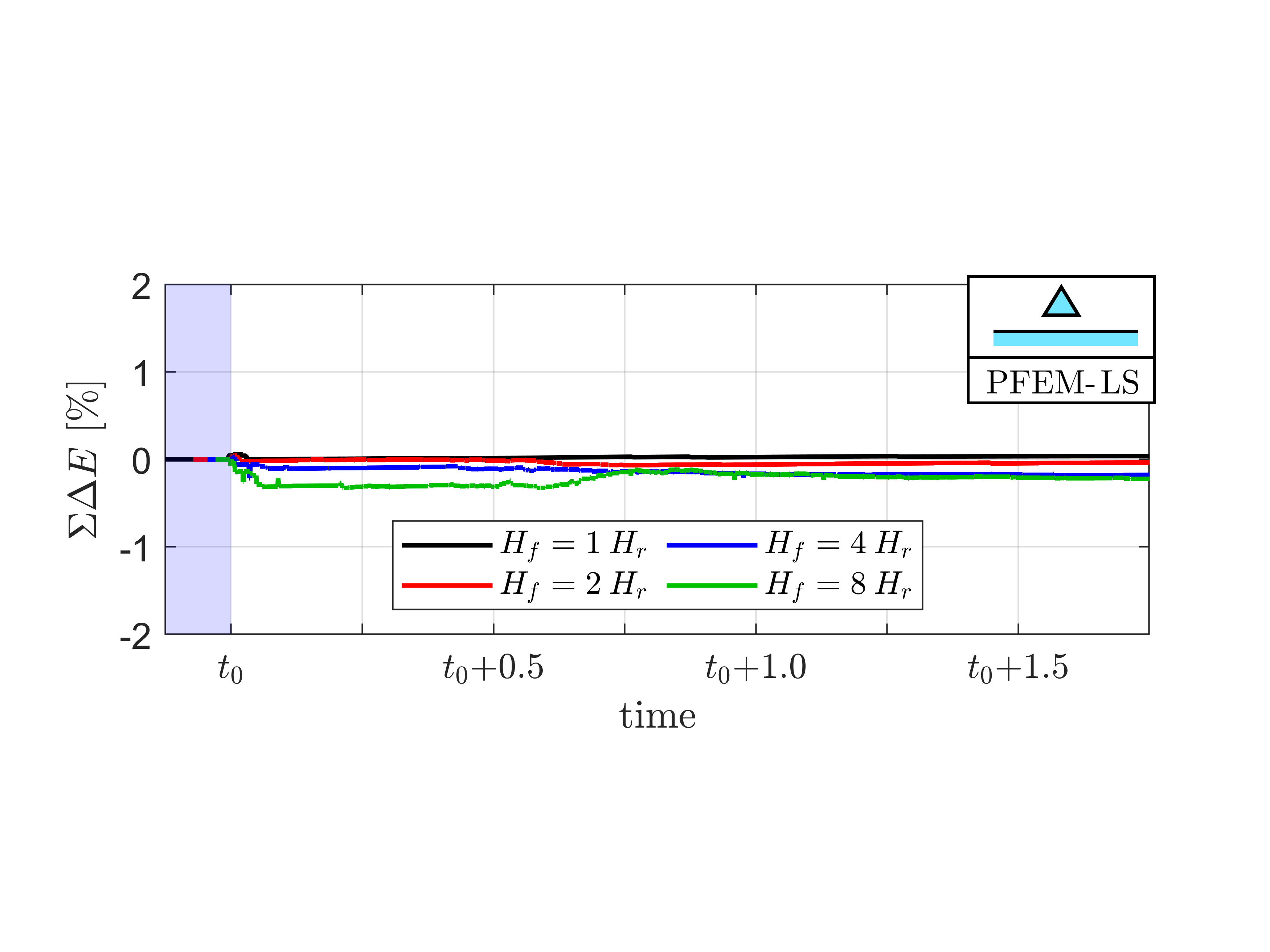}
		\caption{}
		\label{Fig:dropE_b}
	\end{subfigure}
	\vspace{-2mm}\\
	\begin{subfigure}[b]{0.48\textwidth}
		\includegraphics[trim=20 72 20 90,clip,width=1.00\linewidth]{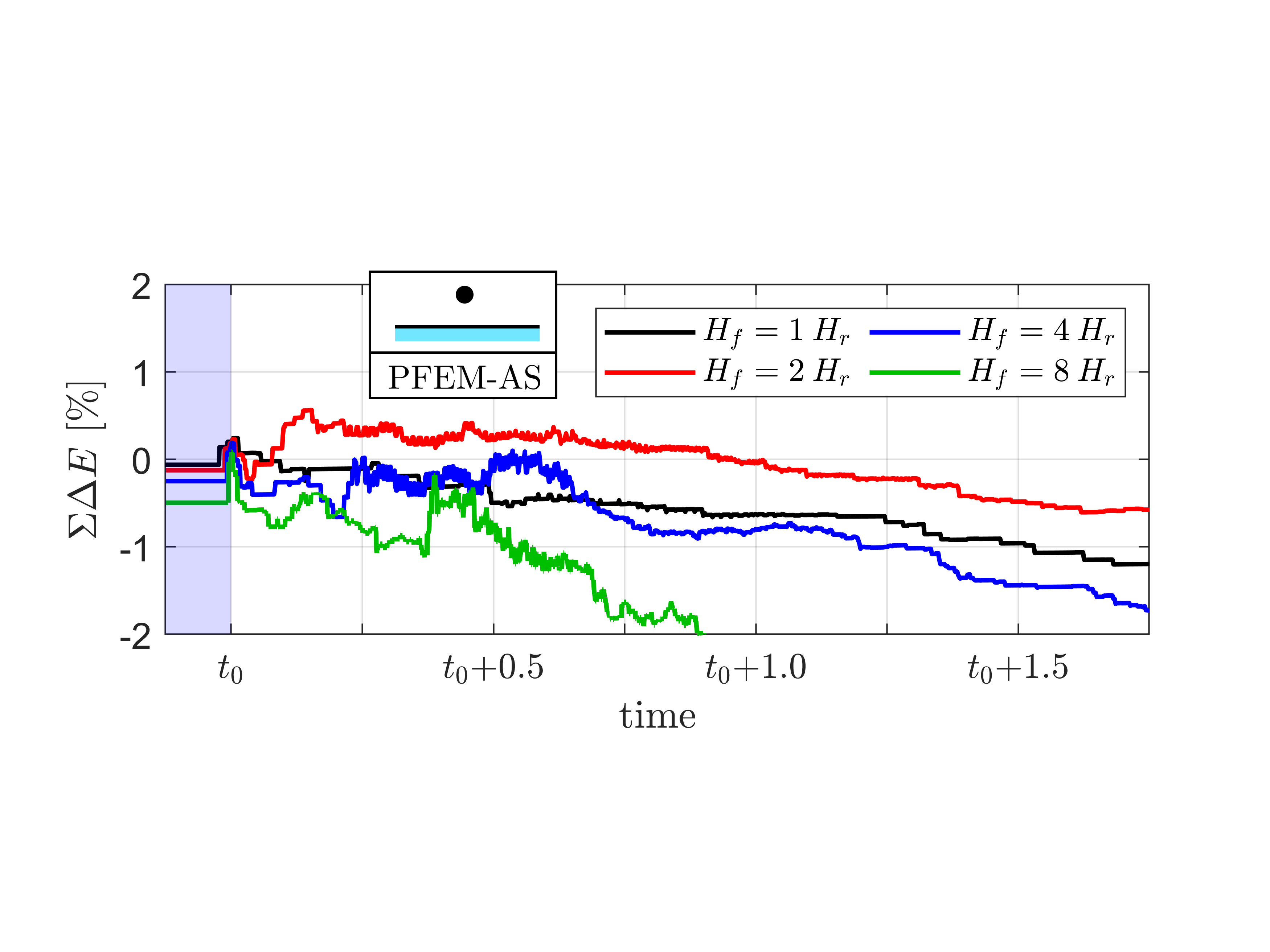}
		\caption{}
		\label{Fig:dropE_e}
	\end{subfigure}
	~
	\begin{subfigure}[b]{0.48\textwidth}	
		\includegraphics[trim=20 72 20 90,clip,width=1.00\linewidth]{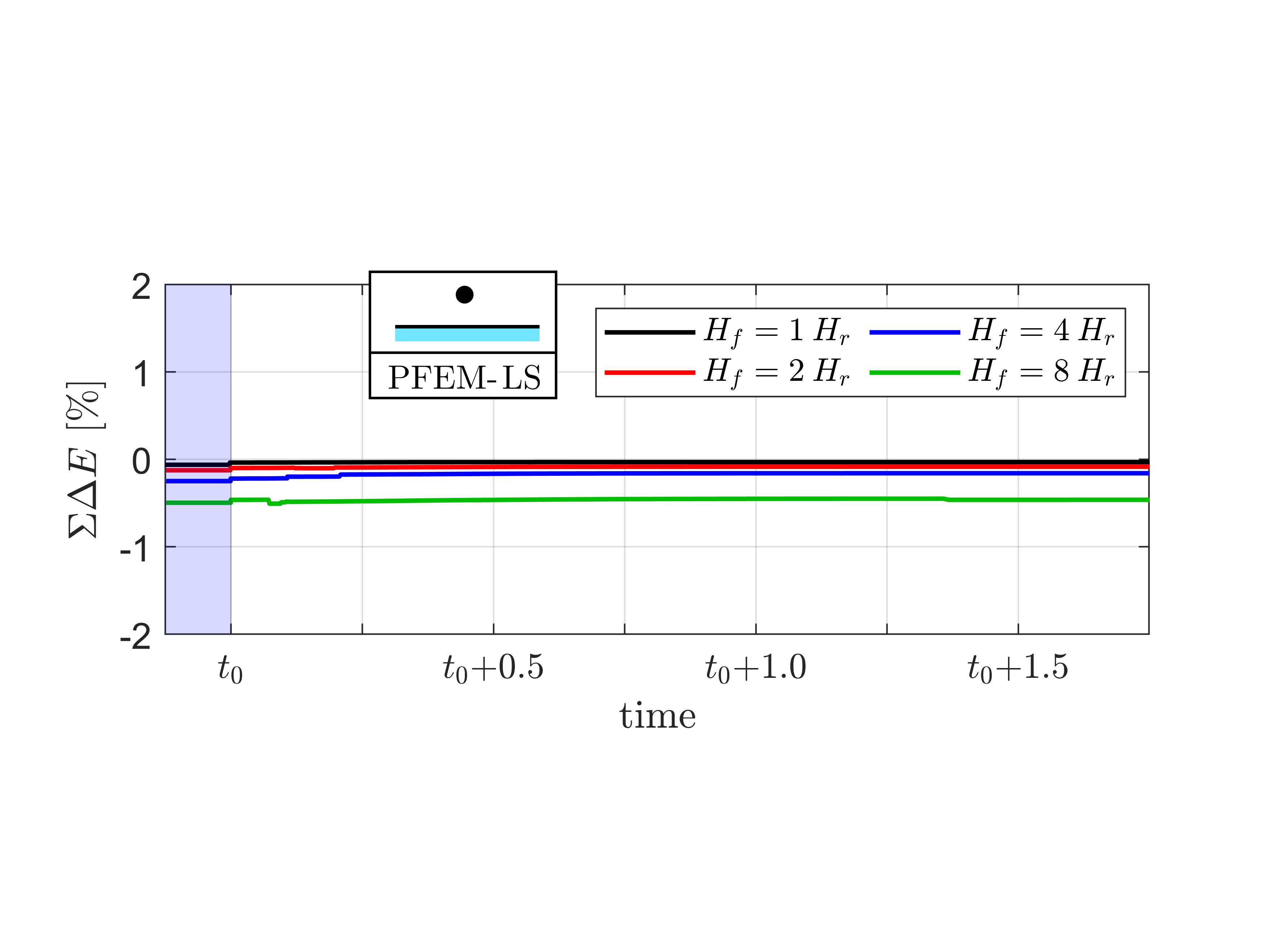}
		\caption{}
		\label{Fig:dropE_f}
	\end{subfigure}
	\\
	\vspace{-73mm}
	\hspace{-80mm} \footnotesize{(\textbf{a})} \hspace{75mm} (\textbf{b}) 
	\\
	\vspace{32mm}
	\hspace{-80mm} (\textbf{c}) \hspace{75mm} (\textbf{d})  
	\vspace{29mm}
\caption{Energy variation during the simulation of a fluid drop impacting a fluid. PFEM simulations using (a,c) Alpha--Shape and (b,d) Level--Set. The drop is discretised using (a,b) a triangular element, and (e,f) a single node. The zone coloured in purple represents the time before the analytical impact ($t_0$).}
\label{Fig:dropE}
\end{figure}

\begin{figure}[t!]
\captionsetup[subfigure]{labelformat=empty}
\centering 
	\begin{subfigure}[b]{0.48\textwidth}
		\includegraphics[trim=20 85 20 95,clip,width=1.00\linewidth]{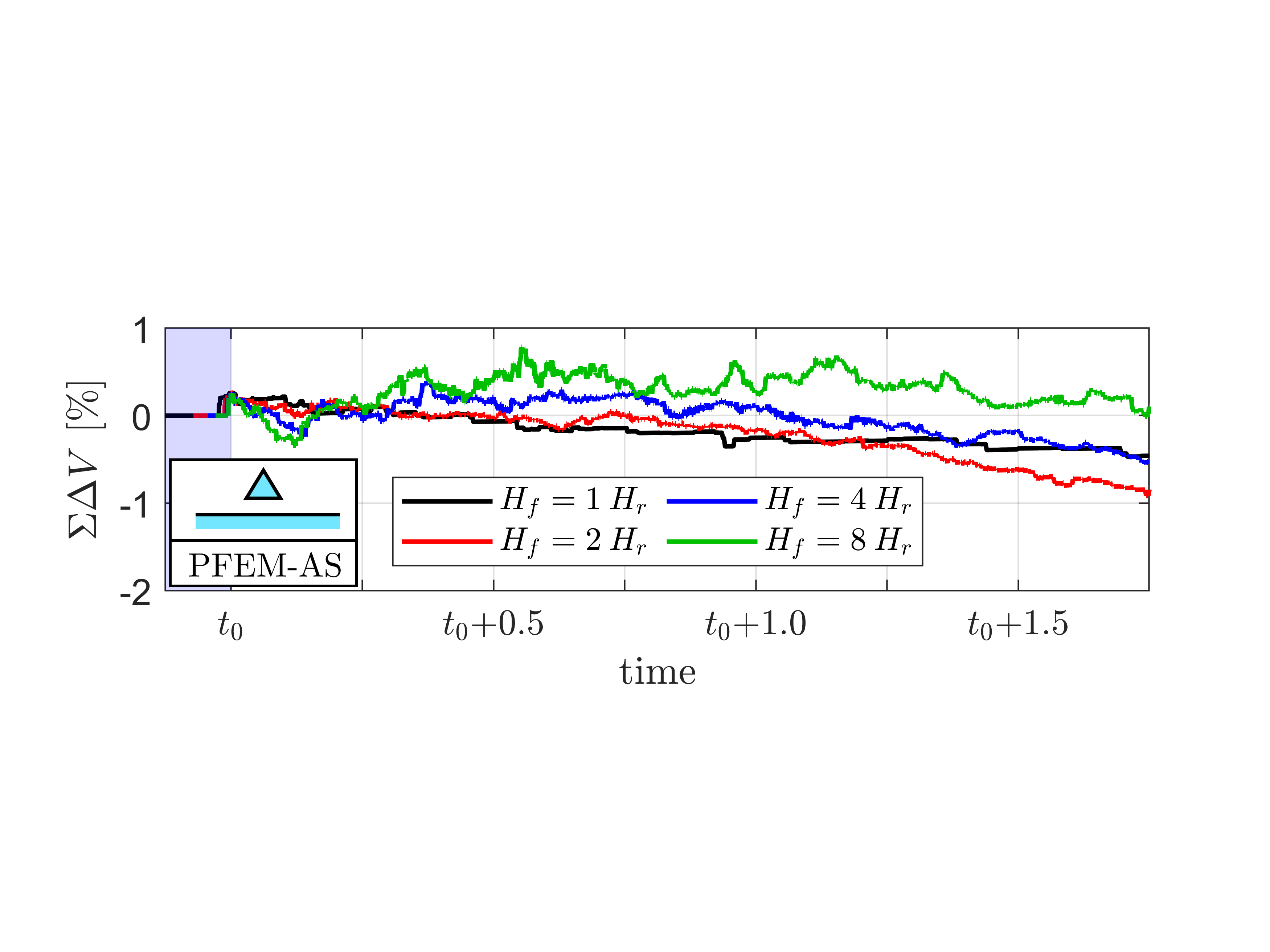}
		\caption{}
		\label{Fig:dropV_a}
	\end{subfigure}
	~
	\begin{subfigure}[b]{0.48\textwidth}	
		\includegraphics[trim=20 85 20 95,clip,width=1.00\linewidth]{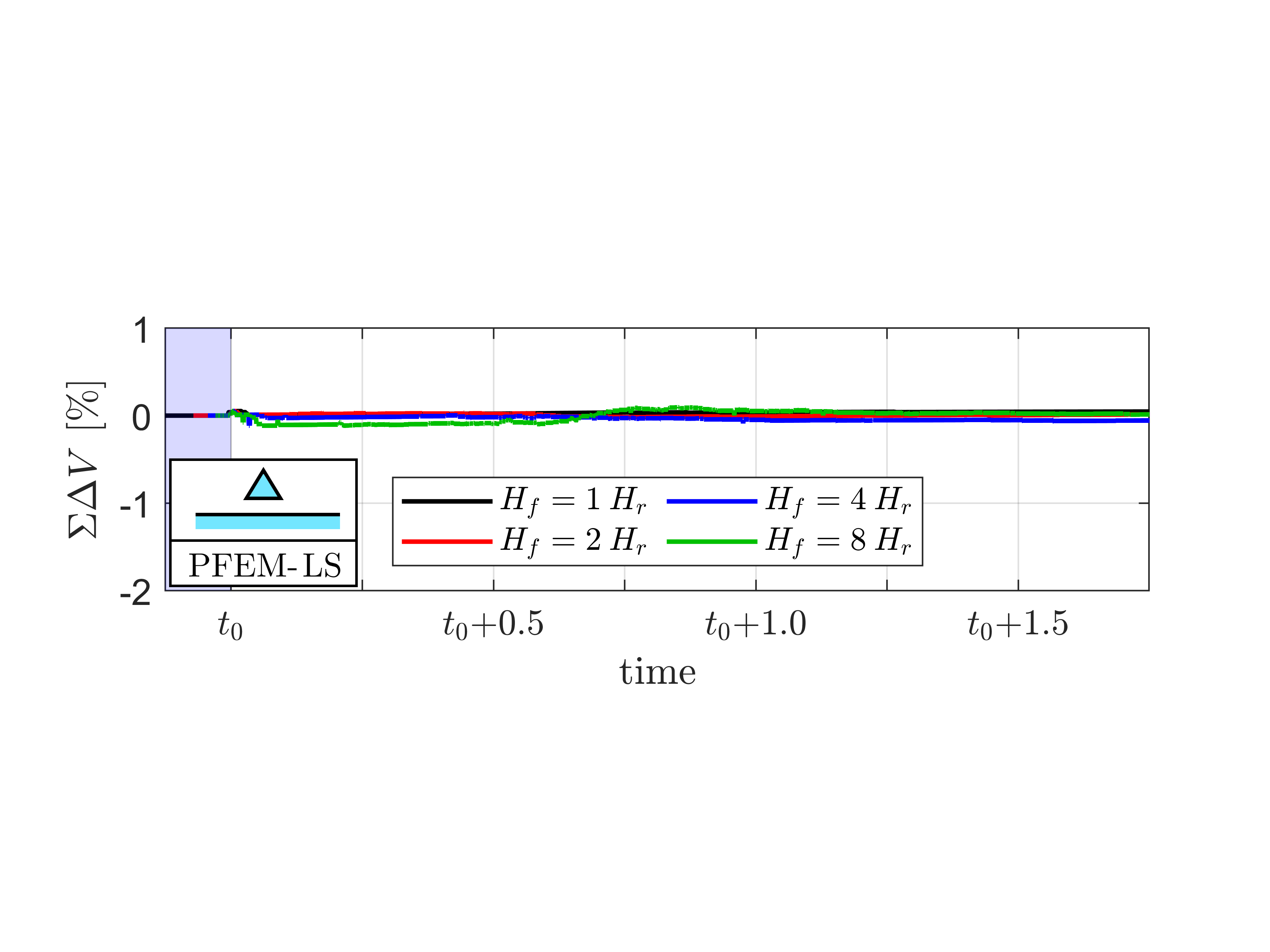}
		\caption{}
		\label{Fig:dropV_b}
	\end{subfigure}
	\vspace{-4mm}\\
	\begin{subfigure}[b]{0.48\textwidth}
		\includegraphics[trim=20 85 20 95,clip,width=1.00\linewidth]{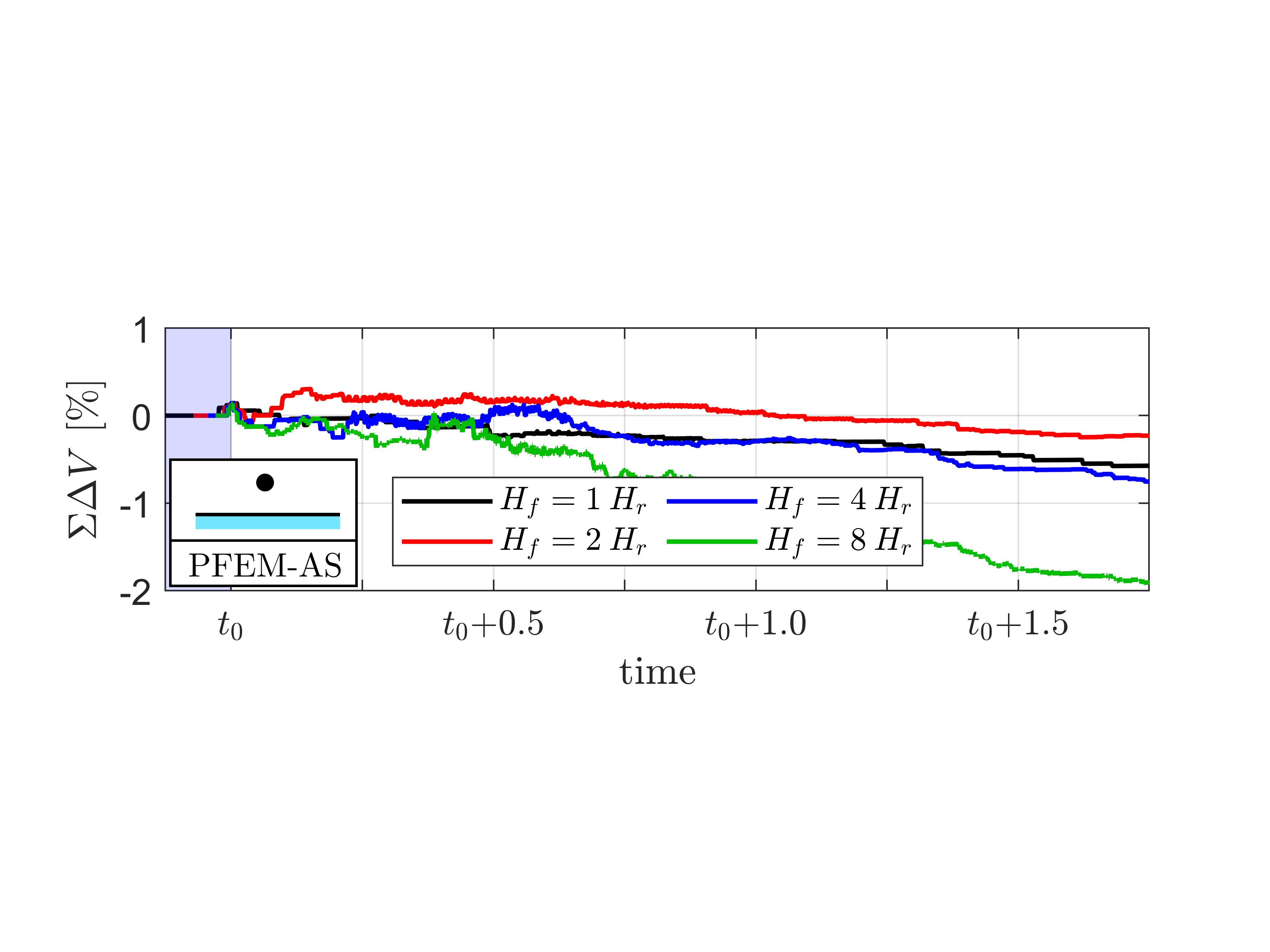}
		\caption{}
		\label{Fig:dropV_e}
	\end{subfigure}
	~
	\begin{subfigure}[b]{0.48\textwidth}	
		\includegraphics[trim=20 85 20 95,clip,width=1.00\linewidth]{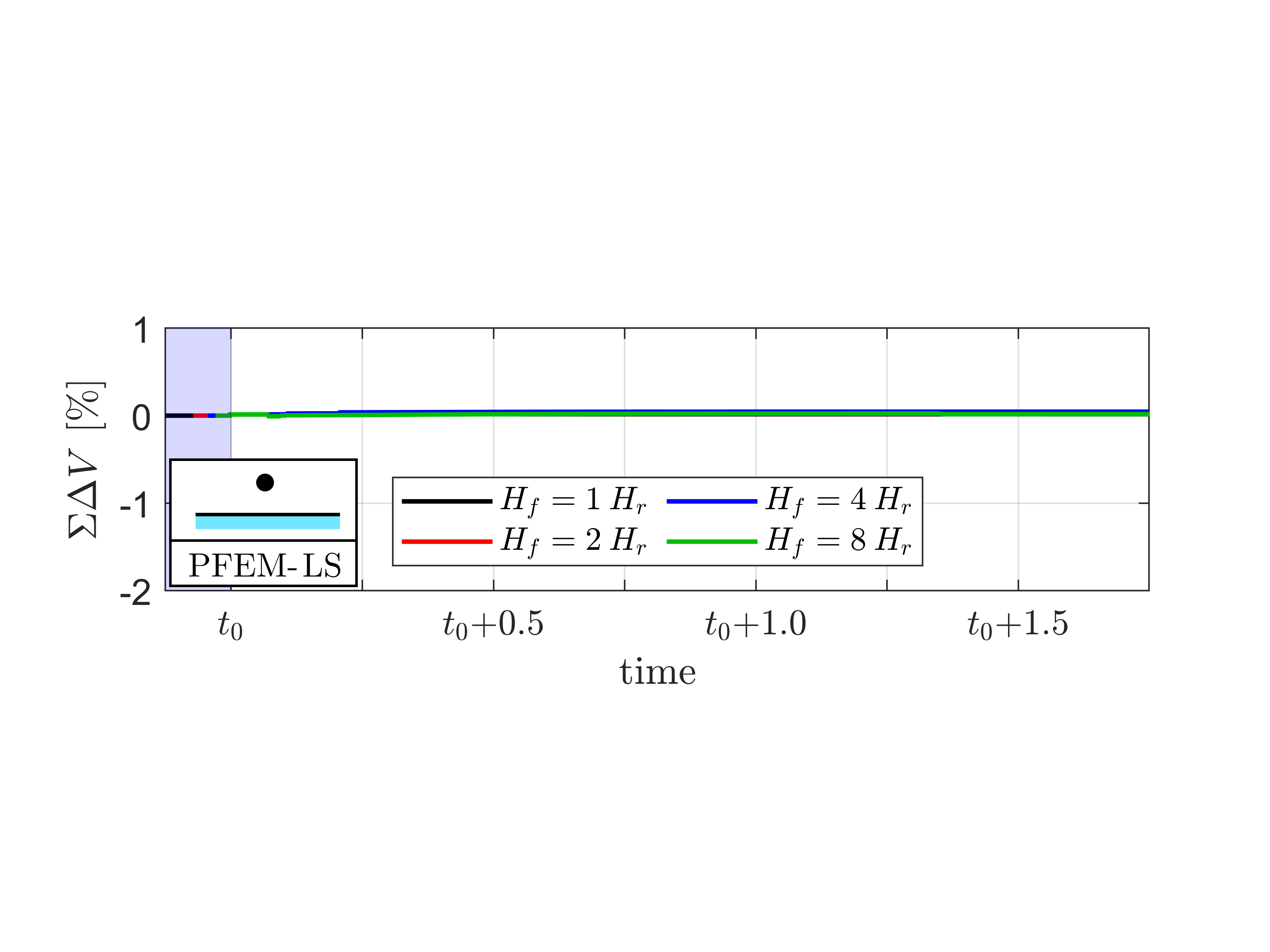}
		\caption{}
		\label{Fig:dropV_f}
	\end{subfigure}
	\\
	\vspace{-62mm}
	\hspace{-80mm} \footnotesize{(\textbf{a})} \hspace{75mm} (\textbf{b}) 
	\\
	\vspace{26mm}
	\hspace{-80mm} (\textbf{c}) \hspace{75mm} (\textbf{d}) 
	\\
	\vspace{23mm}
\caption{Volume variation during the simulation of a water drop hitting a fluid. PFEM simulations using (a,c) Alpha--Shape (AS) and (b,d) Level--Set (LS). The drop is discretised using (a,b) a triangular element, and (c,d) a single node. The zone coloured in purple represents the time before the analytical impact ($t_0$).}
\label{Fig:dropV}
\end{figure}

The drop is represented by an \EF{equilateral triangle of side equal} to $h_\mathrm{FS}$. The triangular drop points upwards, as illustrated within the circle in Fig.~\ref{Fig:drop_geometry_a}. In addition, a second case is taken into account where the initial drop is discretised by a single node (or particle), which is a standard practice in PFEM. Each problem is solved with PFEM--AS and PFEM--LS. For each case, four release heights are considered, $H_f$ = $[H_r \:,\: 2H_r \:,\: 4H_r \:,\: 8H_r]$. For high release heights, splashes and drops may develop. These sub--drops are not meshed in PFEM--AS problems, but they are in PFEM--LS following the scheme presented in the previous section. In summary, 16 problems are solved in total (2 drops $\times$ 4 heights $\times$ 2 PFEM schemes). Problems are simulated for 1.5 physical seconds after the first drop/fluid contact. The time of contact is obtained analytically and is denoted as $t_0$. \EF{For each problem, the variation of energy (kinetic + potential) and mass are computed, and the results are shown Figs.~\ref{Fig:dropE} and \ref{Fig:dropV}, respectively. The variations are given in percentage with respect to the mass and energy of the reference system. As reference system we consider the fluid in the container and the meshed drop, because, even for the drop discretised with a single node, the mass of the reference system should account for the element that retained the particle attached to the free surface (as was illustrated in Fig.~\ref{Fig:mani_cares_LS}a) }

With focus on the meshed drop, an increase of the total energy is observed during the impact, especially in simulations performed with PFEM--AS (Fig.~\ref{Fig:dropE_a}). The creation of energy is due to the creation of elements slightly before the impact, as quantified in Fig.~\ref{Fig:dropV} and illustrated in the first row of Table \ref{Tab:drop}. These elements not only increase the volume and energy of the system, but also anticipate the impact time. Although these drawbacks are present in both PFEM schemes, they are much less pronounced in PFEM--LS.

\begin{figure}[t!]
\captionsetup[subfigure]{labelformat=empty}
	\begin{subfigure}[b]{1.0\textwidth}
		\includegraphics[trim=0 0 0 0,clip,width=1.00\linewidth]{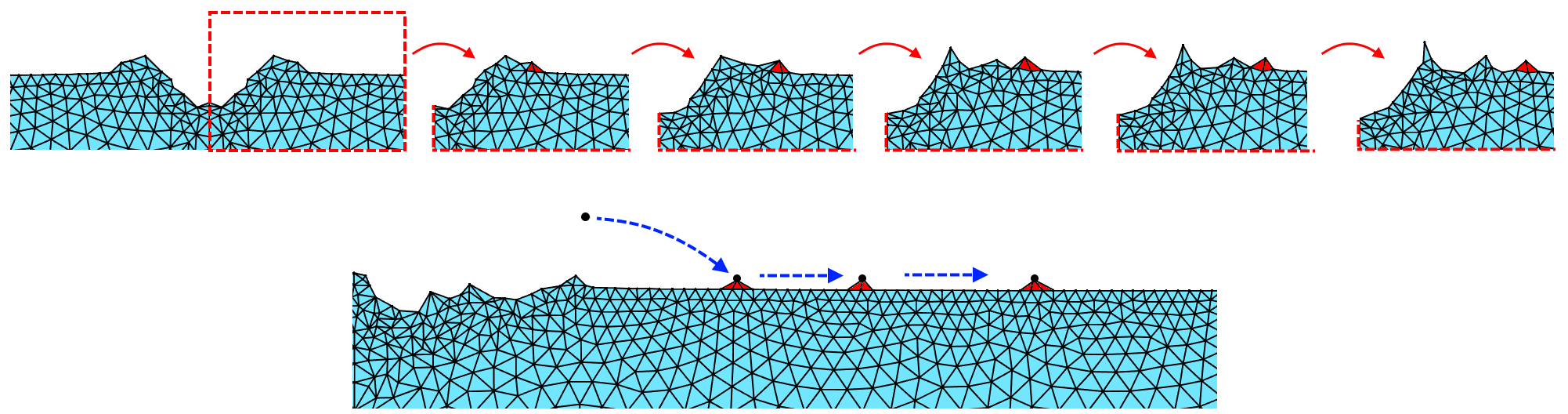}
		\caption{}
		\label{Fig:drop_walking_a}
	\end{subfigure}
	~
	\begin{subfigure}[b]{0.0\textwidth}\caption{}
		\label{Fig:drop_walking_b}
	\end{subfigure}
	\vspace{-53mm}\\
	\hspace{0mm} \footnotesize{}  (\textbf{a})
	\vspace{20mm}\\
	\hspace*{30mm} (\textbf{b}) 
	\vspace{14mm}\\
\caption{Illustration of mechanisms generating \EF{surfing} particles. Generation due to (a) free--surface degradation and (b) impact of a non-meshed drop. Red elements represent those attached to a \EF{surfing} particle. Animations of these snapshots can be found in \citep{YoutubeAll}.}  
\label{Fig:drop_walking}
\end{figure}

After the impact, the system begins to progressively lose energy and volume, as shown in Fig.~\ref{Fig:dropV}. This is clearly observed in PFEM--AS simulations, especially those that consider high release heights ($H_f \geq 2H_r$). The reason is that the PFEM--AS implementation does not prevent the smoothness deterioration of the free surface, so as the free surface stretches, elements are removed during remeshing. The non-smoothness of the free--surface can be noted in Table \ref{Tab:drop} by comparing impact creaters between PFEM--AS and PFEM--LS. In addition, after the impact, the energy and volume variation curves present more noise in PFEM--AS than in PFEM--LS. Some of this noise is due to the travelling of particles on the free surface, as schematized in Fig.~\ref{Fig:drop_walking}. \EF{In our view, these ``\EF{surfing}" particles, as they are called hereafter, originate from two sources. On one side, they are created from a degradation of the free surface, where highly distorted elements are removed by AS, leaving a single particle standing on top of the free surface, as illustrated in Fig.~\ref{Fig:drop_walking_a}. The second source is a flying particle that impacts the free surface with small enough normal velocity to not penetrate into the fluid, but enough tangential velocity to remain on the free surface, as illustrated in Fig.~\ref{Fig:drop_walking_b}. From this perspective, these \EF{surfing} particles are a numerical artifice resulting from discretising drops as nodes and the use of the AS-based remeshing process, which deteriorates the smoothness of the free surface.} On the contrary, PFEM--LS simulations exhibit less energy and mass variation after the drop impact, and do not feature \EF{surfing} particles (see \citep{YoutubeAll} for a comparison on video). \EF{This is due to the fact that PFEM--LS preserves the smoothness of the free surface, which avoids the possibility of obtaining a single particle above the surface. Moreover, even if a particle is placed above the free surface, it will be attached to the surface only if the distance between them is less than $\varepsilon\:h_\mathrm{FS}$, resulting in small elements that prevent the particle from surfing.}

\begin{table}[t!] 
\caption{ \EF{Impact of a drop using PFEM--AS and PFEM--LS. Two discretisations are considered for the drop: a triangular element and a single node. The first row shows the instant at which the contact is captured during remeshing. The second and third rows show the crater produced by the impact from two release heights. The reader is referred to \citep{YoutubeAll} for simulation videos.}}\label{Tab:drop}
\centering
  \begin{tabular}{c cc cc}
  \toprule
  	  & \multicolumn{2}{c}{PFEM--AS} & \multicolumn{2}{c}{PFEM--LS}
  	  \\
  	  \cmidrule(rl){2-3} \cmidrule(rl){4-5}
  	    & \includegraphics[trim=0 0 0 0,clip,width=0.030\linewidth]{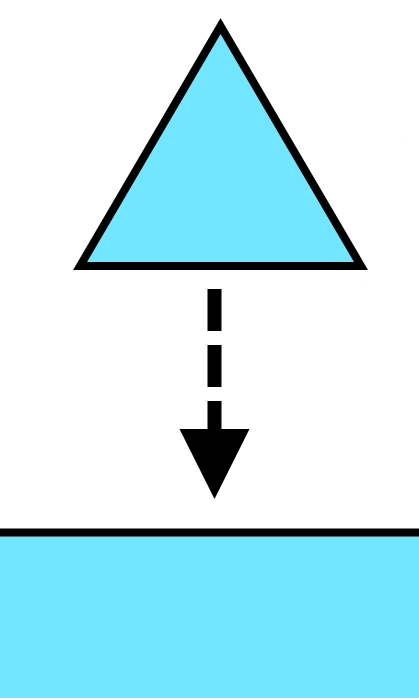} & \includegraphics[trim=0 0 0 0,clip,width=0.030\linewidth]{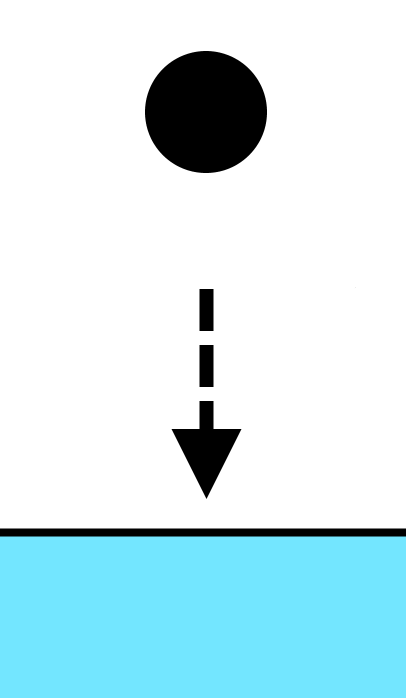} & \includegraphics[trim=0 0 0 0,clip,width=0.030\linewidth]{S42_DF_Drop_m.png} & \includegraphics[trim=0 0 0 0,clip,width=0.030\linewidth]{S42_DF_Drop_p.png}
  	  \\
  	  \cmidrule(rl){2-3} \cmidrule(rl){4-5}
  	  \begin{minipage}{1.5cm}
	  \centering
	  \vspace{-7mm}
  	  Contact 
  	  \end{minipage}  	    	  
  	  &
	  \includegraphics[trim=85 90 70 120,clip,width=0.20\linewidth]{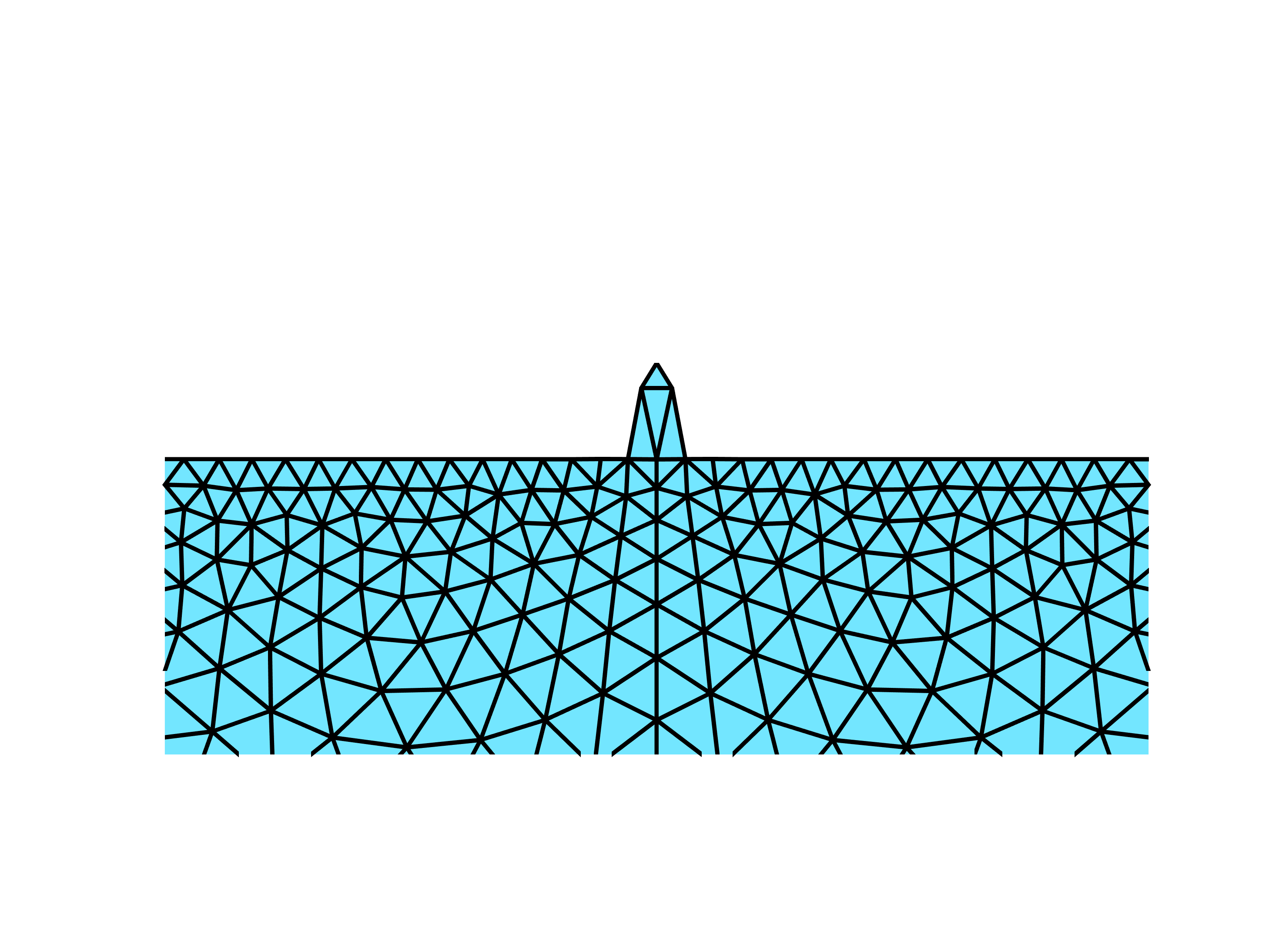}
	  & \hspace{-5mm}
	  \includegraphics[trim=85 90 70 120,clip,width=0.20\linewidth]{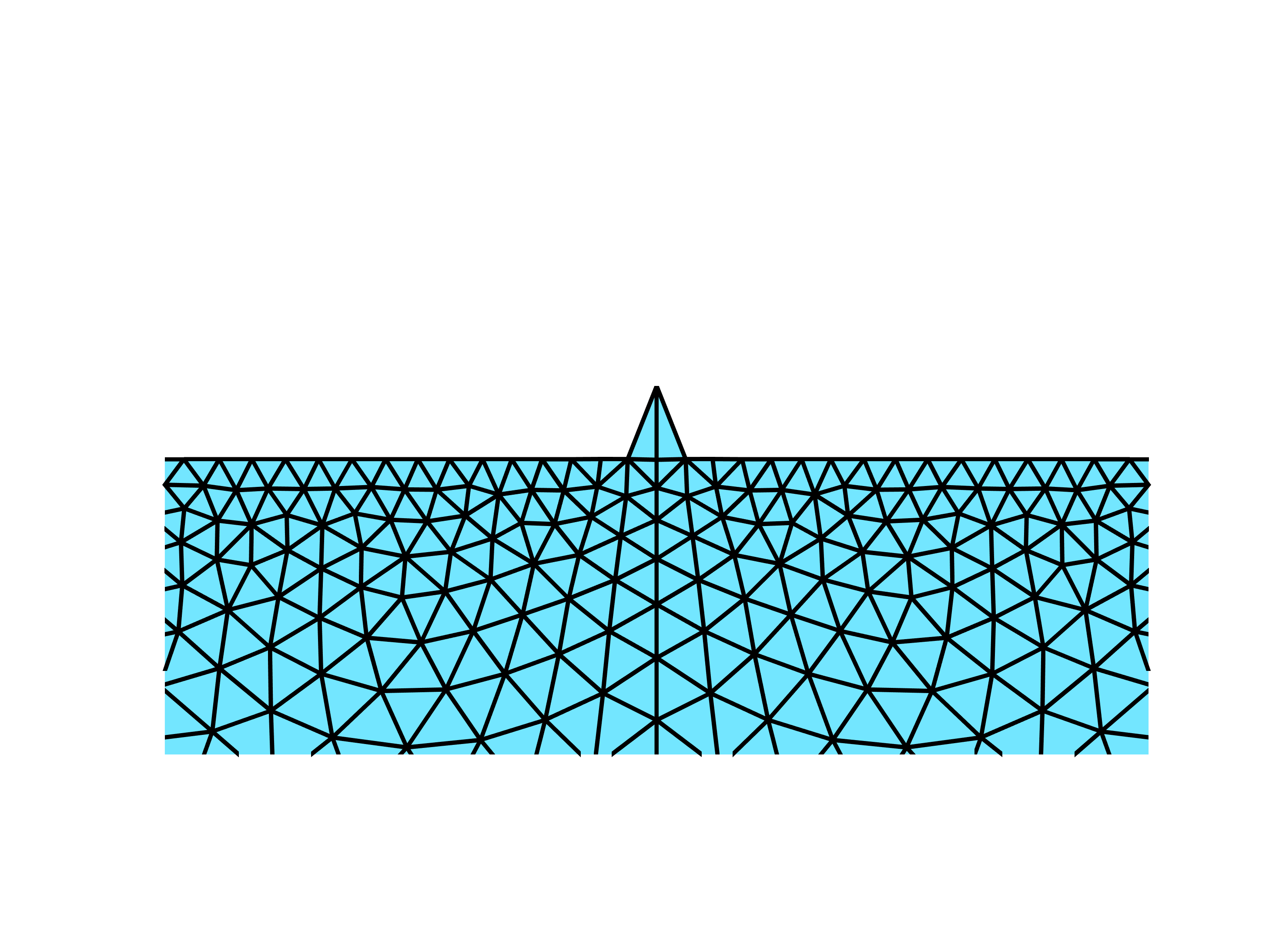}
	  &
	  \includegraphics[trim=85 90 70 120,clip,width=0.20\linewidth]{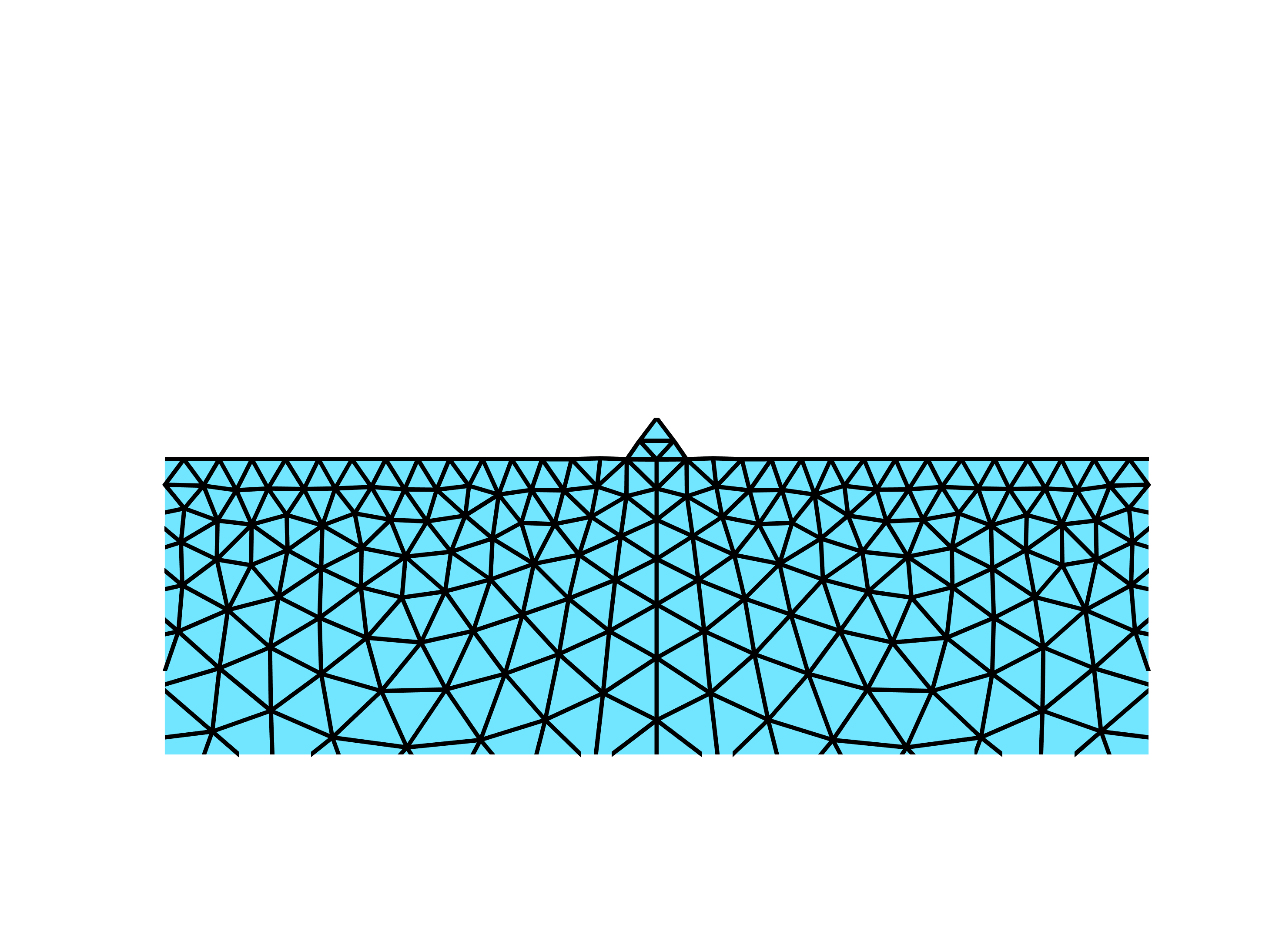}
	  &\hspace{-5mm}
	  \includegraphics[trim=85 90 70 120,clip,width=0.20\linewidth]{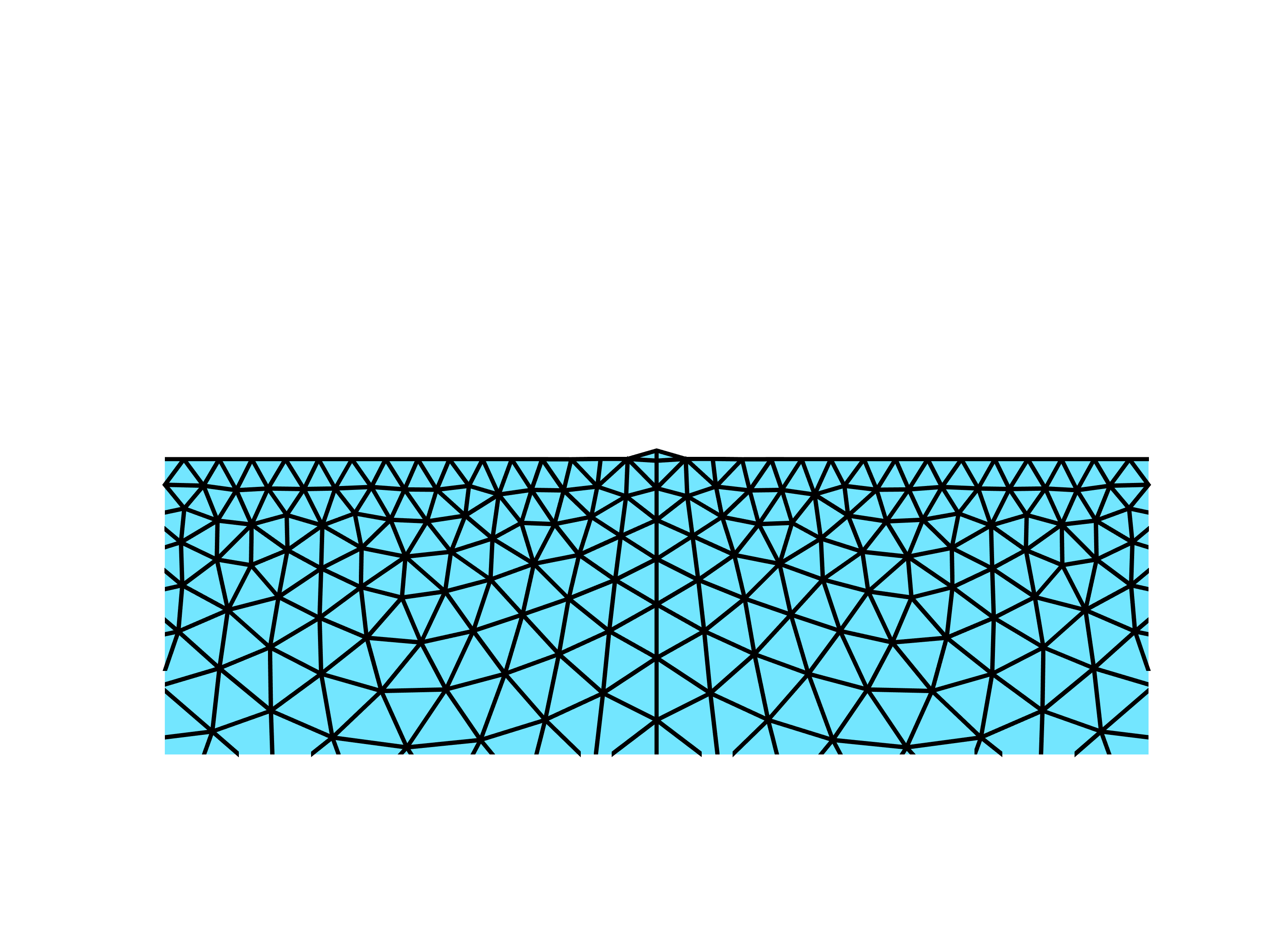}
	  \vspace{-0mm}\\
	  \begin{minipage}{1.5cm}
	  \centering
	  \vspace{-7mm}
  	  Crater 
  	  
  	  ($H_f = H_r$)
  	  \end{minipage}
	  &
	  \includegraphics[trim=85 90 70 120,clip,width=0.20\linewidth]{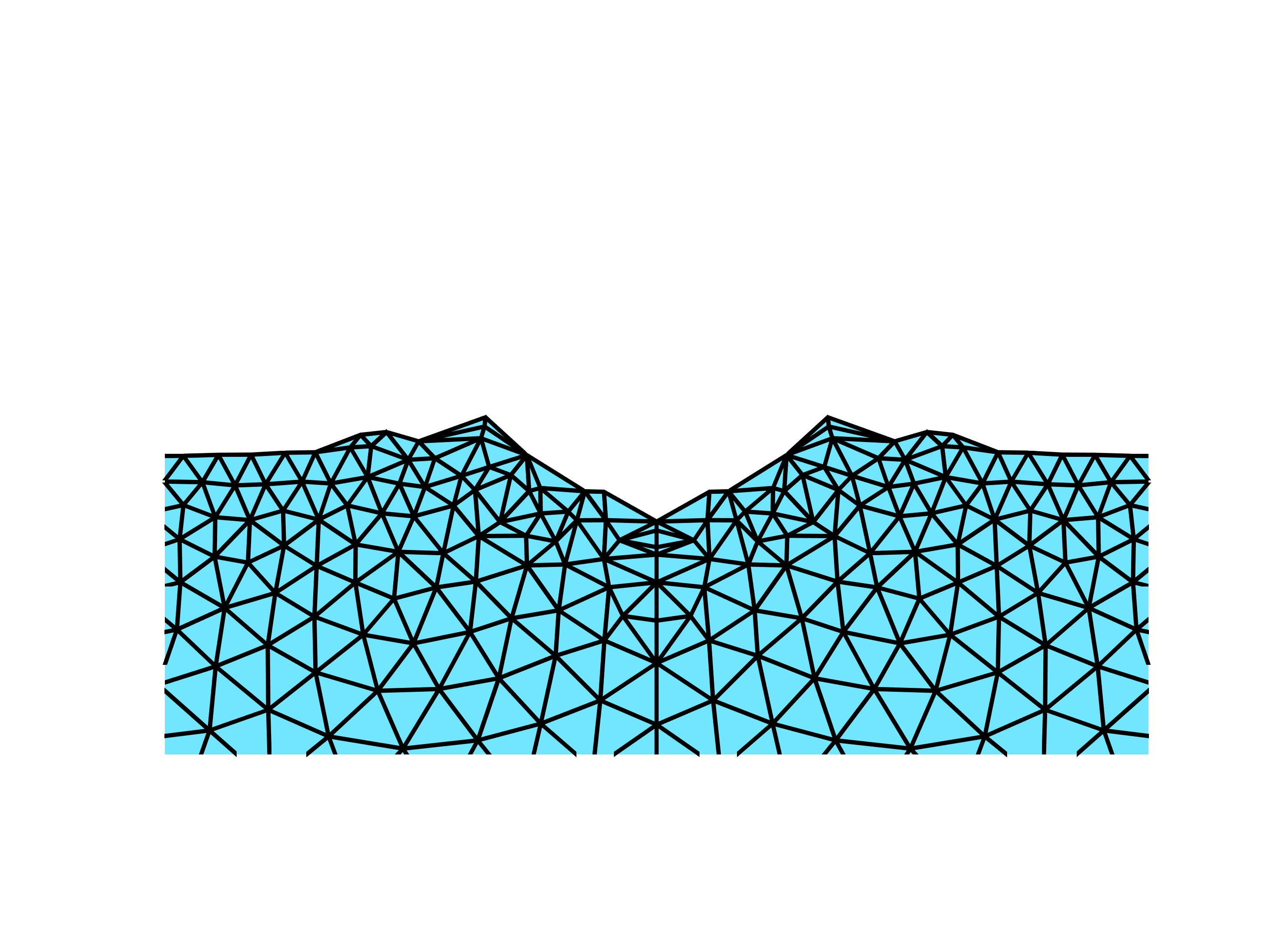}
	  &\hspace{-4mm}
	  \includegraphics[trim=85 90 70 130,clip,width=0.20\linewidth]{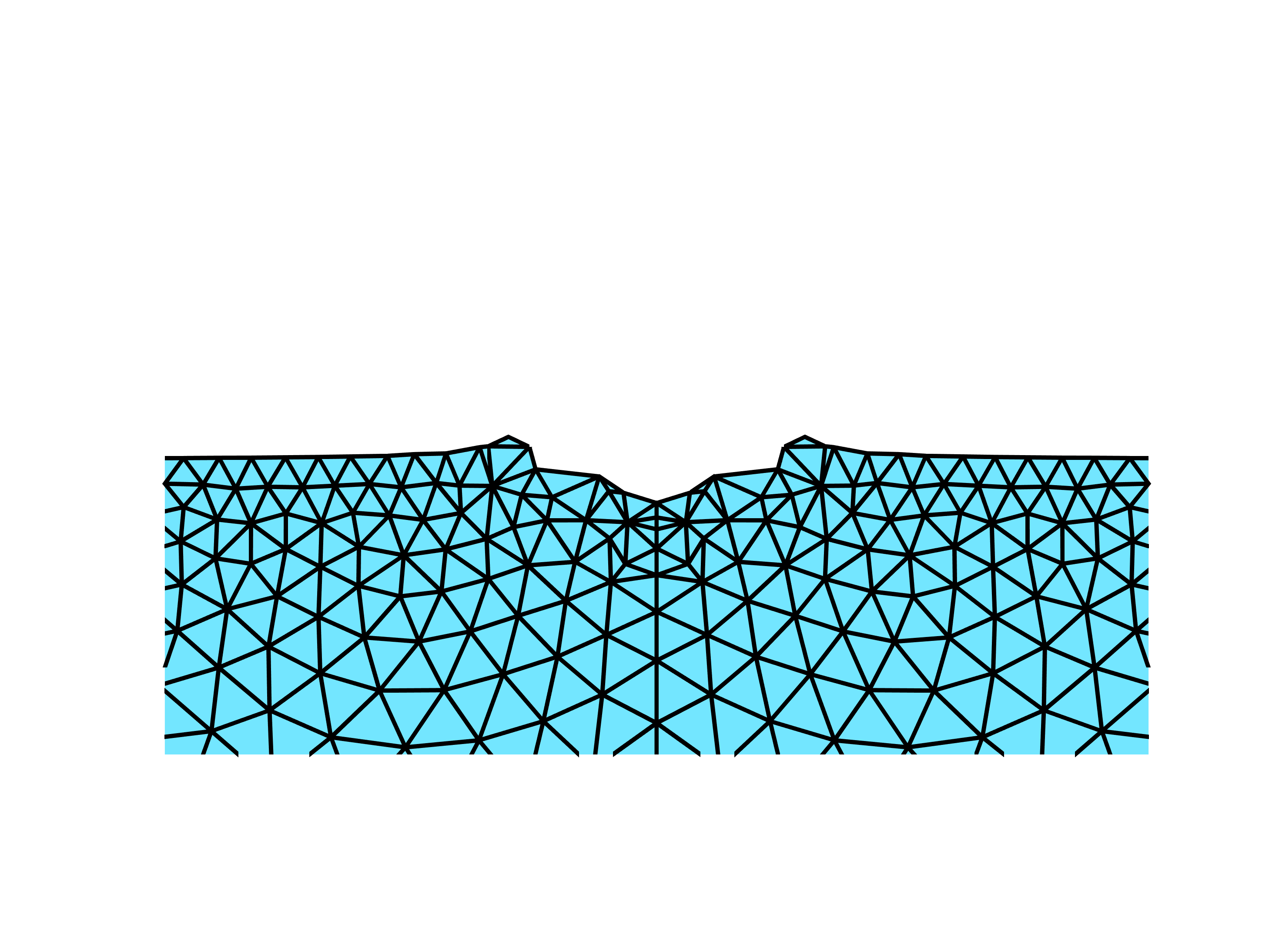}
	  &
	  \includegraphics[trim=85 90 70 120,clip,width=0.20\linewidth]{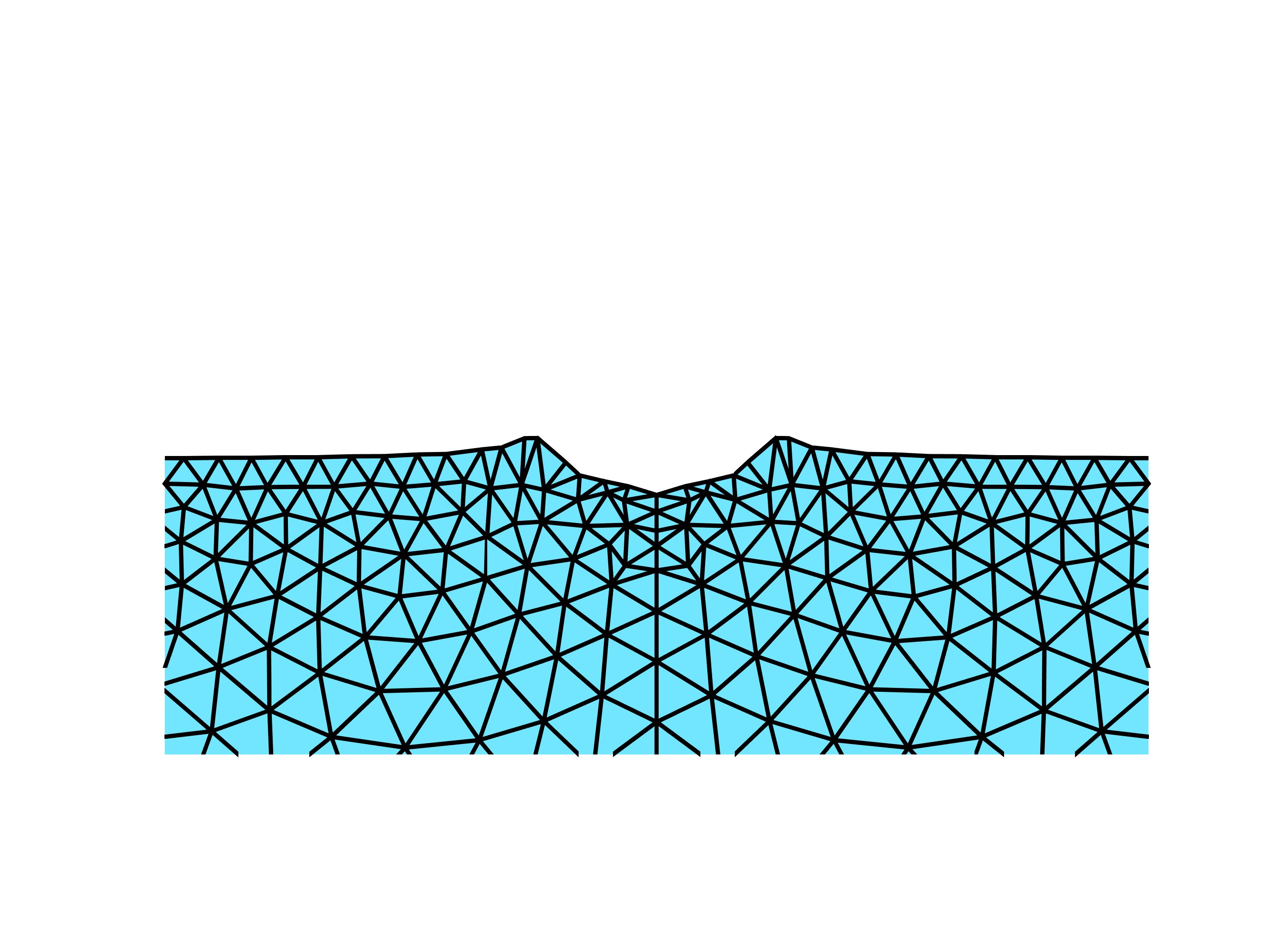}
	  &\hspace{-4mm}
	  \includegraphics[trim=85 90 70 120,clip,width=0.20\linewidth]{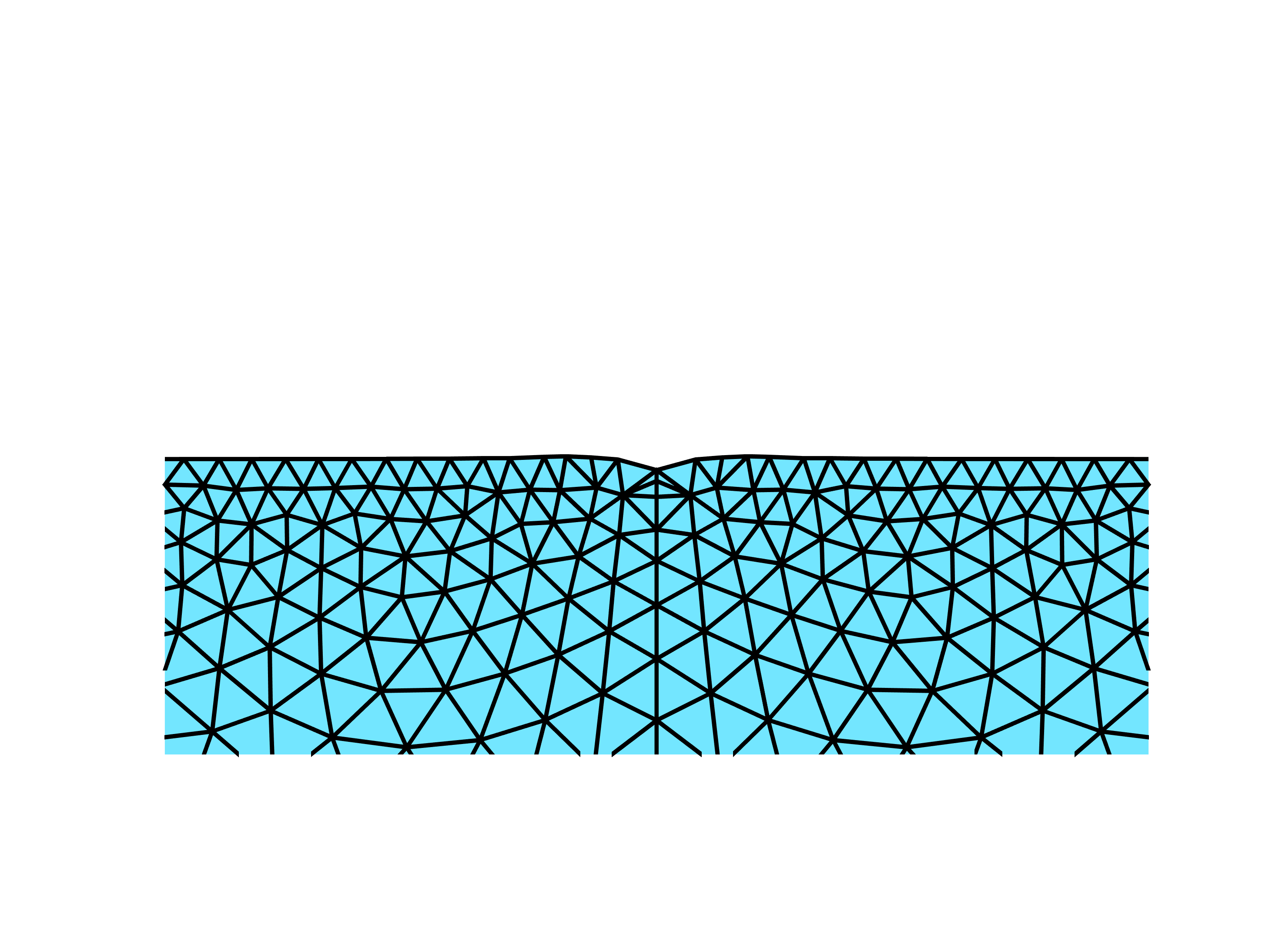}
	  \vspace{-0mm}\\
  	  \begin{minipage}{1.6cm}
  	  \centering
  	  \vspace{-7mm}
  	  Crater
  	  
  	  ($H_f$= 4$H_r$)
  	  \end{minipage}
  	  &
	  \includegraphics[trim=85 90 70 120,clip,width=0.20\linewidth]{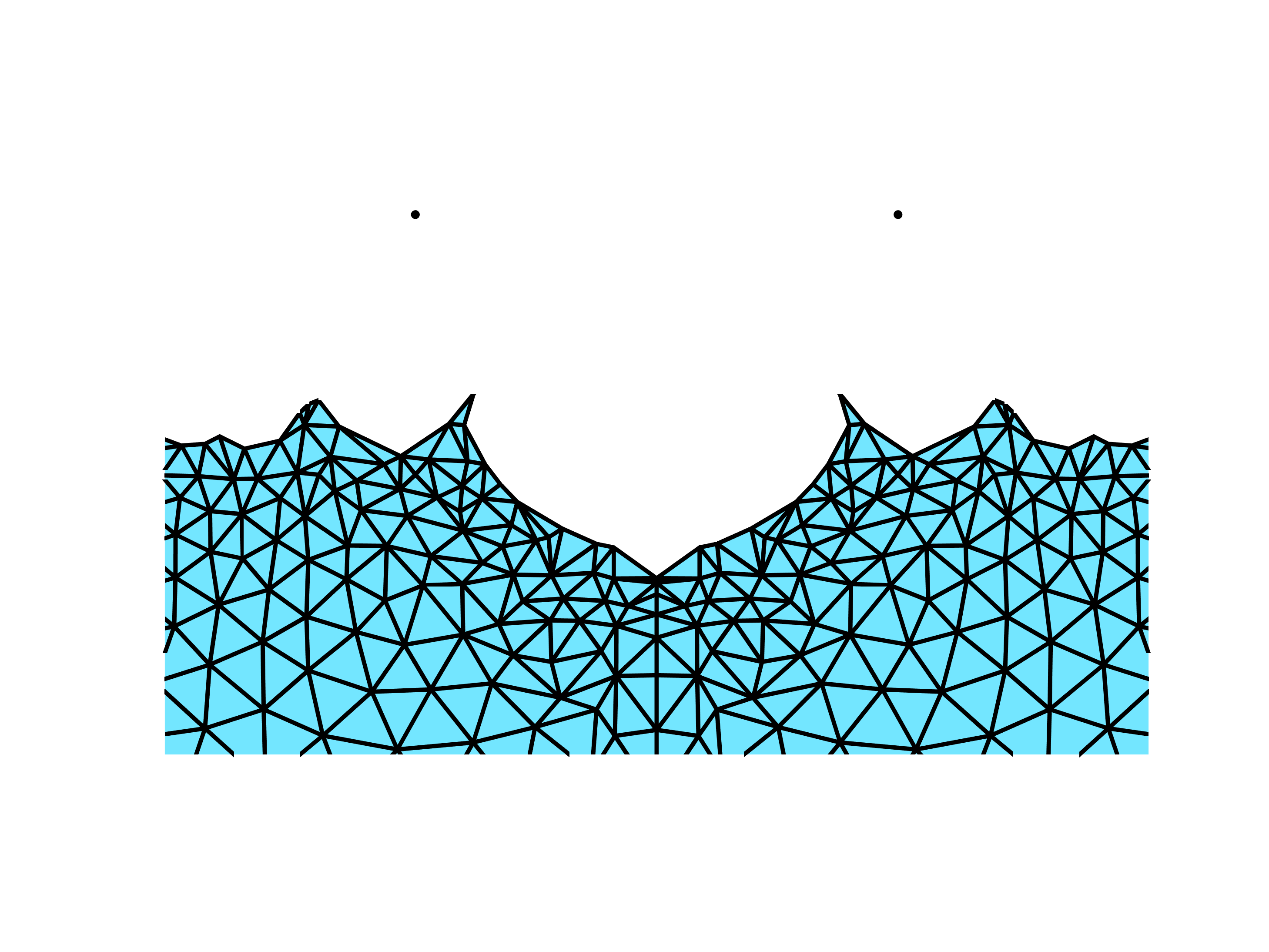}
	  &\hspace{-4mm}
	  \includegraphics[trim=85 90 70 120,clip,width=0.20\linewidth]{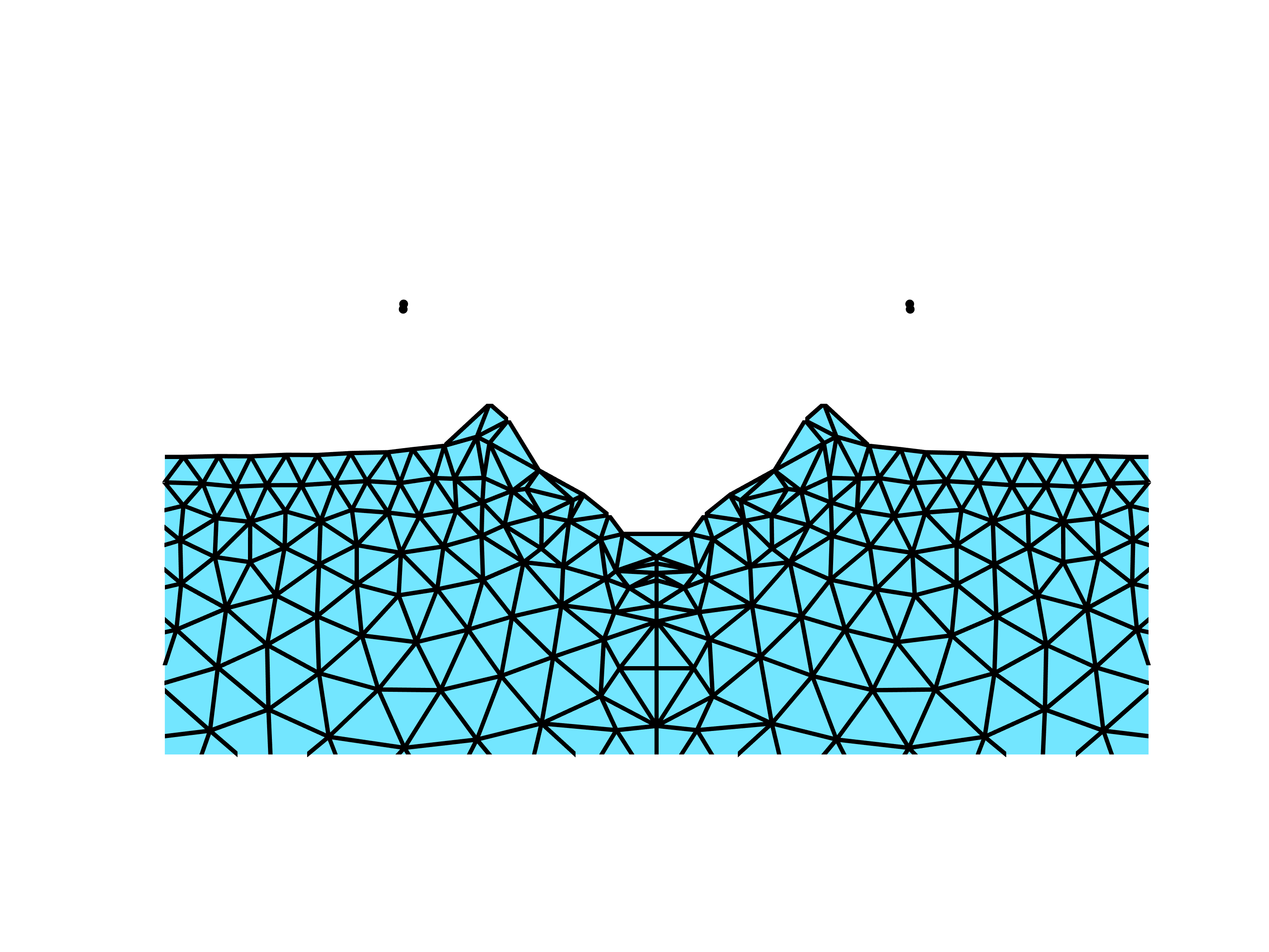}
	  &
	  \includegraphics[trim=85 90 70 120,clip,width=0.20\linewidth]{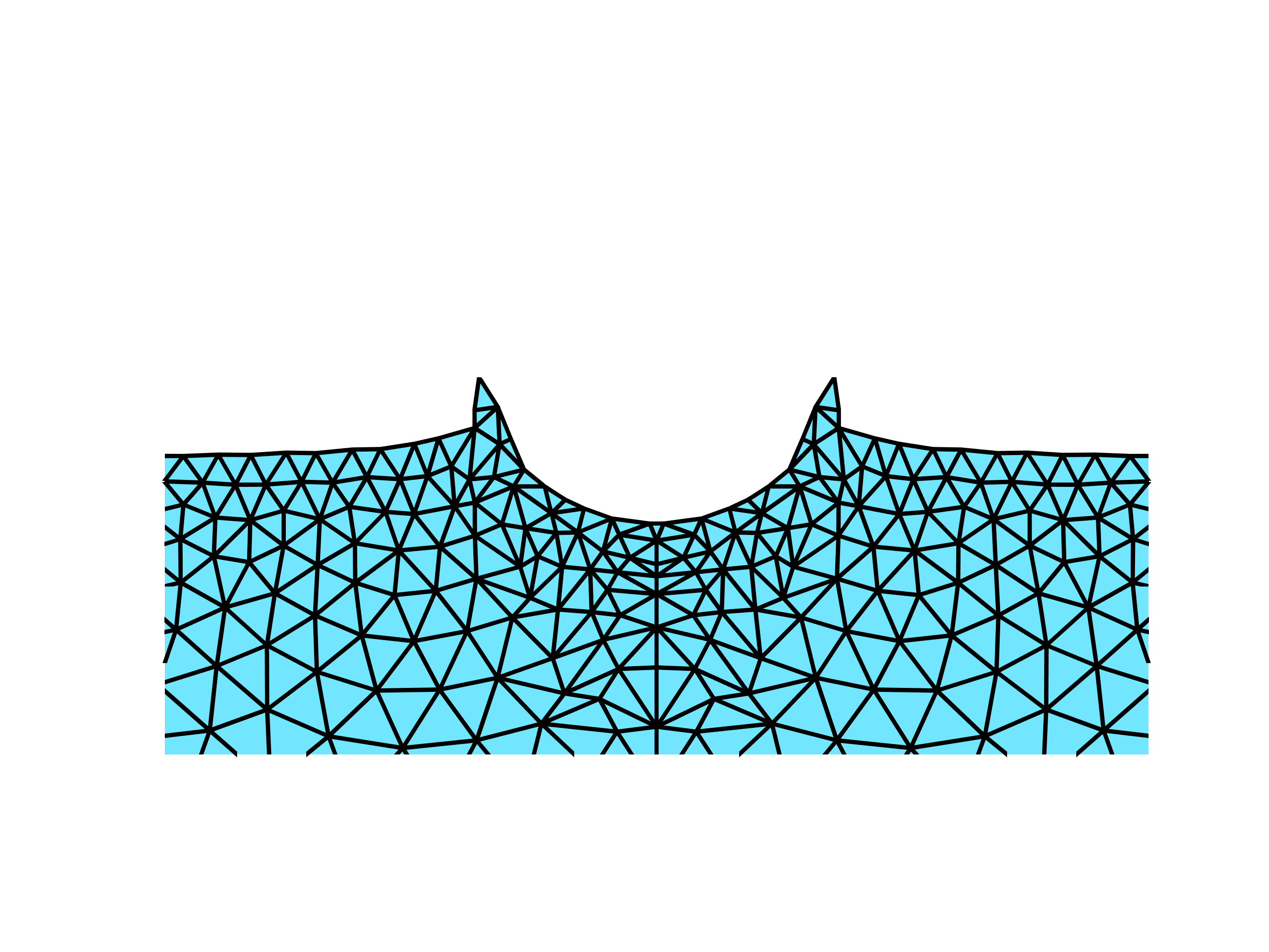}
	  &\hspace{-4mm}
	  \includegraphics[trim=85 90 70 120,clip,width=0.20\linewidth]{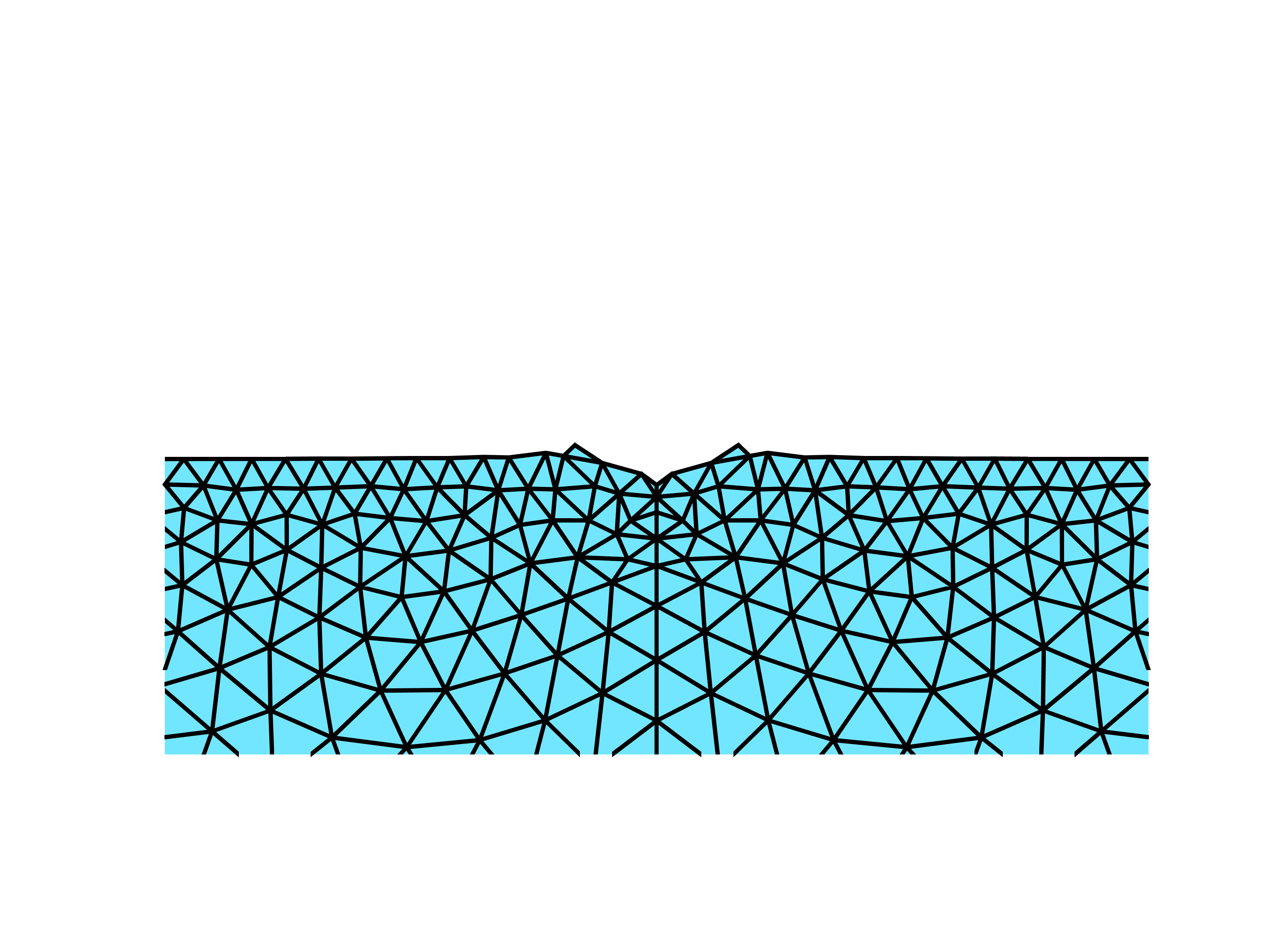}
	  \vspace{-0mm}\\
	  \bottomrule
  \end{tabular}
\end{table}

\EF{
Given that the mass and energy are computed in the finite element mesh, the system with a non-meshed drop presents less mass and energy at the beginning of the simulation than the system with a meshed drop. The difference in mass between systems is negligible and independent of the release height, but the difference in energy is increasingly different with the release height.} Since nodal drops are massless, there is less transfer of energy to the fluid resulting in small impact craters and small volume variations, as can be seen by comparing rows 3 and 6 of Table \ref{Tab:drop}. Certainly, the energy of a meshed drop can be similar to that of a nodal drop in the case of drops having very low masses or being released from very small heights. In such cases, the influence of the drop on the free surface may be negligible and it would be convenient to remove the drop from the model and avoid the impact. For this reason, the PFEM--LS implementation of this work only meshes the drops that are detached from elements having a volume equal to or greater than 30$\%$ of the characteristic element volume.

\subsection{Disk of fluid falls in fluid}

\begin{figure}[t!]
\captionsetup[subfigure]{labelformat=empty}
\centering 
	\begin{subfigure}[b]{0.32\textwidth}
		\includegraphics[trim=0 0 0 0,clip,width=1.00\linewidth]{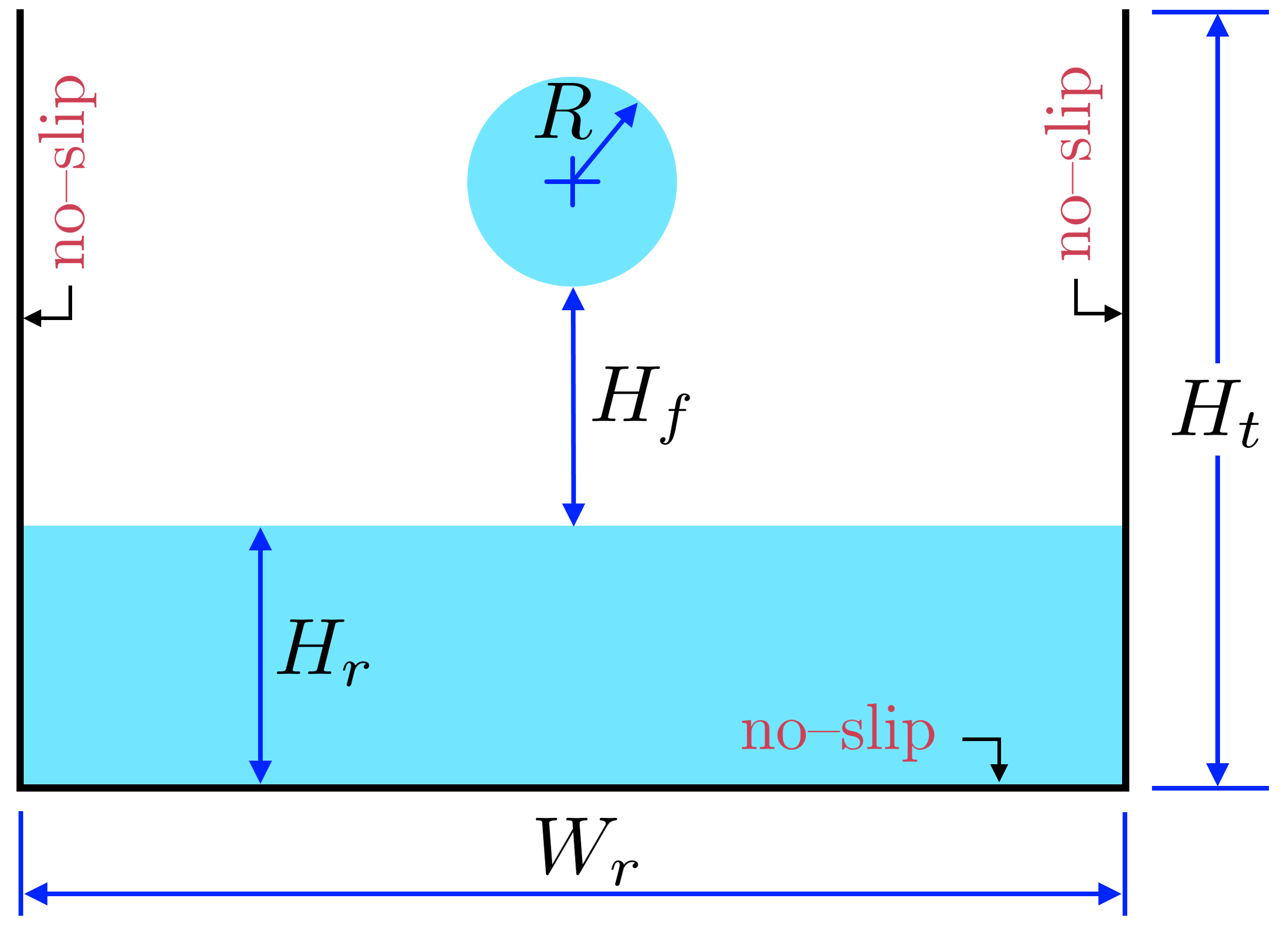}
		\caption{}
		\label{Fig:BFIF_geometry_a}
	\end{subfigure}
	~\hspace{10mm}
	\begin{subfigure}[b]{0.30\textwidth}	
		\includegraphics[trim=0 0 0 0,clip,width=1.00\linewidth]{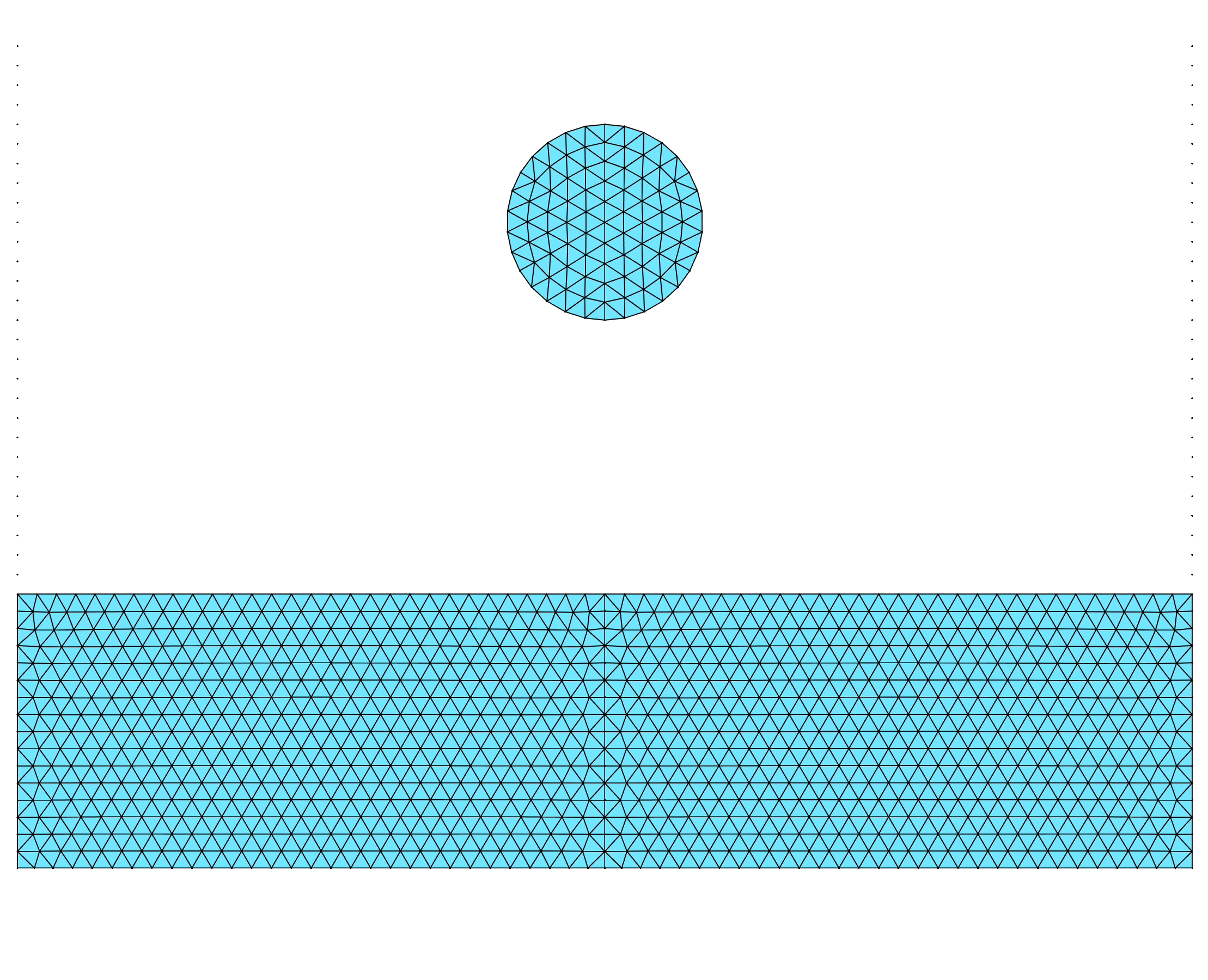}
		\caption{}
		\label{Fig:BFIF_geometry_b}
	\end{subfigure}
	\\
	\vspace{-45mm}
	\hspace{-50mm} \footnotesize{(\textbf{a})} \hspace{70mm} (\textbf{b}) 
	\\
	\vspace{37mm}
\caption{Disk of fluid falling in fluid. (a) Geometry of the problem and (b) initial discretisation using $h_\mathrm{FS} = 5$ mm.
}
\label{Fig:BFIF_geometry}
\end{figure}

This problem consists of a disk of fluid falling and impacting on a fluid reservoir, as illustrated in Fig.~\ref{Fig:BFIF_geometry_a}. The simulation parameters are set according to Franci and Cremonesi \citep{franci2017effect}. Therefore, $H_r = 70$ mm, $W_r = 300$ mm, $H_f = 70$ mm, $R = 25$ mm, and $H_t = 900$ mm. The Newtonian fluid is defined by $\rho = 1000$ kg/m$^3$ and $\mu = 0.1$ Pa s. Following the study of \citep{franci2017effect}, three space discretisations are considered, which set the element size on the free surface as $h_\mathrm{FS}=$ [5.0 , 3.0 , 1.5] mm. The maximum element size is the same in the three discretisations and is equal to $\mathrm{h}_\mathrm{max} = 5$ mm, which is imposed from a distance $\mathrm{d}_\mathrm{max} = H_r$ to the free--surface. Thus, a uniform mesh size is used in the case $h_\mathrm{FS}= 5$ mm, as shown in Fig.~\ref{Fig:BFIF_geometry_b}.

\begin{figure}[t!]
\captionsetup[subfigure]{labelformat=empty}
\centering 
	\begin{subfigure}[b]{1.00\textwidth}
		\includegraphics[trim=0 0 0 0,clip,width=1.00\linewidth]{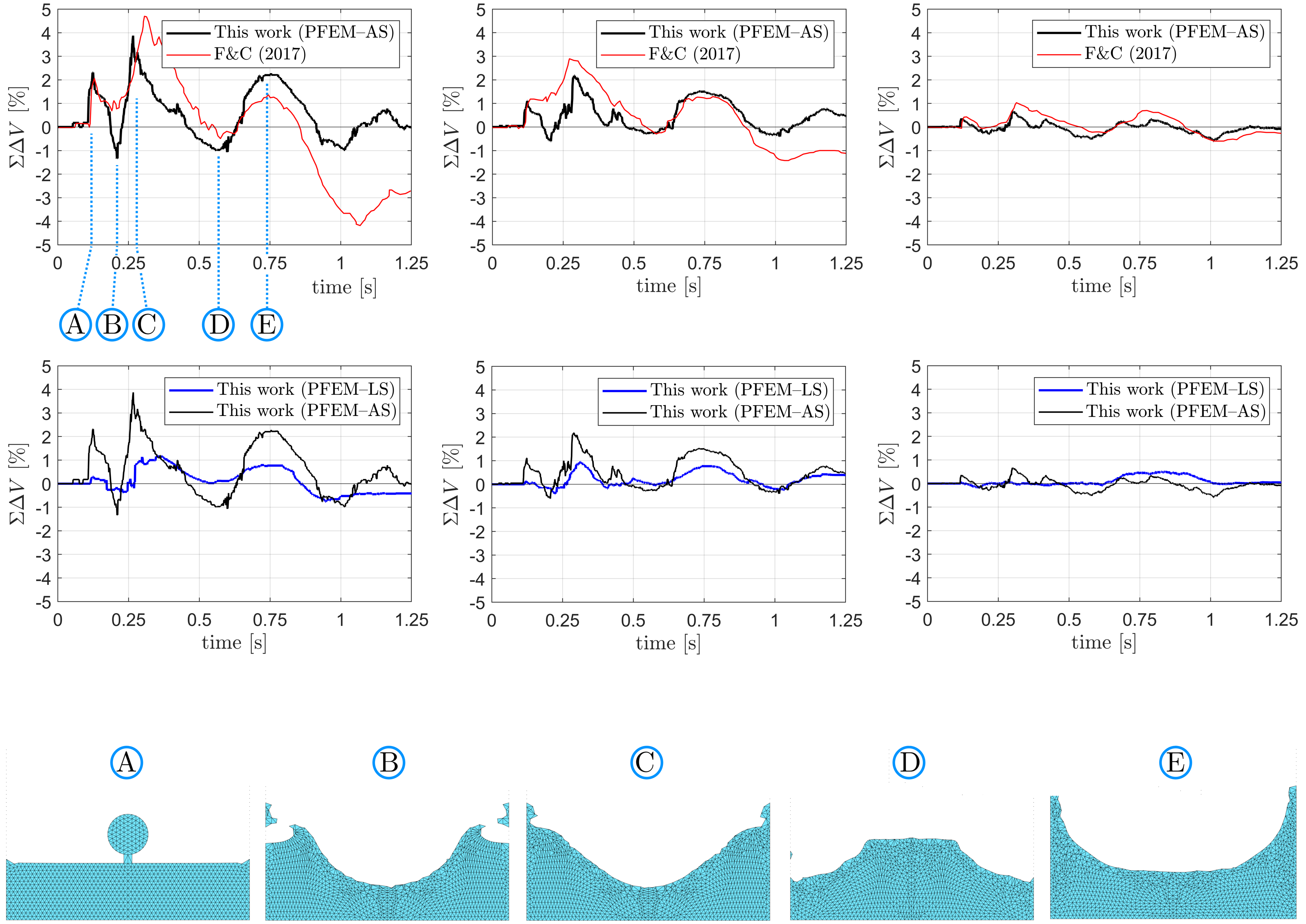}
		\caption{}
		\label{Fig:BFIF_reference_a}
	\end{subfigure}
	\\
	\vspace{-37mm}
	\hspace{5mm} \footnotesize{(\textbf{a}) $h_\mathrm{FS} = 5.0$ mm}
	\hspace{32mm} (\textbf{b}) $h_\mathrm{FS} = 3.0$ mm 
	\hspace{32mm} (\textbf{c}) $h_\mathrm{FS} = 1.5$ mm 
	\vspace{5mm}\\
	\hspace{-160mm} (\textbf{d})
	\vspace{20mm}\\
	\begin{subfigure}[b]{0.00\textwidth}
		\caption{}\label{Fig:BFIF_reference_b}
	\end{subfigure}
	\begin{subfigure}[b]{0.00\textwidth}
		\caption{}\label{Fig:BFIF_reference_c}
	\end{subfigure}
	\begin{subfigure}[b]{0.00\textwidth}
		\caption{}\label{Fig:BFIF_reference_d}
	\end{subfigure}
	\vspace{-6mm}
\caption{In (a-c), volume variation during simulation of a disk of fluid falling in fluid using PFEM--AS (above) and PFEM--LS (below). Results of Franci and Cremonesi \citep{franci2017effect} are also displayed for comparison (F$\&$C (2017)). In (d), snapshots of simulation using PFEM--AS and $h_\mathrm{FS} = 5.0 $mm, at times labelled in (a). 
}
\label{Fig:BFIF_reference}
\end{figure}

\begin{figure}[t!]
\captionsetup[subfigure]{labelformat=empty}
\centering 
	\begin{subfigure}[b]{1.00\textwidth}
		\includegraphics[trim=0 0 0 0,clip,width=1.00\linewidth]{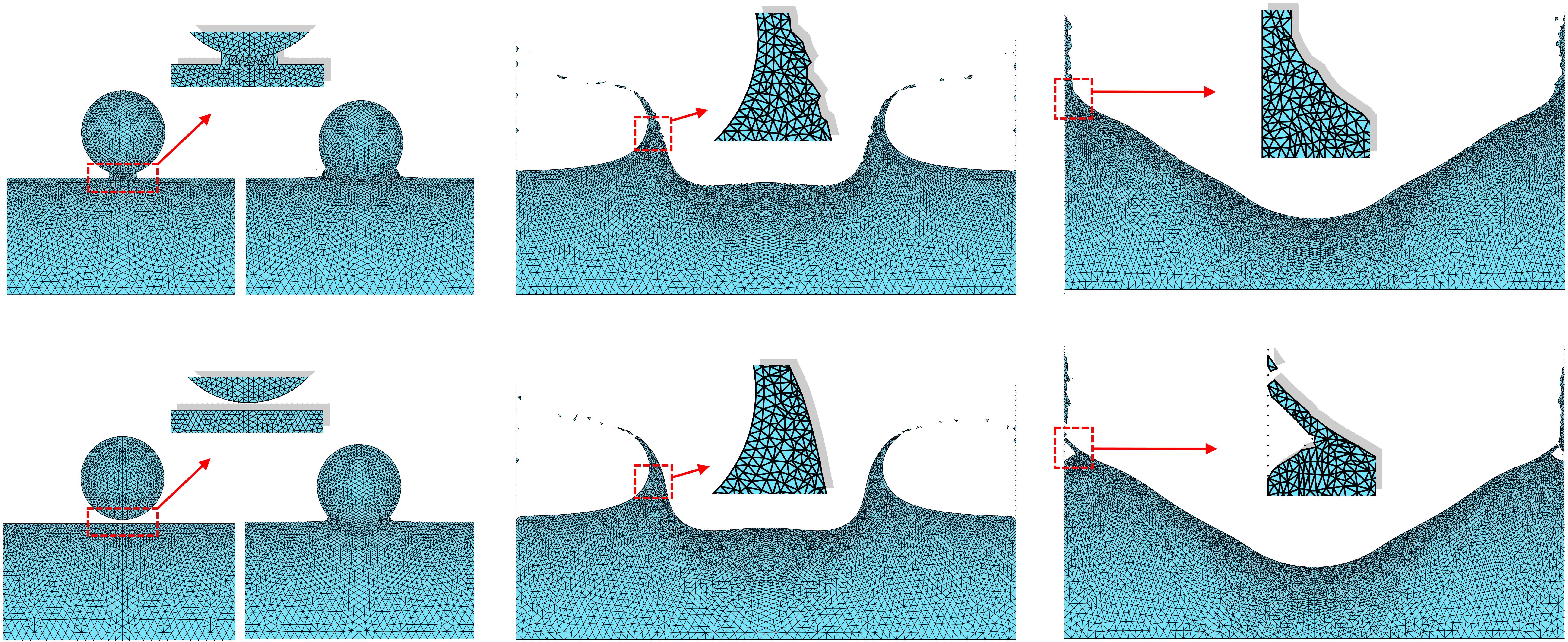}
		\caption{}
		\label{Fig:BFIF_snapshots_a}
	\end{subfigure}
	\\
	\vspace{-41mm}
	\hspace{-13mm} \footnotesize{{(\textbf{a}}) PFEM--AS, time=0.116 ... 0.121}
	\hspace{17mm} {(\textbf{b})} PFEM--AS, time=0.173 
	\hspace{22mm} {(\textbf{c})} PFEM--AS, time=0.299
	\\
	\vspace{33mm}
	\hspace{-13mm} \footnotesize{(\textbf{d}) PFEM--LS, time=0.116 ... 0.121}
	\hspace{17mm} {(\textbf{e})} PFEM--LS, time=0.173 
	\hspace{21mm} {(\textbf{f})} PFEM--LS, time=0.299
	\\
	\begin{subfigure}[b]{0.00\textwidth} \caption{}
		\label{Fig:BFIF_snapshots_b}
	\end{subfigure}
	\begin{subfigure}[b]{0.00\textwidth} \caption{}
		\label{Fig:BFIF_snapshots_c}
	\end{subfigure}
	\begin{subfigure}[b]{0.00\textwidth} \caption{}
		\label{Fig:BFIF_snapshots_d}
	\end{subfigure}
	\begin{subfigure}[b]{0.00\textwidth} \caption{}
		\label{Fig:BFIF_snapshots_e}
	\end{subfigure}
	\begin{subfigure}[b]{0.00\textwidth} \caption{}
		\label{Fig:BFIF_snapshots_f}
	\end{subfigure}
	\vspace{-4mm}
\caption{Snapshots of simulations of a disk of viscous fluid falling on fluid. Simulations using (a-c) PFEM--AS and (d-f) PFEM--LS. Simulation videos are provided in \citep{YoutubeAll}.}
\label{Fig:BFIF_snapshots}
\end{figure}

Fig.~\ref{Fig:BFIF_reference} plots the volume variation during simulation, above using PFEM--AS and below using PFEM--LS. Each plot includes the variation reported by Franci and Cremonesi (F$\&$C (2017)). In general, a good correlation between PFEM--AS and the reference is observed. In all three discretisations, the same order of magnitude and trend of the volume variation curve is exhibited. From these curves, 5 stages can be identified, which are labelled in Fig.~\ref{Fig:BFIF_reference_a} and illustrated in Fig.~\ref{Fig:BFIF_reference_d}. In \circled{A}, the impact of the disk takes place, which features an increase in volume due to the creation of elements. Notably, the volume increase is smaller in PFEM--LS than in PFEM--AS due to the Level--Set function, as shown in Figs.~\ref{Fig:BFIF_snapshots_a} and \ref{Fig:BFIF_snapshots_d}. Between \circled{A} and \circled{B}, a volume reduction occurs, which is due partly to the many splashes generated by the impact, and partly to the degradation of the free surface due to its stretching, as shown in Figs.~\ref{Fig:BFIF_snapshots_b} and \ref{Fig:BFIF_snapshots_e}. Since some drops are meshed in PFEM--LS and the degradation of the free--surface is avoided, a smaller volume reduction is observed between \circled{A} and \circled{B} with PFEM--LS. From \circled{B} onwards, drops begin to reach the walls and the fluid begins to rise up the walls, which creates elements and adds volume to the system. The maximum volume addition in PFEM--AS is observed at \circled{C}, that is the instant at which the bent wave closes. Since PFEM--AS has poor cavity resolution, it prematurely closes the wave against the wall and adds volume to the system, as illustrated in Fig.~\ref{Fig:BFIF_snapshots_c}. From \circled{C} to \circled{D}, the fluid flows down the walls, which removes elements from the system and reduces its volume. From \circled{D} to \circled{E} the fluid rises again along the walls and from \circled{E} onwards remains in a sloshing state.

\begin{figure}[t!]
\captionsetup[subfigure]{labelformat=empty}
\centering 
	\begin{subfigure}[b]{0.48\textwidth}
		\includegraphics[trim=10 35 30 90,clip,width=1.00\linewidth]{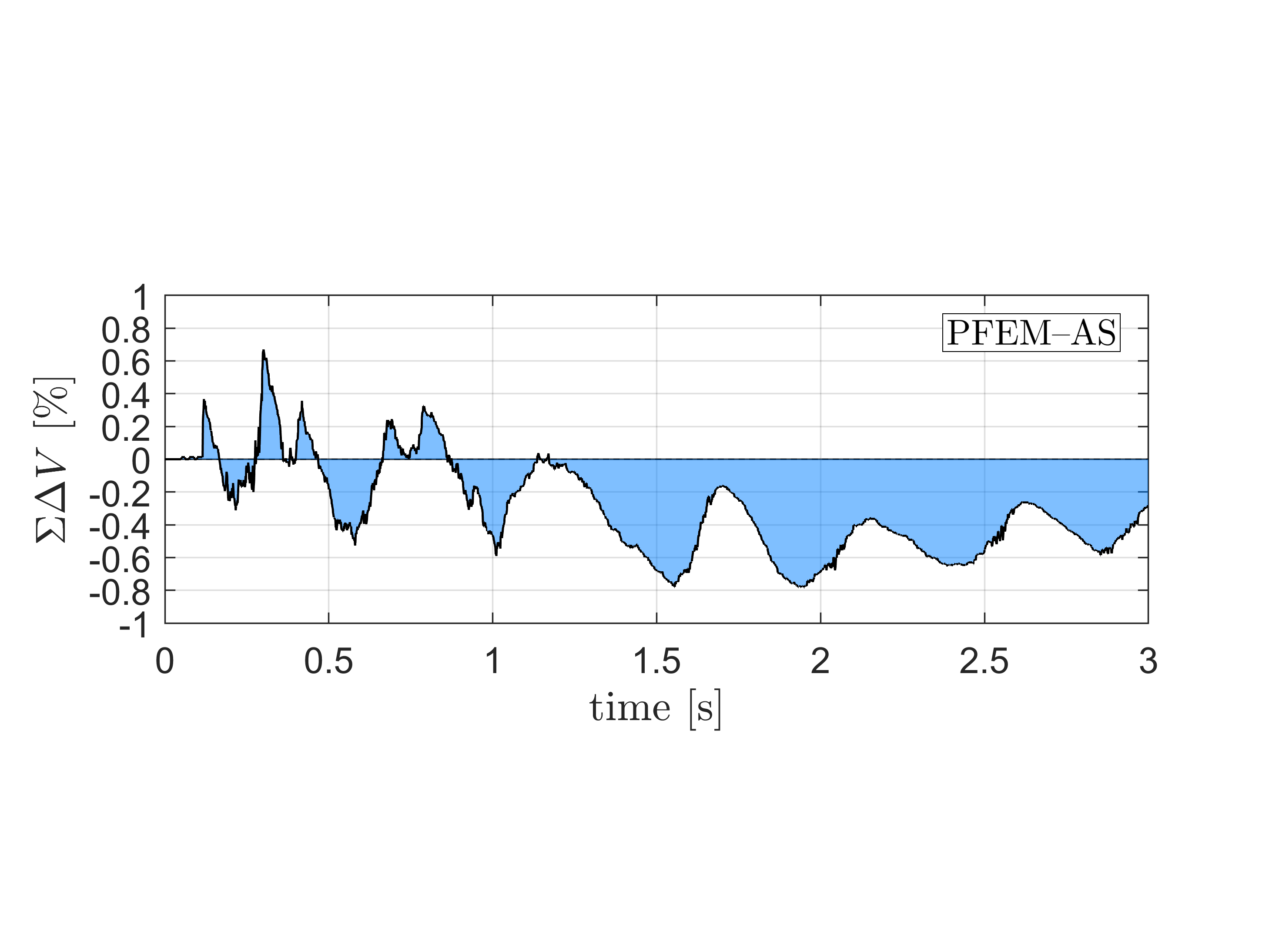}
		\caption{}
		\label{Fig:BFIF_3s_a}
	\end{subfigure}
	~
	\begin{subfigure}[b]{0.48\textwidth}	
		\includegraphics[trim=10 35 35 10,clip,width=1.00\linewidth]{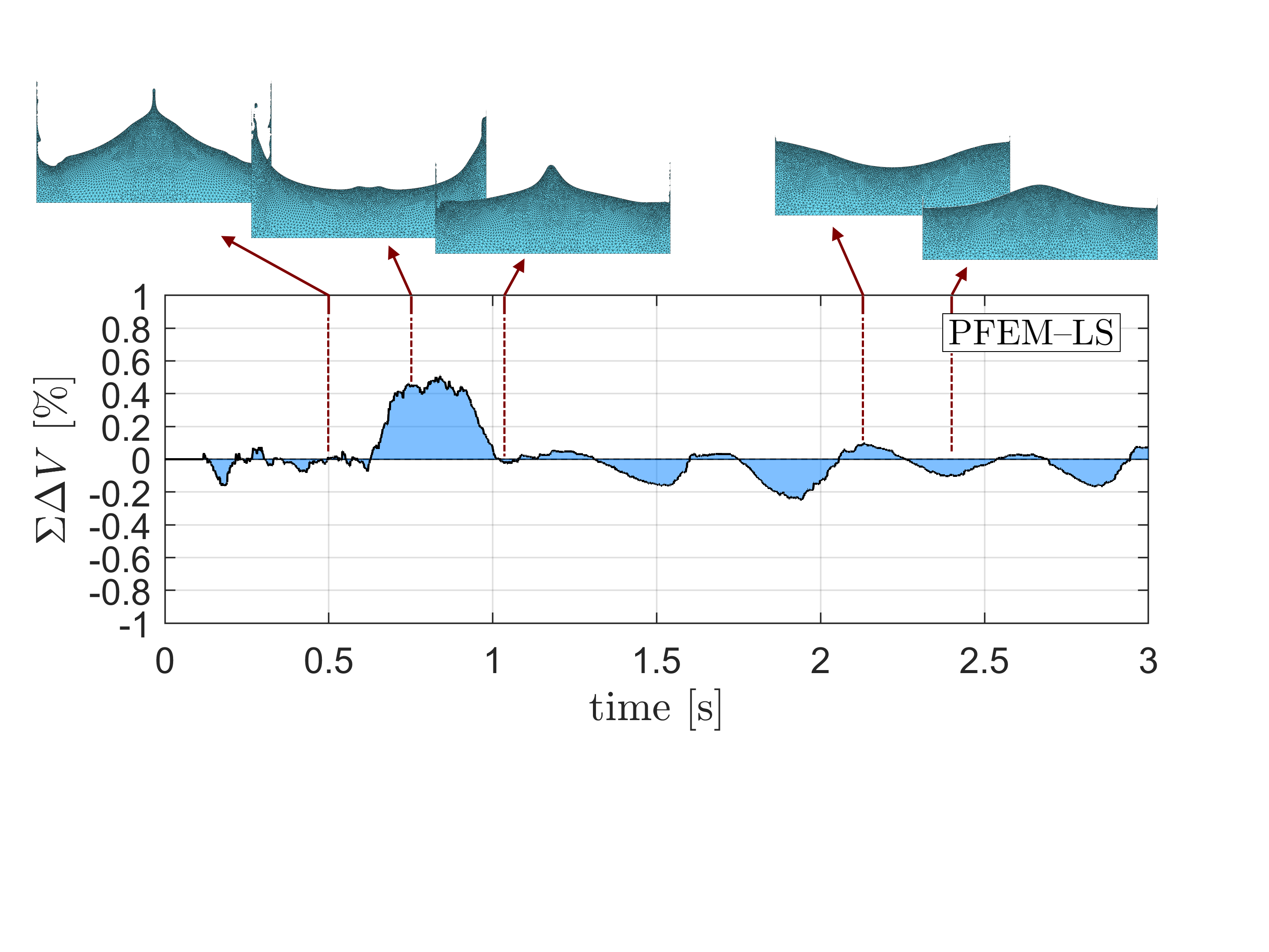}
		\caption{}
		\label{Fig:BFIF_3s_b}
	\end{subfigure}
	\\
	\vspace{-45mm}
	\hspace{-78mm} \footnotesize{(\textbf{a})} \hspace{77mm} (\textbf{b}) 
	\\
	\vspace{27mm}
\caption{ Volume variation in the impact of a viscous fluid disk using PFEM--AS (left) and PFEM--LS (right). The blue region is used to compute the volume variation per unit time, which is reported in Table \ref{Tab:BFIF_dv}.
}
\label{Fig:BFIF_3s}
\end{figure}

The mechanisms of volume addition and removal can be easily identified in the three discretisations when using PFEM--AS, and are equally clear in the curves of Franci and Cremonesi \citep{franci2017effect}. Although PFEM--LS reflects the same volume addition/removal mechanisms, it does so to a much lesser degree. The remarkable case is when $h_\mathrm{FS} = 1.5$ mm (blue curve in Fig.~\ref{Fig:BFIF_reference_c}), since on the scale of the graph, it is not possible to easily distinguish the aforementioned mechanisms due to the low amplitude of the curve. However, it is noteworthy the elevation of the curve between 0.6 s and 1.0 s. In this period, volume is added to the system and then removed. This is due to the ascending and descending path of the fluid along the walls, as the no-slip flow is modelled with the addition and removal of elements in a wall previously discretised with nodes. This drawback is present throughout the simulation, but the largest flow path along the walls occurs in the mentioned period. This can be easily corroborated by analytically estimating the volume variation, assuming that the no-slip flow along the wall creates elements of size $h_\mathrm{FS}$, as follows:
\begin{equation} \label{EQ:dvWall}
	\Delta V_\mathrm{wall} = \left(  \frac{\mathrm{d}_\mathrm{wall}}{\mathrm{h}_\mathrm{FS}} \right)
	\left(  \mathrm{h}_\mathrm{FS}^2 \frac{1}{2}\right) \left(\frac{1}{V_0} \right) 100 \; [\%]
\end{equation}  
\noindent where $\Delta V_\mathrm{wall}$ is the estimated volume variation due to the no--slip flow along one wall, $\mathrm{d}_\mathrm{wall}$ is the distance travelled by the fluid on one wall and $V_0$ is the initial volume. Thus, the first parenthesis in Eq.~\eqref{EQ:dvWall} quantifies the number of elements that are added, the second parenthesis contains the volume added per element, and the last terms normalize the volume variation with respect to the initial one. From the results with $\mathrm{h}_\mathrm{FS} = 1.5$ mm, it is obtained that $\mathrm{d}_\mathrm{wall} = 0.15 - 0.06$ m. Considering that $V_0 = 0.0230$ m$^2$, then $\Delta V_\mathrm{wall} = 0.29 \%$. This means that the analytical approximation of the expected volume variation due to the no-slip flow along the two walls is 0.58 $\%$, which is very close to the value reported in the PFEM--LS curve. To corroborate that the volume variation in the PFEM--LS curve ($\mathrm{h}_\mathrm{FS} = 1.5$) is due mainly to the no-slip flow, the simulation span is extended up to 3 seconds and results for $h_\mathrm{FS}=1.5$ are reported in Fig.~\ref{Fig:BFIF_3s}. After 1 s, the fluid remains in a sloshing motion with no significant drop detachment or wave breaking, so the main mechanism of volume variation is the no-slip flow along the walls. One can repeat the analytical estimation of the volume variation and verify that these agree with the amplitudes shown in the graph of Fig.~\ref{Fig:BFIF_3s_b}. 

The blue area in Fig.~\ref{Fig:BFIF_3s} is computed for each problem and divided by the simulation span (3 seconds) in order to quantify the average volume variation. Values obtained for each discretisation are summarized in Table \ref{Tab:BFIF_dv}. The non-negligible difference in volume variations between PFEM--AS and PFEM--LS lies in the fact that the latter comprises a remeshing process enriched by the topological information of the previous time step.
 
\begin{table}[t!] 
\captionsetup{width=.75\textwidth}
\caption{Average volume variation in the impact of a viscous fluid disk, computed as the ratio between the blue area of Fig.~\ref{Fig:BFIF_3s} and the simulation span. All values are computed over 3 seconds of simulation.}\label{Tab:BFIF_dv}
\centering
  \begin{tabular}{l C{3cm} C{3cm} C{3cm}}
  \toprule
  	  PFEM scheme & 
  	  $h_\mathrm{FS} = 5.0$ mm & 
  	  $h_\mathrm{FS} = 3.0$ mm &
  	  $h_\mathrm{FS} = 1.5$ mm 
  	  \\
  	  \cmidrule(rl){1-4} 
  	  PFEM--AS & 1.07 $\%$ & 0.41 $\%$ & 0.35 $\%$\\[1ex]
  	  PFEM--LS & 0.64 $\%$ & 0.19 $\%$ & 0.09 $\%$
	  \\
	  \bottomrule
  \end{tabular}
\end{table}

\newpage
\subsection{Fluid discharged from two nozzles impacts on fluid}

\begin{figure}[t!]
\captionsetup[subfigure]{labelformat=empty}
\centering 
	\begin{subfigure}[b]{0.33\textwidth}
		\includegraphics[trim=0 0 0 0,clip,width=1.00\linewidth]{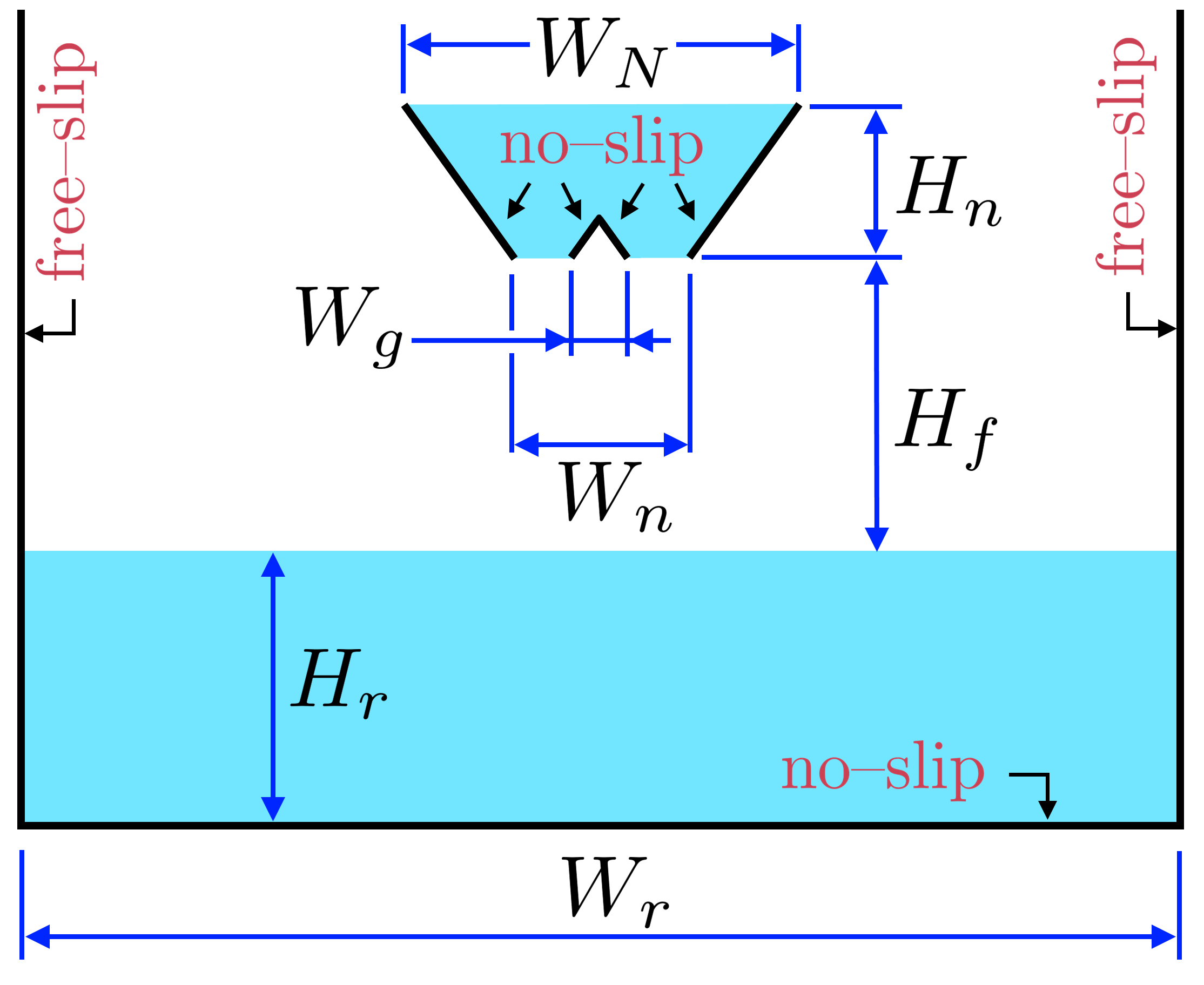}
		\caption{}
		\label{Fig:TN_geo_a}
	\end{subfigure}
	~\hspace{10mm}
	\begin{subfigure}[b]{0.36\textwidth}	
		\includegraphics[trim=0 0 0 0,clip,width=1.00\linewidth]{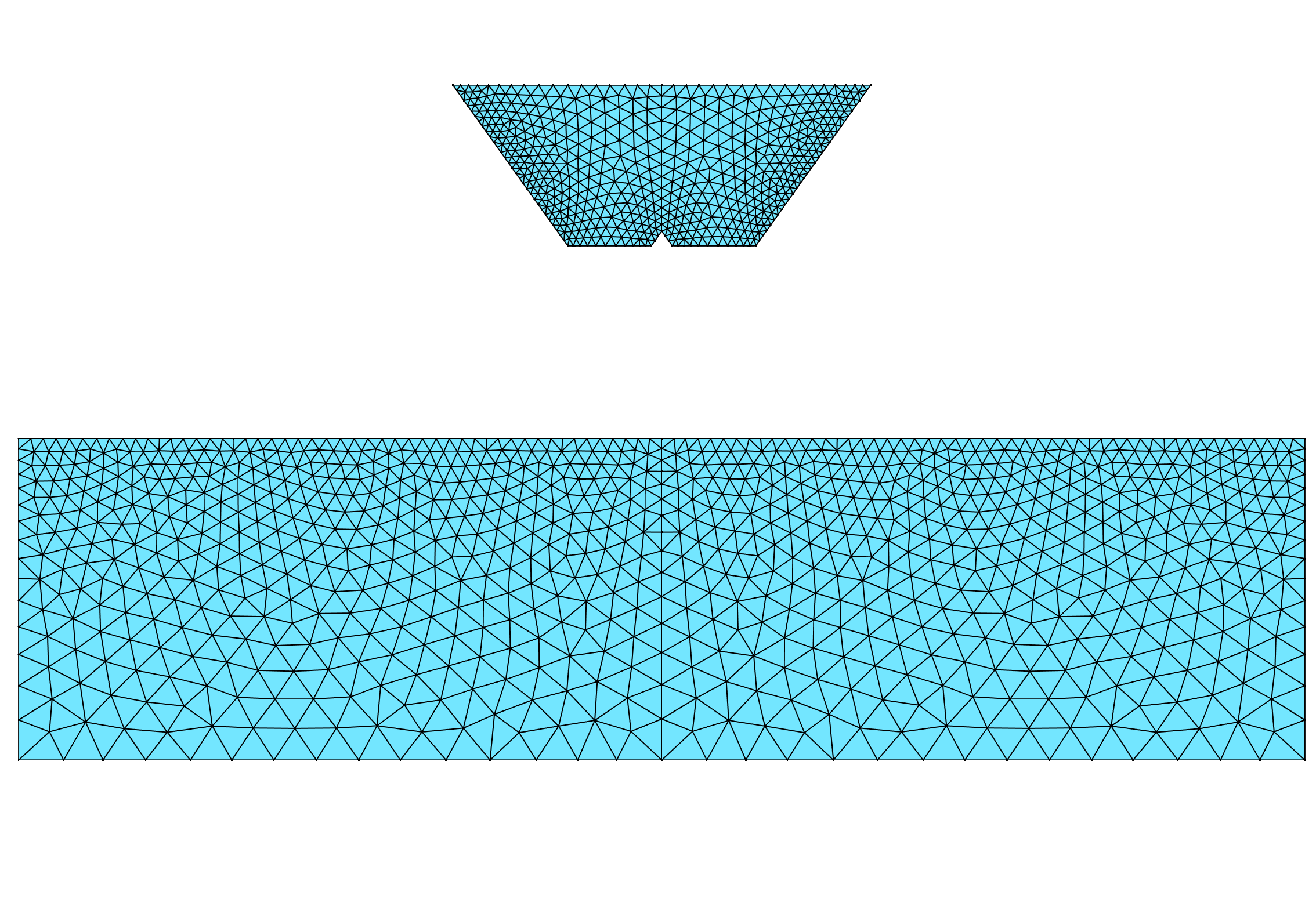}
		\caption{}
		\label{Fig:TN_geo_b}
	\end{subfigure}
	\\
	\vspace{-50mm}
	\hspace{-60mm} \footnotesize{(\textbf{a})} \hspace{70mm} (\textbf{b}) 
	\\
	\vspace{42mm}
\caption{Viscous fluid passing two nozzles. (a) Initial geometry, where $W_N = $, $W_n = 2.925$, $W_g= 0.325$, $H_n = 2.5$, $H_f =3$, $H_r = 5$ and $W_r=20$, all units in m. (b) Space discretisation using $h_\mathrm{FS} = 0.2$ m.
}
\label{Fig:TN_geo}
\end{figure}

This last example consists of a storage tank equipped with two nozzles that is initially filled with fluid, as illustrated in Fig.~\ref{Fig:TN_geo_a}. The problem is inspired in the ``filling of an elastic container" problem that is well reported in the PFEM literature \citep{franci2016bunified}. The tank is located at a height $H_f = 3$ m from the free surface of a fluid within a bigger reservoir. The tank and reservoir contain the same fluid characterized by $\rho = 1000$ kg/m$^3$ and $\mu = 50$ Pa s. Between the fluid and the storage tank equipped with two nozzles, a no--slip condition is imposed on the four sloped walls. Between the fluid and the big reservoir, a free--slip condition is imposed on the walls and no--slip on the bottom of the reservoir. The dimensions of the tank and reservoir are detailed in the caption of Fig.~\ref{Fig:TN_geo}. 
 
The problem is solved using PFEM--AS and PFEM--LS considering three spatial discretisations, where $h_\mathrm{FS}$ = [0.20 , 0.10 , 0.05] m. The maximum element size is $h_\mathrm{max}$ = 3 $h_\mathrm{FS}$ and is imposed from a distance $\mathrm{d}_\mathrm{max} = H_r$. Within the small storage tank, a mesh refinement is performed in the vicinity of the walls where element size is set to $0.5 h_\mathrm{FS}$, simply to reduce the volume variation due to the no-slip condition. The analytical estimation for volume variation reveals that upon complete draining of the storage tank, the volume reduction are 0.075 $\%$, 0.15 $\%$ and 0.30 $\%$ for the discretisations using $h_\mathrm{FS}$ = 0.05 m, $h_\mathrm{FS}$ = 0.1 m, and $h_\mathrm{FS}$ = 0.2 m, respectively.  

\begin{figure}[t!]
\captionsetup[subfigure]{labelformat=empty}
\centering 
	\begin{subfigure}[b]{1.0\textwidth}
		\includegraphics[trim=0 0 0 0,clip,width=1.00\linewidth]{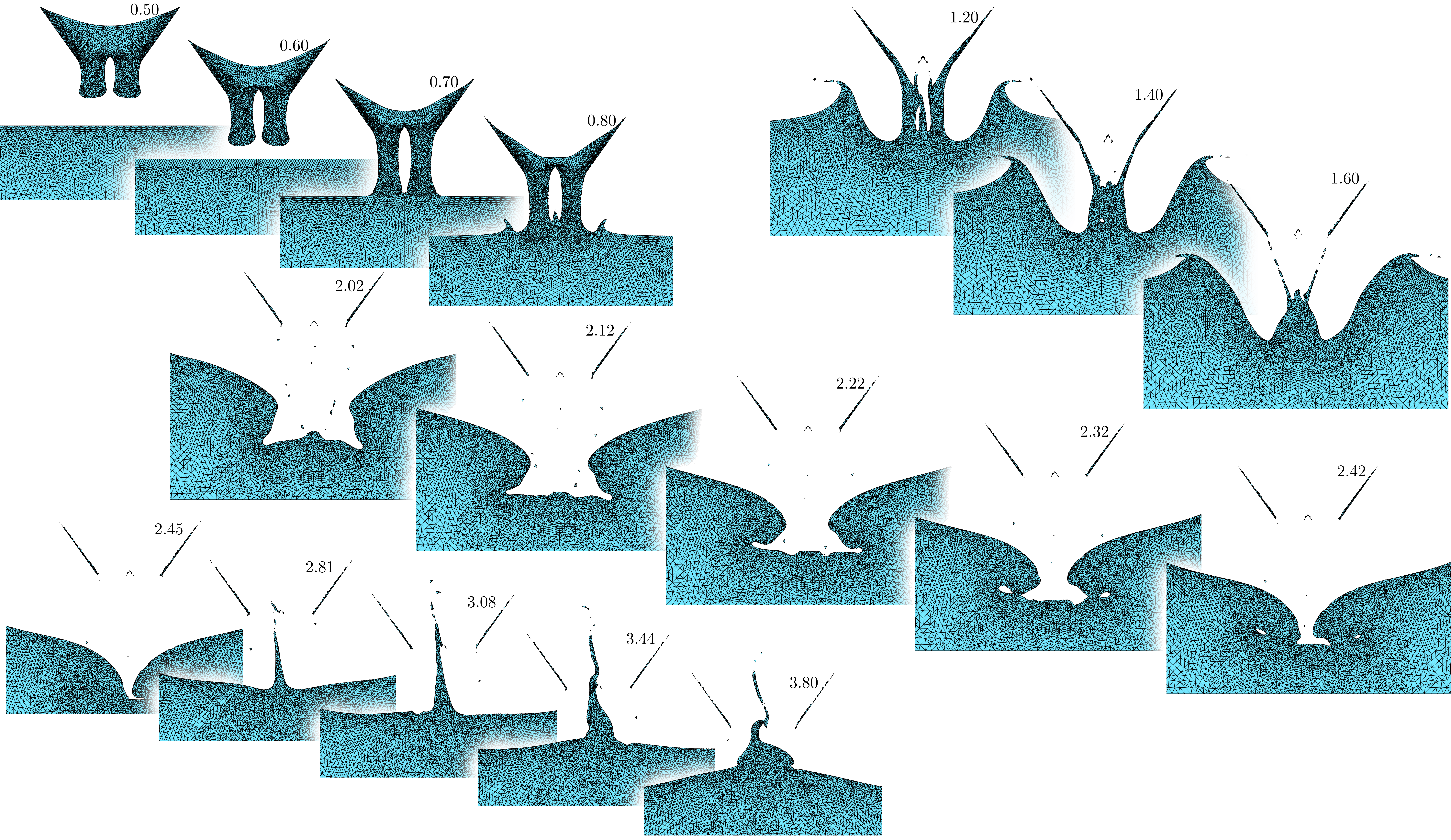}
		\caption{}
		\label{Fig:TNs_a}
	\end{subfigure}
	\\
	\begin{subfigure}[b]{1.0\textwidth} \caption{} \label{Fig:TNs_b}
	\end{subfigure}
	~
	\begin{subfigure}[b]{1.0\textwidth} \caption{} \label{Fig:TNs_c}
	\end{subfigure} 
	~
	\begin{subfigure}[b]{1.0\textwidth} \caption{} \label{Fig:TNs_d}
	\end{subfigure}
	\\
	\vspace{-115mm}
	\hspace{-75mm} 
	\footnotesize{(\textbf{a})}
	\hspace{85mm} 
	(\textbf{b})
	\\
	\vspace{30mm}
	\hspace*{-127mm}	(\textbf{c})
	\\
	\vspace{22mm}
	\hspace*{-160mm}  (\textbf{d})
	\\
	\vspace{32mm}
\caption{Four stages of a fluid discharged from two nozzles impacting on fluid. The time of the snapshot is shown above each picture and is given in s. Snapshots are obtained from the simulation using PFEM--LS and $h_\mathrm{FS} = 0.1$ m.}
\label{Fig:TNs}
\end{figure}

Prior to the volume variation analysis, the simulation is described. In a time span of 4.5 seconds, four stages can be identified, which are shown sequentially in Fig.~\ref{Fig:TNs}. The first represents the flow through the two nozzles and the impact of two columns with the stationary fluid (Fig.~\ref{Fig:TNs_a}). The second phase consists of the complete draining of the small tank, the merging of the two fluid columns and the displacement of a large volume of fluid towards the walls (Fig.~\ref{Fig:TNs_b}). The third phase consists of the fluid flowing back towards the center, which generates the collision of two large waves (Fig.~\ref{Fig:TNs_c}). Finally, a large column of fluid is pushed up and impacts with the walls of the small tank (Fig.~\ref{Fig:TNs_d}). After that, the fluid remains in a sloshing mode without splashing or significant wave breaking. From the PFEM perspective, the first 4 stages present a challenge for volume conservation due to significant fluid/fluid contact (the first 3 stages) and fluid/solid contact (last stage).

\begin{figure}[h!]
\captionsetup[subfigure]{labelformat=empty}
\centering 
	\begin{subfigure}[b]{0.48\textwidth}
		\includegraphics[trim=80 10 100 20,clip,width=1.00\linewidth]{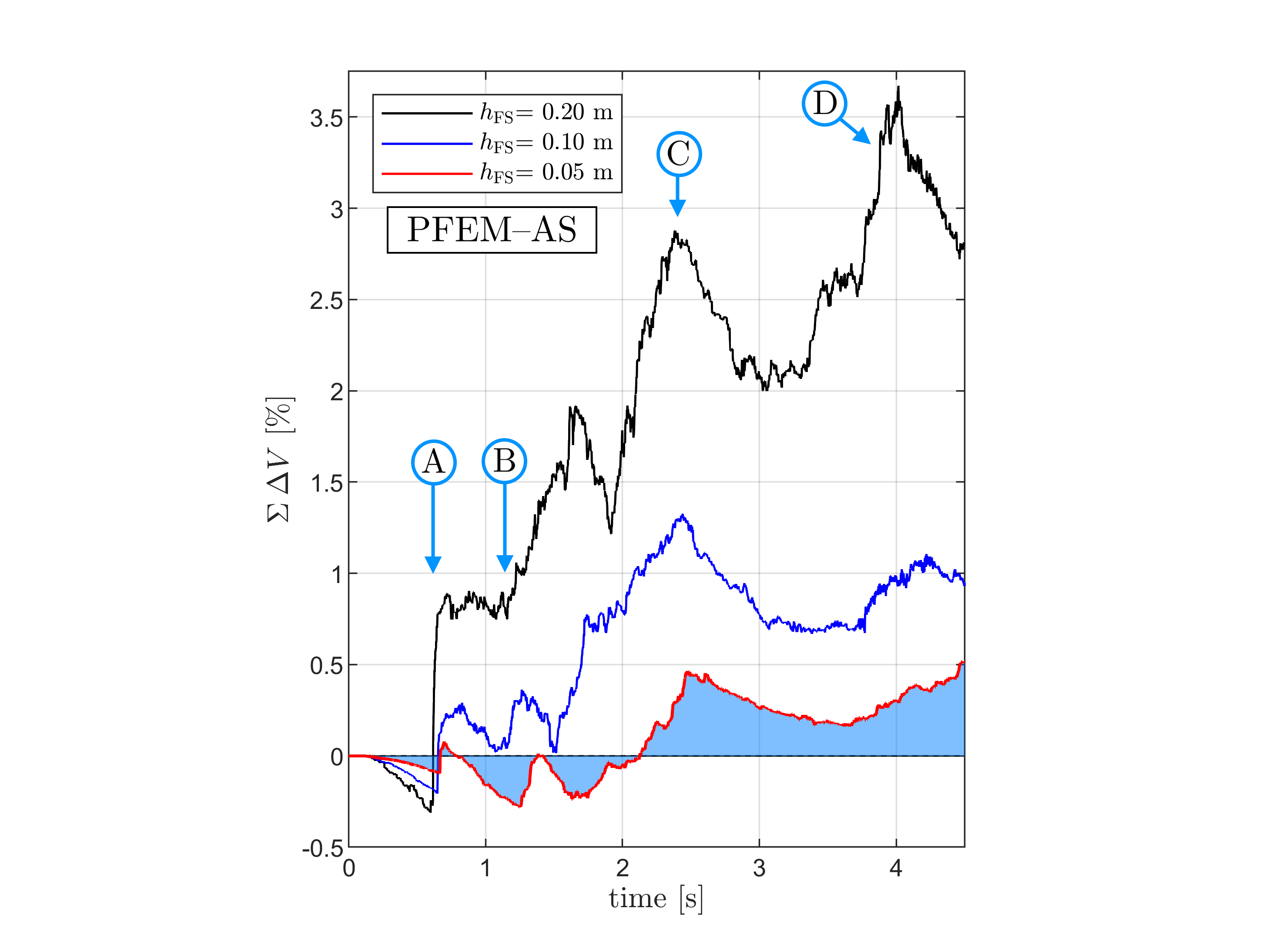}
		\caption{}
		\label{Fig:TN_g_a}
	\end{subfigure}
	~\hspace{4mm}
	\begin{subfigure}[b]{0.48\textwidth}	
		\includegraphics[trim=80 10 100 20,clip,width=1.00\linewidth]{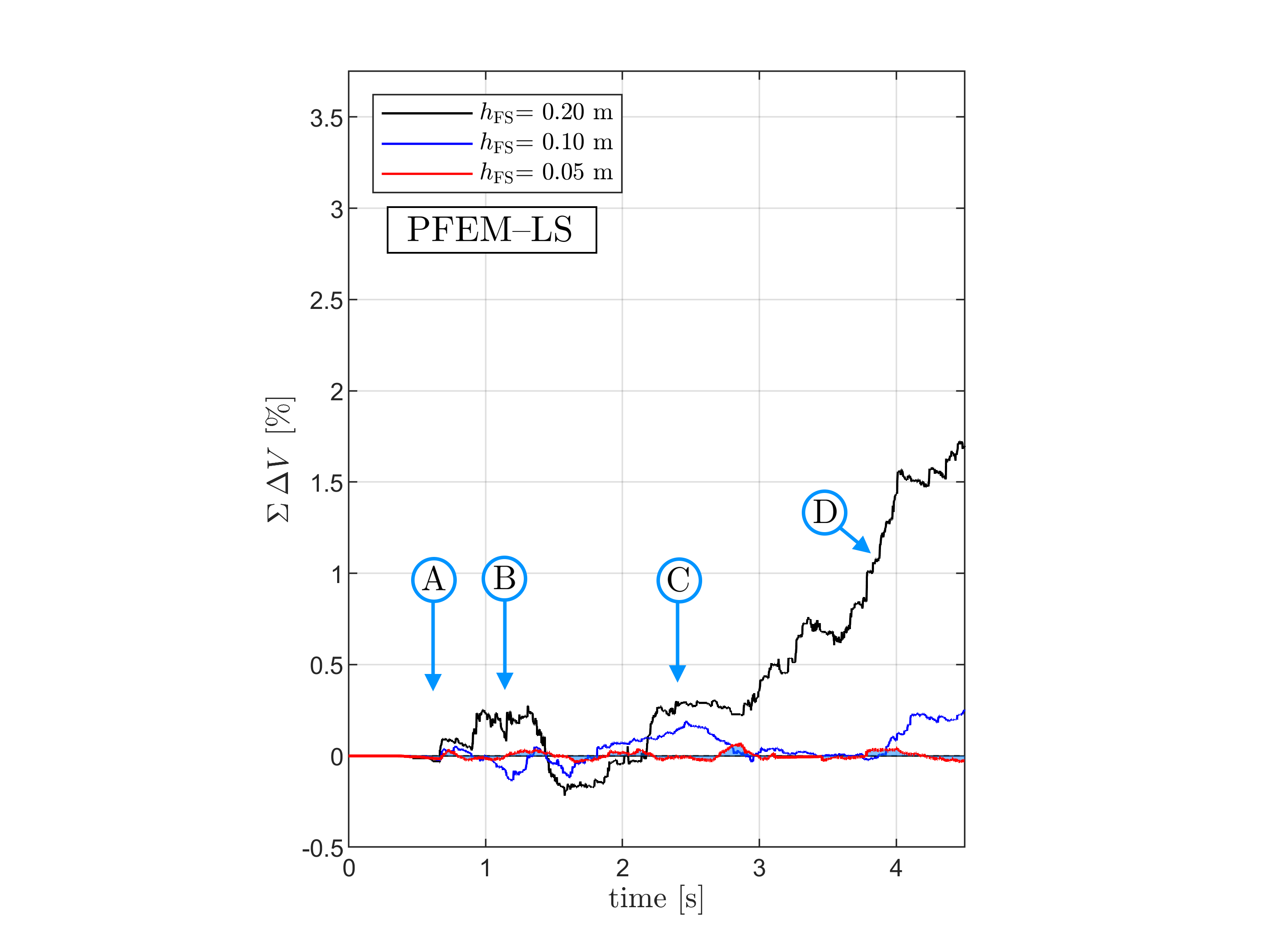}
		\caption{}
		\label{Fig:TN_g_b}
	\end{subfigure}
	\\
	\vspace{-100mm}
	\hspace{-77mm} \footnotesize{(\textbf{a})} \hspace{80mm} (\textbf{b}) 
	\\
	\vspace{91mm}
\caption{Volume variation for the viscous fluid passing two nozzles problem, (a) using PFEM--AS and (b) using PFEM--LS. The blue coloured zone represents the area between the curve with $h_\mathrm{FS}=0.05$ m and the curve $\Sigma\Delta V = 0$.}
\label{Fig:TN_g}
\end{figure}

\begin{figure}[t!]
\captionsetup[subfigure]{labelformat=empty}
\centering 
	\begin{subfigure}[b]{1.0\textwidth}
		\includegraphics[trim=0 0 0 0,clip,width=1.00\linewidth]{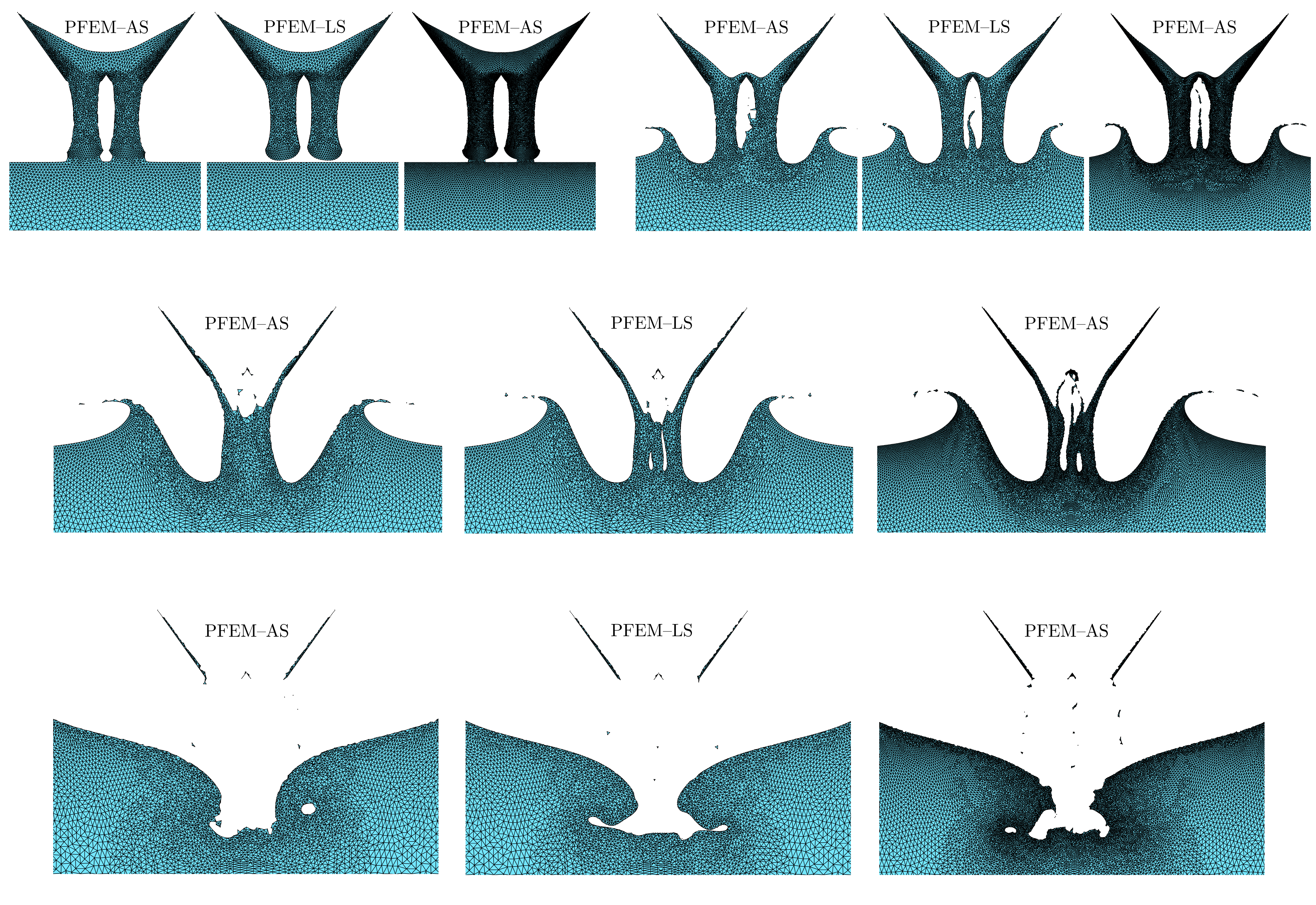}
		\caption{}
		\label{Fig:TN_a}
	\end{subfigure}
	~
	\begin{subfigure}[b]{0.0\textwidth}\caption{}\label{Fig:TN_b}
	\end{subfigure}
	~
	\begin{subfigure}[b]{0.0\textwidth}\caption{}\label{Fig:TN_c}
	\end{subfigure}
	~
	\begin{subfigure}[b]{0.0\textwidth}\caption{}\label{Fig:TN_d}
	\end{subfigure}
	\\
	\vspace{-93mm}
	\hspace{-10mm} 
	\footnotesize{(\textbf{a}) $t$= 0.67 , $\Sigma \Delta V= [\;\;0.128 \;\;,\;\; -0.001 \;\;,\;\; 0.016 \;\;]$ $\%$}
	\hspace{20mm} 
	(\textbf{b}) $t$= 0.94 , $\Sigma \Delta V= [\;\;0.159 \;\;,\;\; -0.011 \;\;,\;\; -0.102 \;\;]$ $\%$
	\\
	\vspace{35mm}
	\hspace{-4mm} 
	(\textbf{c}) $t$= 1.29 , $\Sigma \Delta V= [\;\;0.320 \;\;,\;\; -0.091 \;\;,\;\; -0.184 \;\;]$ $\%$ 
	\vspace{40mm}
	\\
	\hspace{-5mm} 
	(\textbf{d}) $t$= 2.29 , $\Sigma \Delta V= [\;\;1.105 \;\;,\;\; 0.101 \;\;,\;\; 0.169 \;\;]$ $\%$ 	 
	\vspace{2mm}
\caption{Simulation snapshots of a fluid discharged from two nozzles impacting on a fluid at rest. Each subfigure compares simulations obtained with PFEM--AS and $h_\mathrm{FS} = 0.1$ m (left), PFEM--LS and $h_\mathrm{FS} = 0.1$ m (middle), and PFEM--AS and $h_\mathrm{FS} = 0.05$ m (right). The snapshot time and total volume variation due to remeshing are detailed below the pictures. Components of the volume variation vector refer to the snapshot in the [left , middle , right]. The reader is referred to \citep{YoutubeAll} for simulation videos.
}
\label{Fig:TN}
\end{figure}

The volume variation for the 3 discretisations and the 2 PFEM schemes are summarized in Fig.~\ref{Fig:TN_g}. As in the previous examples, a clear difference is observed between PFEM--AS and PFEM--LS, where the latter stands out with significantly smaller volume variations. To underline the benefits of PFEM--LS over PFEM--AS, the different sources of volume variation during the simulation are detailed. For this purpose, pertinent times are labelled in Fig.~\ref{Fig:TN_g} with letters from \circled{A} to \circled{D}. In addition, comparative snapshots are given in Fig.~\ref{Fig:TN}. This figure compares 3 simulations, PFEM--AS using $h_\mathrm{FS}$ = 0.10 m, PFEM--LS using $h_\mathrm{FS}$ = 0.10 m, and PFEM--AS using $h_\mathrm{FS}$ = 0.05 m. The latter is included to show that topological details captured by LS and $h_\mathrm{FS}$ = 0.10 m, are similar to those captured by AS and $h_\mathrm{FS}$ = 0.05 m. The sources of volume variation in the labelled times are listed below.

\begin{itemize}
\item[0 to \circled{A} :] From  the beginning of the simulation until A, a mass reduction is observed in PFEM--AS due to the stretching of the free surface in the two columns of fluid flowing out from the nozzles. Since PFEM--LS avoids the degradation of the free surface \EF{by refining the highly stretched facets}, it presents smaller volume loss in this period.

\item[\circled{A} :] The impact of the two fluid columns against the free surface of the stationary fluid takes place, which increases the volume of the system due to the creation of elements modelling the contact. This is less noticeable in the Level-Set based scheme, as illustrated in Fig.~\ref{Fig:TN_a}.  

\item[\circled{A} to \circled{B} :] Significant volume loss occurs due to the stretching of the free surface in the two fluid columns and the generation of two large waves. However, volume creation is also present due to a third column of fluid that merges to the other two due to the small gap between them, as shown in Fig.~\ref{Fig:TN_b}. These volume addition and removal mechanisms can compensate each other in some cases, as in PFEM--AS and $h_\mathrm{FS} = 0.20 $ m.

\item[\circled{B} to \circled{C} :] The volume addition from the merging of the three fluid columns dominates, as illustrated in Fig.~\ref{Fig:TN_c}. Since PFEM--LS is better at capturing topological details, it shows less mass variation at this stage than the PFEM--AS scheme.

\item[\circled{C} :]  The breaking of two waves and the impact between them takes place. This process creates volume due to significant fluid/fluid contact. In addition, the collapse of the waves occurs early in the PFEM--AS, as shown in Fig.~\ref{Fig:TN_d}. Since PFEM--LS manages the fluid/fluid contact better, it has considerably less volume addition in this period.

\item[\circled{C} to \circled{D} :] The collision of the two waves lifts a large column of fluid that impacts the central part of the small tank. Here, volume addition and removal mechanisms are present. On the one hand, the stretching of the free surface of the large column reduces the volume, and on the other hand the impact of the column with the small tank increases the volume due to the no-slip condition.

\item[\circled{D} to 4.5 :] The large fluid column descends and impacts again with the small tank, creating volume due to the no-slip condition. This is noted in both PFEM schemes.
\end{itemize}

In summary, the significant reduction in volume variation when using a Level--Set function during the remeshing process is due to an improved control of the fluid/fluid contact. To quantify the improvement obtained with PFEM--LS, the average volume variations are reported for each simulation of the fluid discharged from two nozzles. These are gathered in Table \ref{Tab:TN_dv}.

\begin{table}[t!] 
\captionsetup{width=.75\textwidth}
\caption{Average volume variation during the discharge of fluid from two nozzles impacting on fluid. All values are computed over 4.5 seconds of simulation.}\label{Tab:TN_dv}
\centering
  \begin{tabular}{l C{3cm} C{3cm} C{3cm}}
  \toprule
  	  PFEM scheme & 
  	  $h_\mathrm{FS} = 0.20$ m & 
  	  $h_\mathrm{FS} = 0.10$ m &
  	  $h_\mathrm{FS} = 0.05$ m 
  	  \\
  	  \cmidrule(rl){1-4} 
  	  PFEM--AS & 1.820 $\%$ & 0.618 $\%$ & 0.192 $\%$\\[1ex]
  	  PFEM--LS & 0.431 $\%$ & 0.061 $\%$ & 0.016 $\%$
	  \\
	  \bottomrule
  \end{tabular}
\end{table}

\section{Conclusions}\label{sec:Conclusions}

The Particle Finite Element Method (PFEM) has demonstrated adaptability to various dynamic problems involving large deformations and topological changes that are difficult to predict. The essence of its success is the combination of the well-developed FEM in a Lagrangian framework and a remeshing strategy that avoids mesh degradation. To keep a low computational cost during remeshing, PFEM resorts to a Delaunay triangulation and the Alpha-Shape (AS) algorithm. However, AS introduces a list of shortcomings during simulation, where the non-conservation of mass stands out. This work has proposed the use of a Level-Set (LS) function during the remeshing process, which allows to reduce mass variations in PFEM owing to three main improvements with respect to the classical AS--based approach. First, the use of a LS function allows to avoid the degradation of the free surface and the associated problems, such as the mass loss and numerical artifacts like the ``\EF{surfing}" particles. Second, the LS function allows to reduce the creation of elements during the contact between bodies, in particular the fluid/fluid contact. Third, by providing a better fluid/fluid contact, the LS function helps to represent drops using elements instead of massless particles, which allows for better conservation of mass and energy. 

\EF{Despite the significant improvements, the proposed LS-based remeshing approach does not reduce the mass variation in the no-slip flow of a fluid over a solid surface. This intrinsic issue of the PFEM formulation along with the accuracy of the LS function, computational performance,} and the 3D extension of the method will be the subject of future research.

\section*{Financial disclosure}
This work was supported by the ALFEWELD project --- \textit{Amélioration et modélisation du FMB (Friction Melt Bonding) pour le soudage par recouvrement de l’aluminium et de l’acier} --- (convention 1710162) funded by the WALInnov program of the Walloon Region of Belgium.

\section*{Conflict of interest}
The authors declare no potential conflict of interests.



\bibliographystyle{elsarticle-num-names} 
\bibliography{references.bib}

\begin{thebibliography}{32}
\expandafter\ifx\csname natexlab\endcsname\relax\def\natexlab#1{#1}\fi
\providecommand{\url}[1]{\texttt{#1}}
\providecommand{\href}[2]{#2}
\providecommand{\path}[1]{#1}
\providecommand{\DOIprefix}{doi:}
\providecommand{\ArXivprefix}{arXiv:}
\providecommand{\URLprefix}{URL: }
\providecommand{\Pubmedprefix}{pmid:}
\providecommand{\doi}[1]{\href{http://dx.doi.org/#1}{\path{#1}}}
\providecommand{\Pubmed}[1]{\href{pmid:#1}{\path{#1}}}
\providecommand{\bibinfo}[2]{#2}
\ifx\xfnm\relax \def\xfnm[#1]{\unskip,\space#1}\fi
\bibitem[{Idelsohn et~al.(2004)Idelsohn, O{\~n}ate, and
  Pin}]{idelsohn2004particle}
\bibinfo{author}{S.~R. Idelsohn}, \bibinfo{author}{E.~O{\~n}ate},
  \bibinfo{author}{F.~D. Pin},
\newblock \bibinfo{title}{The particle finite element method: a powerful tool
  to solve incompressible flows with free-surfaces and breaking waves},
\newblock \bibinfo{journal}{International journal for numerical methods in
  engineering} \bibinfo{volume}{61} (\bibinfo{year}{2004})
  \bibinfo{pages}{964--989}.
\bibitem[{Edelsbrunner and M{\"u}cke(1994)}]{edelsbrunner1994three}
\bibinfo{author}{H.~Edelsbrunner}, \bibinfo{author}{E.~P. M{\"u}cke},
\newblock \bibinfo{title}{Three-dimensional alpha shapes},
\newblock \bibinfo{journal}{ACM Transactions on Graphics (TOG)}
  \bibinfo{volume}{13} (\bibinfo{year}{1994}) \bibinfo{pages}{43--72}.
\bibitem[{Carbonell et~al.(2020)Carbonell, Rodr{\'\i}guez, and
  O{\~n}ate}]{carbonell2020modelling}
\bibinfo{author}{J.~M. Carbonell}, \bibinfo{author}{J.~Rodr{\'\i}guez},
  \bibinfo{author}{E.~O{\~n}ate},
\newblock \bibinfo{title}{Modelling {3D} metal cutting problems with the
  particle finite element method},
\newblock \bibinfo{journal}{Computational Mechanics} \bibinfo{volume}{66}
  (\bibinfo{year}{2020}) \bibinfo{pages}{603--624}.
\bibitem[{Cerquaglia et~al.(2019)Cerquaglia, Thomas, Boman, Terrapon, and
  Ponthot}]{cerquaglia2019fully}
\bibinfo{author}{M.~L. Cerquaglia}, \bibinfo{author}{D.~Thomas},
  \bibinfo{author}{R.~Boman}, \bibinfo{author}{V.~Terrapon},
  \bibinfo{author}{J.-P. Ponthot},
\newblock \bibinfo{title}{A fully partitioned {Lagrangian} framework for {FSI}
  problems characterized by free surfaces, large solid deformations and
  displacements, and strong added-mass effects},
\newblock \bibinfo{journal}{Computer Methods in Applied Mechanics and
  Engineering} \bibinfo{volume}{348} (\bibinfo{year}{2019})
  \bibinfo{pages}{409--442}.
\bibitem[{Meduri et~al.(2022)Meduri, Cremonesi, Frangi, and
  Perego}]{meduri2022lagrangian}
\bibinfo{author}{S.~Meduri}, \bibinfo{author}{M.~Cremonesi},
  \bibinfo{author}{A.~Frangi}, \bibinfo{author}{U.~Perego},
\newblock \bibinfo{title}{A {Lagrangian} fluid--structure interaction approach
  for the simulation of airbag deployment},
\newblock \bibinfo{journal}{Finite Elements in Analysis and Design}
  \bibinfo{volume}{198} (\bibinfo{year}{2022}) \bibinfo{pages}{103659}.
\bibitem[{Bobach et~al.(2021)Bobach, Boman, Celentano, Terrapon, and
  Ponthot}]{bobach2021phase}
\bibinfo{author}{B.-J. Bobach}, \bibinfo{author}{R.~Boman},
  \bibinfo{author}{D.~Celentano}, \bibinfo{author}{V.~Terrapon},
  \bibinfo{author}{J.-P. Ponthot},
\newblock \bibinfo{title}{Simulation of the {Marangoni} effect and phase change
  using the particle finite element method},
\newblock \bibinfo{journal}{Applied Sciences} \bibinfo{volume}{11}
  (\bibinfo{year}{2021}) \bibinfo{pages}{11893}.
\bibitem[{Sengani and Mulenga(2020)}]{sengani2020review}
\bibinfo{author}{F.~Sengani}, \bibinfo{author}{F.~Mulenga},
\newblock \bibinfo{title}{A review on the application of particle finite
  element methods ({PFEM}) to cases of landslides},
\newblock \bibinfo{journal}{International Journal of Geotechnical Engineering}
  (\bibinfo{year}{2020}) \bibinfo{pages}{1--15}.
\bibitem[{Cremonesi et~al.(2020)Cremonesi, Franci, Idelsohn, and
  O{\~n}ate}]{cremonesi2020state}
\bibinfo{author}{M.~Cremonesi}, \bibinfo{author}{A.~Franci},
  \bibinfo{author}{S.~Idelsohn}, \bibinfo{author}{E.~O{\~n}ate},
\newblock \bibinfo{title}{A state of the art review of the particle finite
  element method ({PFEM})},
\newblock \bibinfo{journal}{Archives of Computational Methods in Engineering}
  \bibinfo{volume}{27} (\bibinfo{year}{2020}) \bibinfo{pages}{1709--1735}.
\bibitem[{Franci and Cremonesi(2017)}]{franci2017effect}
\bibinfo{author}{A.~Franci}, \bibinfo{author}{M.~Cremonesi},
\newblock \bibinfo{title}{On the effect of standard {PFEM} remeshing on volume
  conservation in free-surface fluid flow problems},
\newblock \bibinfo{journal}{Computational Particle Mechanics}
  \bibinfo{volume}{4} (\bibinfo{year}{2017}) \bibinfo{pages}{331--343}.
\bibitem[{Cerquaglia(2019)}]{cerquaglia2019development}
\bibinfo{author}{M.-L. Cerquaglia}, \bibinfo{title}{Development of a
  fully-partitioned {PFEM-FEM} approach for fluid-structure interaction
  problems characterized by free surfaces, large solid deformations, and strong
  added-mass effects}, volume \bibinfo{volume}{{PhD} thesis, University of
  Liege}, \bibinfo{year}{2019}.
  \bibinfo{note}{\url{https://hdl.handle.net/2268/233166}}.
\bibitem[{Cerquaglia et~al.(2017)Cerquaglia, Deli{\'e}ge, Boman, Terrapon, and
  Ponthot}]{cerquaglia2017free}
\bibinfo{author}{M.~L. Cerquaglia}, \bibinfo{author}{G.~Deli{\'e}ge},
  \bibinfo{author}{R.~Boman}, \bibinfo{author}{V.~Terrapon},
  \bibinfo{author}{J.-P. Ponthot},
\newblock \bibinfo{title}{Free-slip boundary conditions for simulating
  free-surface incompressible flows through the particle finite element
  method},
\newblock \bibinfo{journal}{International Journal for Numerical Methods in
  Engineering} \bibinfo{volume}{110} (\bibinfo{year}{2017})
  \bibinfo{pages}{921--946}.
\bibitem[{Rodr{\'\i}guez et~al.(2017)Rodr{\'\i}guez, Carbonell, Cante, and
  Oliver}]{rodriguez2017continuous}
\bibinfo{author}{J.~M. Rodr{\'\i}guez}, \bibinfo{author}{J.~M. Carbonell},
  \bibinfo{author}{J.~Cante}, \bibinfo{author}{J.~Oliver},
\newblock \bibinfo{title}{Continuous chip formation in metal cutting processes
  using the particle finite element method ({PFEM})},
\newblock \bibinfo{journal}{International Journal of Solids and Structures}
  \bibinfo{volume}{120} (\bibinfo{year}{2017}) \bibinfo{pages}{81--102}.
\bibitem[{Falla et~al.(2023)Falla, Bobach, Boman, Ponthot, and
  Terrapon}]{falla2022Mesh}
\bibinfo{author}{R.~Falla}, \bibinfo{author}{B.-J. Bobach},
  \bibinfo{author}{R.~Boman}, \bibinfo{author}{J.-P. Ponthot},
  \bibinfo{author}{V.~E. Terrapon},
\newblock \bibinfo{title}{Mesh adaption for two-dimensional bounded and
  free-surface flows with the particle finite element method},
\newblock \bibinfo{journal}{Computational Particle Mechanics}
  (\bibinfo{year}{2023}) \bibinfo{pages}{1--28}.
\bibitem[{Osher and Sethian(1988)}]{osher1988fronts}
\bibinfo{author}{S.~Osher}, \bibinfo{author}{J.~A. Sethian},
\newblock \bibinfo{title}{Fronts propagating with curvature-dependent speed:
  Algorithms based on hamilton-jacobi formulations},
\newblock \bibinfo{journal}{Journal of computational physics}
  \bibinfo{volume}{79} (\bibinfo{year}{1988}) \bibinfo{pages}{12--49}.
\bibitem[{Cremers et~al.(2007)Cremers, Rousson, and
  Deriche}]{cremers2007review}
\bibinfo{author}{D.~Cremers}, \bibinfo{author}{M.~Rousson},
  \bibinfo{author}{R.~Deriche},
\newblock \bibinfo{title}{A review of statistical approaches to level set
  segmentation: integrating color, texture, motion and shape},
\newblock \bibinfo{journal}{International journal of computer vision}
  \bibinfo{volume}{72} (\bibinfo{year}{2007}) \bibinfo{pages}{195--215}.
\bibitem[{Osher and Paragios(2003)}]{osher2003geometric}
\bibinfo{author}{S.~Osher}, \bibinfo{author}{N.~Paragios},
  \bibinfo{title}{Geometric level set methods in imaging, vision, and
  graphics}, \bibinfo{publisher}{Springer Science \& Business Media},
  \bibinfo{year}{2003}.
\bibitem[{Van~Dijk et~al.(2013)Van~Dijk, Maute, Langelaar, and
  Van~Keulen}]{van2013level}
\bibinfo{author}{N.~P. Van~Dijk}, \bibinfo{author}{K.~Maute},
  \bibinfo{author}{M.~Langelaar}, \bibinfo{author}{F.~Van~Keulen},
\newblock \bibinfo{title}{Level-set methods for structural topology
  optimization: a review},
\newblock \bibinfo{journal}{Structural and Multidisciplinary Optimization}
  \bibinfo{volume}{48} (\bibinfo{year}{2013}) \bibinfo{pages}{437--472}.
\bibitem[{Stolarska et~al.(2001)Stolarska, Chopp, Mo{\"e}s, and
  Belytschko}]{stolarska2001modelling}
\bibinfo{author}{M.~Stolarska}, \bibinfo{author}{D.~L. Chopp},
  \bibinfo{author}{N.~Mo{\"e}s}, \bibinfo{author}{T.~Belytschko},
\newblock \bibinfo{title}{Modelling crack growth by level sets in the extended
  finite element method},
\newblock \bibinfo{journal}{International journal for numerical methods in
  Engineering} \bibinfo{volume}{51} (\bibinfo{year}{2001})
  \bibinfo{pages}{943--960}.
\bibitem[{Becker et~al.(2015)Becker, Idelsohn, and
  O{\~n}ate}]{becker2015unified}
\bibinfo{author}{P.~Becker}, \bibinfo{author}{S.~R. Idelsohn},
  \bibinfo{author}{E.~O{\~n}ate},
\newblock \bibinfo{title}{A unified monolithic approach for multi-fluid flows
  and fluid--structure interaction using the particle finite element method
  with fixed mesh},
\newblock \bibinfo{journal}{Computational Mechanics} \bibinfo{volume}{55}
  (\bibinfo{year}{2015}) \bibinfo{pages}{1091--1104}.
\bibitem[{Grooss and Hesthaven(2006)}]{grooss2006level}
\bibinfo{author}{J.~Grooss}, \bibinfo{author}{J.~S. Hesthaven},
\newblock \bibinfo{title}{A level set discontinuous galerkin method for free
  surface flows},
\newblock \bibinfo{journal}{Computer Methods in Applied Mechanics and
  Engineering} \bibinfo{volume}{195} (\bibinfo{year}{2006})
  \bibinfo{pages}{3406--3429}.
\bibitem[{Gibou et~al.(2018)Gibou, Fedkiw, and Osher}]{gibou2018review}
\bibinfo{author}{F.~Gibou}, \bibinfo{author}{R.~Fedkiw},
  \bibinfo{author}{S.~Osher},
\newblock \bibinfo{title}{A review of level-set methods and some recent
  applications},
\newblock \bibinfo{journal}{Journal of Computational Physics}
  \bibinfo{volume}{353} (\bibinfo{year}{2018}) \bibinfo{pages}{82--109}.
\bibitem[{Chen(2018)}]{chen2018thermomechanical}
\bibinfo{author}{Q.~Chen}, \bibinfo{title}{Thermomechanical numerical modelling
  of additive manufacturing by selective laser melting of powder bed:
  Application to ceramic materials}, Ph.D. thesis, Universit{\'e} Paris
  sciences et lettres, \bibinfo{year}{2018}.
\bibitem[{Fern{\'a}ndez et~al.(2023)Fern{\'a}ndez, F{\'e}vrier, Lacroix, Boman,
  and Ponthot}]{fernandez2022}
\bibinfo{author}{E.~Fern{\'a}ndez}, \bibinfo{author}{S.~F{\'e}vrier},
  \bibinfo{author}{M.~Lacroix}, \bibinfo{author}{R.~Boman},
  \bibinfo{author}{J.-P. Ponthot},
\newblock \bibinfo{title}{Generalized-$\alpha$ scheme in the pfem for
  velocity-pressure and displacement-pressure formulations of the
  incompressible navier--stokes equations},
\newblock \bibinfo{journal}{International Journal for Numerical Methods in
  Engineering} \bibinfo{volume}{124} (\bibinfo{year}{2023})
  \bibinfo{pages}{40--79}.
\bibitem[{Carr et~al.(2001)Carr, Beatson, Cherrie, Mitchell, Fright, McCallum,
  and Evans}]{carr2001reconstruction}
\bibinfo{author}{J.~C. Carr}, \bibinfo{author}{R.~K. Beatson},
  \bibinfo{author}{J.~B. Cherrie}, \bibinfo{author}{T.~J. Mitchell},
  \bibinfo{author}{W.~R. Fright}, \bibinfo{author}{B.~C. McCallum},
  \bibinfo{author}{T.~R. Evans},
\newblock \bibinfo{title}{Reconstruction and representation of {3D} objects
  with radial basis functions},
\newblock in: \bibinfo{booktitle}{Proceedings of the 28th annual conference on
  Computer graphics and interactive techniques}, \bibinfo{year}{2001}, pp.
  \bibinfo{pages}{67--76}.
\bibitem[{Belytschko et~al.(2003)Belytschko, Parimi, Mo{\"e}s, Sukumar, and
  Usui}]{belytschko2003structured}
\bibinfo{author}{T.~Belytschko}, \bibinfo{author}{C.~Parimi},
  \bibinfo{author}{N.~Mo{\"e}s}, \bibinfo{author}{N.~Sukumar},
  \bibinfo{author}{S.~Usui},
\newblock \bibinfo{title}{Structured extended finite element methods for solids
  defined by implicit surfaces},
\newblock \bibinfo{journal}{International journal for numerical methods in
  engineering} \bibinfo{volume}{56} (\bibinfo{year}{2003})
  \bibinfo{pages}{609--635}.
\bibitem[{Zalesak(1979)}]{zalesak1979fully}
\bibinfo{author}{S.~T. Zalesak},
\newblock \bibinfo{title}{Fully multidimensional flux-corrected transport
  algorithms for fluids},
\newblock \bibinfo{journal}{Journal of computational physics}
  \bibinfo{volume}{31} (\bibinfo{year}{1979}) \bibinfo{pages}{335--362}.
\bibitem[{Hartmann et~al.(2010)Hartmann, Meinke, and
  Schr{\"o}der}]{hartmann2010constrained}
\bibinfo{author}{D.~Hartmann}, \bibinfo{author}{M.~Meinke},
  \bibinfo{author}{W.~Schr{\"o}der},
\newblock \bibinfo{title}{The constrained reinitialization equation for level
  set methods},
\newblock \bibinfo{journal}{Journal of computational physics}
  \bibinfo{volume}{229} (\bibinfo{year}{2010}) \bibinfo{pages}{1514--1535}.
\bibitem[{Ryzhakov et~al.(2017)Ryzhakov, Jarauta, Secanell, and
  Pons-Prats}]{ryzhakov2017application}
\bibinfo{author}{P.~B. Ryzhakov}, \bibinfo{author}{A.~Jarauta},
  \bibinfo{author}{M.~Secanell}, \bibinfo{author}{J.~Pons-Prats},
\newblock \bibinfo{title}{On the application of the pfem to droplet dynamics
  modeling in fuel cells},
\newblock \bibinfo{journal}{Computational Particle Mechanics}
  \bibinfo{volume}{4} (\bibinfo{year}{2017}) \bibinfo{pages}{285--295}.
\bibitem[{Hieber and Koumoutsakos(2005)}]{hieber2005lagrangian}
\bibinfo{author}{S.~E. Hieber}, \bibinfo{author}{P.~Koumoutsakos},
\newblock \bibinfo{title}{A {Lagrangian} particle level set method},
\newblock \bibinfo{journal}{Journal of Computational Physics}
  \bibinfo{volume}{210} (\bibinfo{year}{2005}) \bibinfo{pages}{342--367}.
\bibitem[{Henri et~al.(2022)Henri, Coquerelle, and
  Lubin}]{henri2022geometrical}
\bibinfo{author}{F.~Henri}, \bibinfo{author}{M.~Coquerelle},
  \bibinfo{author}{P.~Lubin},
\newblock \bibinfo{title}{Geometrical level set reinitialization using closest
  point method and kink detection for thin filaments, topology changes and
  two-phase flows},
\newblock \bibinfo{journal}{Journal of Computational Physics}
  \bibinfo{volume}{448} (\bibinfo{year}{2022}) \bibinfo{pages}{110704}.
\bibitem[{Fern\'{a}ndez(2023)}]{YoutubeAll}
\bibinfo{author}{E.~Fern\'{a}ndez}, \bibinfo{title}{\textit{PFEM\_LS} [video
  list]}, \bibinfo{year}{Accessed February 20, 2023}.
  \bibinfo{note}{\url{https://youtube.com/playlist?list=PLDU2T_jKI__8_voiYUwkZq3YGM7e9TVXF}}.
\bibitem[{Franci et~al.(2016)Franci, O{\~n}ate, and
  Carbonell}]{franci2016bunified}
\bibinfo{author}{A.~Franci}, \bibinfo{author}{E.~O{\~n}ate},
  \bibinfo{author}{J.~M. Carbonell},
\newblock \bibinfo{title}{Unified {Lagrangian} formulation for solid and fluid
  mechanics and {FSI} problems},
\newblock \bibinfo{journal}{Computer Methods in Applied Mechanics and
  Engineering} \bibinfo{volume}{298} (\bibinfo{year}{2016})
  \bibinfo{pages}{520--547}.

\end{thebibliography}





\end{document}